\documentclass[times]{aastex63}
\usepackage{amsmath}
\usepackage{color}
\usepackage{hyperref}

\begin{document}
\title{NuSTAR Observations of 52 Compton-thick Active Galactic Nuclei Selected by the Swift/BAT All-sky Hard X-Ray Survey}
\correspondingauthor{Atsushi Tanimoto}
\email{atsushi.tanimoto@phys.s.u-tokyo.ac.jp}
\author[0000-0002-0114-5581]{Atsushi Tanimoto}
\affiliation{Department of Physics, The University of Tokyo, Tokyo 113--0033, Japan}
\author[0000-0001-7821-6715]{Yoshihiro Ueda}
\affiliation{Department of Astronomy, Kyoto University, Kyoto 606--8502, Japan}
\author[0000-0003-2670-6936]{Hirokazu Odaka}
\affiliation{Department of Physics, The University of Tokyo, Tokyo 113--0033, Japan}
\affiliation{Research Center for the Early Universe, The University of Tokyo, Tokyo 113-0033, Japan}
\affiliation{Kavli IPMU (WPI), UTIAS, The University of Tokyo, Chiba 277--8583, Japan}
\author[0000-0002-9754-3081]{Satoshi Yamada}
\affiliation{Department of Astronomy, Kyoto University, Kyoto 606--8502, Japan}
\author[0000-0001-5231-2645]{Claudio Ricci}
\affiliation{Núcleo de Astronomía de la Facultad de Ingeniería, Universidad Diego Portales, Av. Ejército Libertador 441, Santiago, Chile}
\affiliation{Kavli Institute for Astronomy and Astrophysics, Peking University, Beijing 100871, China}
\begin{abstract}
We present the systematic broadband X-ray spectral analysis of 52 Compton-thick ($24 \leq \log N_{\mathrm{H}}^{\mathrm{LOS}}/\mathrm{cm}^{-2}$) active galactic nucleus (CTAGN) candidates selected by the Swift/BAT all-sky hard X-ray survey observed with Chandra, XMM--Newton, Swift/XRT, Suzaku, and NuSTAR. The XMM--Newton data of 10 objects and the NuSTAR data of 15 objects are published for the first time. We use an X-ray spectral model from a clumpy torus (XClumpy) to determine the torus properties. As a result, the hydrogen column density along the line of sight $N_{\mathrm{H}}^{\mathrm{LOS}}$ obtained from the XClumpy model indicate that 24 objects are Compton-thin AGNs and 28 objects are Compton-thick AGNs in the 90\% confidence interval. The main reason is the difference in the torus model applied. The hydrogen column density along the equatorial direction $N_{\mathrm{H}}^{\mathrm{Equ}}$ of CTAGNs inferred from the XClumpy model is larger than that of less obscured AGNs. The Compton-thin torus covering factor $C_{22}$ obtained from the XClumpy model is consistent with that of \cite{Ricci17c} in the low Eddington ratio ($\log R_{\mathrm{Edd}} \leq -1.0$), whereas $C_{22}$ inferred from the XClumpy model is larger than that of \cite{Ricci17c} in the high Eddington ratio ($-1.0 \leq \log R_{\mathrm{Edd}}$). The average value of the Compton-thick torus covering factor $C_{24}$ obtained from the XClumpy model is $36_{-4}^{+4}$\%. This value is larger than that of \cite{Ricci15} ($C_{24} \simeq 27_{-4}^{+4}$\%) based on the assumption that all AGNs have intrinsically the same torus structure. These results suggest that the structure of CTAGN may be intrinsically different from that of less obscured AGN. 
\end{abstract}

\section{Introduction}\label{Section01}
Elucidating the nature of Compton-thick ($24 \leq \log N_{\mathrm{H}}^{\mathrm{LOS}}/\mathrm{cm}^{-2}$) active galactic nuclei (CTAGN) is essential for understanding the coevolution between supermassive black holes (SMBH) and their host galaxies \citep[e.g.,][]{Kormendy13}. CTAGNs are also important to explain the origin of the cosmic X-ray background \citep[e.g.,][]{Ueda98, Ueda99, Ueda03, Gilli07, Ueda14, Harrison16, Ananna19, Ananna20}. Recent studies suggested that CTAGNs appear to be at the stage of intense star formation and rapid mass accretion due to major mergers \citep[e.g.,][]{Hopkins06, Ricci17b, Ricci21, Yamada21}. Hence CTAGNs may be at a different evolutionary stage compared to less obscured AGNs, rather than simply having a higher hydrogen column density along the line of sight. However, it is still unclear whether CTAGNs are intrinsically different from other objects. The reason is that it is difficult to observe CTAGNs due to their heavy obscuration.

Hard X-ray observations above 10~keV are one of the best ways to detect CTAGNs due to their high penetration power against obscuration. Recently, the Neil Gehrels Swift Observatory \citep{Gehrels04} has performed an all-sky hard X-ray survey and produced local AGN catalogs \citep[e.g.,][]{Markwardt05, Tueller08, Tueller10, Baumgartner13, Oh18}. Using the Swift/Burst Alert Telescope (BAT) 70-month catalog \citep{Baumgartner13}, \cite{Ricci17a} systematically analyzed broadband X-ray spectra of 838 AGNs mainly observed with Swift/X-Ray Telescope (XRT) and Swift/BAT. As a result, \cite{Ricci15} identified 55 CTAGN candidates. To accurately determine the properties of CTAGNs, however, more sensitive hard X-ray observation data are required.

The Nuclear Spectroscopic Telescope Array (NuSTAR: \citealt{Harrison13}) is the first orbiting telescopes to focus hard X-rays. The NuSTAR consists of two co-aligned grazing incidence telescopes with two Focal Plane Modules (FPM: 3--79~keV). It provides a combination of sensitivity, spatial resolution, and spectral resolution factors of 10 to 100 improved over previous X-ray satellites in the hard X-ray band \citep{Harrison13}. Recently, NuSTAR observed a lot of local CTAGNs \citep[e.g.,][]{Arevalo14, Balokovic14, Puccetti14, Bauer15, Rivers15, Guainazzi16, Koss16, Puccetti16, Oda17, Balokovic18, Marchesi19b, Marchesi19a, Zhao19b, Zhao19a, Zhao20}.

To estimate the properties of CTAGNs, we have to assume an X-ray spectral model from an AGN torus. The X-ray spectrum of the AGN mainly consists of three components. (1) The direct power-law component from the center due to inverse Compton scattering by a hot corona. (2) The reflection component from the accretion disk and the torus. (3) The fluorescence Fe K$\alpha$ line at 6.4~keV from the accretion disk and the torus. Especially in the case of CTAGNs, the modeling of the AGN torus is important. This is because the direct component is strongly absorbed by the torus and the reflection component is dominant \citep[e.g.,][]{Tanimoto18}.

Many studies indicated that the AGN torus consists of dusty clumps (clumpy torus: \citealt{Krolik88, Laor93, Honig06, Honig07, Nenkova08a, Nenkova08b}). Nevertheless, few X-ray spectral models from the clumpy torus have been constructed. Extending the earlier works \citep{Liu14, Furui16}, \cite{Tanimoto19} generated a new X-ray spectral model from the clumpy torus (XClumpy) utilizing the Monte Carlo simulation for astrophysics and cosmology (MONACO: \citealt{Odaka11, Odaka16}). In the XClumpy model, we assumed the power-law distribution of clumps in the radial direction and the Gaussian distribution in the elevation direction\footnote{The inner radius is 0.05~pc, the outer radius is 1.00~pc, and the radius of each clump is 0.002~pc. The absolute values of these three parameters are freely selected and only their ratios are important. This is because a self-similar geometry produces identical results.}. This XClumpy model has been used to analyze the X-ray spectra of AGNs \citep[e.g.,][]{Miyaji19, Ogawa19, Tanimoto19, Tanimoto20, Toba20, Yamada20, Ogawa21, Uematsu21, Yamada21}. Recently, \cite{Buchner19} also constructed an X-ray spectral model from the clumpy torus called UXCLUMPY.

This paper presents the results of applications of the XClumpy model to the broadband X-ray spectra of 52 CTAGN candidates detected by the Swift/BAT and observed with Chandra, XMM--Newton, Swift/XRT, Suzaku, and NuSTAR. The structure of this paper is the following. \hyperref[Section02]{Section~2} describes the sample selection. In \hyperref[Section03]{Section~3}, we present the data analysis. \hyperref[Section04]{Section~4} describes the X-ray spectral analysis with the XClumpy model. In \hyperref[Section05]{Section~5}, we examine the fraction of CTAGN and validate the radiation-regulated AGN unification model. We assume the solar abundances by \cite{Anders89}. To estimate the luminosity, we adopt the cosmological parameters ($H_{0} = 70.0$ km s$^{-1}$ Mpc$^{-1}$, $\Omega_{\mathrm{m}} = 0.30$, $\Omega_{\lambda} = 0.70$).

\section{Sample Selection}\label{Section02}
We select 52 CTAGN candidates observed with Chandra, XMM--Newton, Swift, Suzaku, and NuSTAR from the sample of \cite{Ricci15}. \hyperref[Table01]{Table~1} shows the information of the 52 CTAGN candidates in our sample. Here we exclude three objects (NGC~1068, Circinus~Galaxy, and NGC~6921). The reasons are the following:
\begin{enumerate}
\item \textbf{NGC~1068}. There is a lot of contamination from the host galaxy and it is difficult to determine the parameters of the torus \citep{Bauer15}.
\item \textbf{Circinus~Galaxy}. \cite{Tanimoto19} have already analyzed the broadband X-ray spectra of the Circinus~Galaxy with the XClumpy model.
\item \textbf{NGC~6921}. \cite{Yamada21} have already analyzed the broadband X-ray spectra of the NGC~6921 with the XClumpy model.
\end{enumerate}

To determine the torus parameters, we need to analyze both soft X-ray data (Chandra, XMM--Newton, Swift/XRT, and Suzaku) and hard X-ray data (Swift/BAT and NuSTAR). \hyperref[Table02]{Table~2} summarizes details of the observations. In this study, we select the soft X-ray data in the following order of priority:
\begin{enumerate}
\item Simultaneous observational data with the XMM--Newton and NuSTAR.
\item Non-simultaneous observational data with the XMM--Newton or Suzaku and NuSTAR. Here we select the one with the longest observation time among the XMM--Newton and Suzaku data.
\item Non-simultaneous observational data with the Chandra and NuSTAR.
\item Simultaneous observation data with the Swift/XRT and NuSTAR.
\end{enumerate}
Here we prioritize non-simultaneous observations with Chandra, XMM--Newton, Suzaku and NuSTAR over simultaneous observations with the Swift/XRT and NuSTAR. In the case of CTAGN candidates, the X-ray flux in 2--10~keV is very small due to their strong absorption. To determine the hydrogen column density along the line of sight ($\log N_{\mathrm{H}}^{\mathrm{LOS}}/\mathrm{cm}^{-2}$), we have to observe the X-ray spectral shape in 2--10~keV. Especially, XMM--Newton and Suzaku have an effective area about 10 times larger than Swift/XRT in 2--10~keV, and can determine $\log N_{\mathrm{H}}^{\mathrm{LOS}}/\mathrm{cm}^{-2}$. In fact, we confirmed that the results did not change if we included the Swift/XRT data. \clearpage
\startlongtable
\begin{deluxetable*}{llllllll}\label{Table01}
\tablecaption{Summary of Objects}
\tablehead
{                                                                                                                       
Swift Name                                                  & Counterpart                                               &
RA                                                          & Dec                                                       &
Redshift                                                    & $N_{\mathrm{H}}^{\mathrm{Gal}}$                           &
$\log M_{\mathrm{BH}}/M_{\sun}$                             & Reference                                                 \\
(1)                                                         & (2)                                                       &
(3)                                                         & (4)                                                       &
(5)                                                         & (6)                                                       &
(7)                                                         & (8)                                                       }
\startdata
SWIFT J0030.0--5904                                         & ESO 112--G006                                             &
00h30m43.83s                                                & --59d00m25.9s                                             &
0.02885                                                     & 0.01360                                                   &
\nodata                                                     & \nodata                                                   \\
SWIFT J0105.5--4213                                         & MCG --07--03--007                                         &
01h05m26.82s                                                & --42d12m58.3s                                             &
0.02988                                                     & 0.02660                                                   &
8.56                                                        & (02)                                                      \\
SWIFT J0111.4--3808                                         & NGC 0424                                                  &
01h11m27.63s                                                & --38d05m00.5s                                             &
0.01176                                                     & 0.01630                                                   &
7.78                                                        & (01)                                                      \\
SWIFT J0122.8+5003                                          & MCG +08--03--018                                          &
01h22m34.43s                                                & +50d03m18.0s                                              &
0.02040                                                     & 0.21800                                                   &
8.41                                                        & (02)                                                      \\
\enddata
\tablecomments{Column (1): Swift name. Column (2): counterpart name based on the BAT AGN Spectroscopic Survey (BASS: \url{https://www.bass-survey.com/}). Column (3)-(4): position taken from NASA/IPAC Extragalactic Database (NED: \url{https://ned.ipac.caltech.edu/}). Column (5): redshift taken from NED. Column (6): total Galactic HI and H$_{2}$ weighted values in units of 10$^{22}$ cm$^{-2}$ (\url{https://www.swift.ac.uk/analysis/nhtot/index.php}: \citealt{Willingale13}). Column~(7) black hole mass. Column~(8): reference of the black hole mass.}
\tablerefs{(01)~\cite{Bian07}. (02)~\cite{Koss17}. (03)~\cite{Merloni03}. (04)~\cite{Bosch16}.}
\end{deluxetable*}
     \clearpage
\startlongtable
\begin{deluxetable*}{llllllll}\label{Table02}
\tablecaption{Summary of Observations}
\tablehead
{
Galaxy Name                                                 & Observatory                                               &
Observation ID                                              & Exposure                                                  &
Start Date                                                  & End Date                                                  &
Binning                                                     & Reference                                                 \\
(1)                                                         & (2)							                            &
(3)					                                        & (4)							                            &
(5)						                                    & (6)							                            &
(7)					                                        & (8)                                                       }
\startdata
ESO 112--G006                                               & XMM--Newton                                               &
0852180101                                                  & 21                                                        &
2019 Dec 06                                                 & 2019 Dec 06                                               &
50                                                          & (21)                                                      \\
                                                            & NuSTAR                                                    &
60561038002                                                 & 56                                                        &
2019 Dec 06                                                 & 2019 Dec 07                                               &
50                                                          & (21)                                                      \\
MCG --07--03--007                                           & XMM--Newton                                               &
0852180201                                                  & 21                                                        &
2019 Dec 28                                                 & 2019 Dec 28                                               &
50                                                          & (21)                                                      \\
                                                            & NuSTAR                                                    &
60561039002                                                 & 54                                                        &
2019 Dec 27                                                 & 2019 Dec 29                                               &
50                                                          & (21)                                                      \\
\enddata
\tablecomments{Column~(1): galaxy name. Column~(2): satellite name (Chandra, XMM--Newton, Swift, Suzaku, and NuSTAR). Column~(3): observation identification number. Column~(4): exposure time (based on Chandra/ACIS, XMM--Newton/PN, Swift/XRT, Suzaku/XIS0, and NuSTAR/FPMA) in units of ks. Column~(5): start date in units of ymd. Column~(6): end date in units of ymd. Column~(7): binning. Column~(8): reference of the previous work.}
\tablerefs{(01)~\cite{Balokovic14}. (02)~\cite{Balokovic18}. (03)~\cite{Bauer15}. (04)~\cite{Georgantopoulos19}. (05)~\cite{Guainazzi04}. (06)~\cite{Guainazzi05} (07)~\cite{Guainazzi16}. (08)~\cite{Kawamuro16b}. (09)~\cite{Konami12}. (10)~\cite{Koss16}. (11)~\cite{Marchesi19a}. (12)~\cite{Marchesi19b}. (13)~\cite{Netzer05}. (14)~\cite{Oda17}. (15)~\cite{Piconcelli07}. (16)~\cite{Piconcelli11}. (17)~\cite{Ricci17a}. (18)~\cite{Tanimoto16}. (19)~\cite{Tanimoto18}. (20)~\cite{Tanimoto20}. (21)~\cite{Torres21}. (22)~\cite{Ueda07}. (23)~\cite{Zhao20}.}
\end{deluxetable*}
     \clearpage
\section{Data Analysis}\label{Section03}
 
\subsection{Chandra}\label{Section0301}
Chandra (1999--: \citealt{Weisskopf02}) is NASA's X-ray astronomical satellite. It carries two Advanced CCD Imaging Spectrometers (ACIS: \citealt{Garmire03}). We analyzed the ACIS data using the Chandra interactive analysis of observations 4.12.1 and the calibration database 4.9.2.1. We reprocessed the unfiltered ACIS data using chandra\_repro. The source spectra were extracted from a circular region with a 30~arcsec radius centered on the flux peak and the background spectra were taken from a source-free circular region with the same radius using specextract. We generated the redistribution matrix files (RMF) and the ancillary response files (ARF) using specextract.
\subsection{XMM--Newton}\label{Section0302}
XMM--Newton (1999--: \citealt{Jansen01}) is the second ESA's X-ray astronomical satellite. It has three European photon imaging cameras: two MOS (MOS1 and MOS2: \citealt{Turner01}) and one PN \citep{Struder01}. We analyzed the MOS and PN data using the science analysis system 18.00 and the current calibration file released on 2020 April 20. We reprocessed the unfiltered MOS and PN data using emproc and epproc, respectively. The source spectra were extracted from a circular region with a 30~arcsec radius centered on the flux peak and the background spectra were obtained from a source-free circular region with a 30~arcsec radius in the same CCD chip. We generated the RMF and the ARF using rmfgen and arfgen, respectively. We combined the source spectra, background spectra, RMF, and ARF of MOS1 and MOS2 using addascaspec.
\subsection{Swift}\label{Section0303}
The Neil Gehrels Swift Observatory (2004--: \citealt{Gehrels04}) carries one X-Ray Telescope (XRT: \citealt{Burrows05}) and one Burst Alert Telescope (BAT: \citealt{Barthelmy05}). We analyzed the XRT data using the HEAsoft 6.28 and the calibration database released on 2020 July 24. We reprocessed the unfiltered XRT data using xrtpipeline. The source spectra were extracted from a circular region with a 30~arcsec radius centered on the flux peak and the background spectra were obtained from a source-free circular region with the same radius using xrtproducts. We also utilized the time-averaged spectra obtained from the Swift/BAT 105-month catalog \citep{Oh18}\footnote{\url{https://swift.gsfc.nasa.gov/results/bs105mon/}}.
\subsection{Suzaku}\label{Section0304}
Suzaku (2005--2015: \citealt{Mitsuda07}) was the fifth Japanese X-ray astronomical satellite. It had four X-ray Imaging Spectrometers (XIS0, XIS1, XIS2, and XIS3: \citealt{Koyama07}). One CCD camera (XIS2) is the back-illuminated XIS (BIXIS: 0.20--12.0~keV) and three CCD cameras (XIS0, XIS1, and XIS3) are the front-illuminated XIS (FIXIS: 0.40--12.0~keV). We analyzed the XIS data using the HEAsoft 6.28 and the calibration database released on 2018 October 10. We reprocessed the unfiltered XIS data using aepipeline. The source spectra were extracted from a circular region with a 60~arcsec radius centered on the flux peak and the background spectra were taken from a source-free circular region with the same radius. In the case of ESO~005--G004, the fluxes are very low and we need to determine the background more precisely \citep{Ueda07}. We extracted the source spectra from a circular region with a 90 arcsec and the background spectra from a source-free circular region with a 180 arcsec for ESO~005--G004. We generated the RMF and the ARF using xisrmfgen and xissimarfgen \citep{Ishisaki07}, respectively. We combined the source spectra, background spectra, RMF, and ARF of FIXIS using addascaspec.
\subsection{NuSTAR}\label{Section0305}
NuSTAR (2012--: \citealt{Harrison13}) is the NASA's X-ray astronomical satellite. It carries two co-aligned grazing incidence telescopes coupled with two Focal Plane Modules (FPM: 3--79 keV). We analyzed FPM data using the HEAsoft 6.28 and the calibration database released on 2020 August 26. We reprocessed the unfiltered FPM data using nupipeline. The source spectra were extracted from a circular region with a 30~arcsec radius centered on the flux peak and the background spectra were obtained from a source-free circular region with a 30~arcsec radius using nuproducts. We combined the source spectra, background spectra, RMF, and ARF using addascaspec.
 \clearpage
\section{Spectral Analysis}\label{Section04}
We performed simultaneous fitting to the Chandra/ACIS (0.5--8.0~keV), XMM--Newton/MOS (0.5--8.0~keV), XMM--Newton/PN (0.5--10.0~keV), Swift/XRT (0.5--6.0~keV), Swift/BAT (14.0--100.0~keV), Suzaku/BIXIS (0.5--8.0~keV), Suzaku/FIXIS (2.0--10.0~keV), NuSTAR/FPM (4.0--60.0~keV). Our X-ray spectral model is the following:

\begin{align}
& \mathrm{const1 \times phabs}                                                                                      \nonumber\\
& \times (\mathrm{const2 \times cabs \times zphabs \times zcutoffpl}                                                \nonumber\\
& +	      \mathrm{const3 \times zcutoffpl+atable\{xclumpy\_v01\_RC.fits\}}                                          \nonumber\\
& +       \mathrm{atable\{xclumpy\_v01\_RL.fits\}+(zgauss)+(apec))}.
\end{align}

Our model consists of six components.
\begin{enumerate}
\item $\mathbf{const1} \times \mathbf{phabs}$. The const1 term represents a cross calibration constant to compensate for differences in absolute flux among instruments. We assume that the cross calibration constant of NuSTAR/FPM to Swift/BAT ($C_{\mathrm{FPM}}$) is unity as a reference. To reduce the number of free parameters, we fix other cross calibration constants ($C_{\mathrm{ACIS}} = 1.10$, $C_{\mathrm{MOS}} = 1.00, C_{\mathrm{PN}} = 0.90, C_{\mathrm{XRT}} = 1.05, C_{\mathrm{BIXIS}} = 0.90, C_{\mathrm{FIXIS}} = 0.95$) based on the results of \cite{Madsen17}, which presented the cross calibration campaigns among Chandra, XMM--Newton, Swift, Suzaku, and NuSTAR. The phabs term represents the Galactic absorption. We utilized the total Galactic $HI$ and $H_{2}$ values provided by \citet{Willingale13}.

\item $\mathbf{const2} \times \mathbf{cabs} \times \mathbf{zphabs} \times \mathbf{zcutoffpl}$. This component represents the transmitted continuum through the torus. The const2 term is a constant to consider the time variability among Chandra, XMM--Newton, Swift, Suzaku, and NuSTAR observations. We fix the time variability constant of Swift/BAT at unity as a reference and make the constant of Chandra, XMM--Newton, Swift/XRT, Suzaku ($T_{\mathrm{Satellite1}}$), and NuSTAR ($T_{\mathrm{Satellite2}}$) free parameters. In the case of simultaneous observations of XMM--Newton and NuSTAR or Swift and NuSTAR, we link $T_{\mathrm{Satellite2}}$ to $T_{\mathrm{Satellite1}}$. We limit $T_{\mathrm{Satellite1}}$ and $T_{\mathrm{Satellite2}}$ values within a range of 0.10--10.0 to avoid unrealistic results. If the time variability constant is not determined in the range of 0.10--10.0, we fix its value at 1.00. We multiply this constant to the transmitted component only. This is because the sizes of the scatterer and reflector are most likely parsec or larger scales and we can ignore the time variability of these components. The zphabs and cabs terms represent the photoelectric absorption and Compton scattering by the torus, respectively. We determine the hydrogen column density along the line of sight ($N_{\mathrm{H}}^{\mathrm{LOS}}$) according to Equation~3. The zcutoffpl represents the power-law with an exponential cutoff. Since it is difficult to determine the cutoff energy, we fix this value at a typical one ($E_{\mathrm{cut}} = 370$ keV: \citealt{Ricci18})\footnote{\cite{Ricci18} obtained $E_{\mathrm{cut}} = 370_{-51}^{+51}$ keV for sources with the low Eddington ratios ($\log R_{\mathrm{Edd}} \leq -1.0$), whereas they estimated $E_{\mathrm{cut}} = 160_{-41}^{+41}$ keV for sources with the high Eddington ratios ($-1.0 \leq \log R_{\mathrm{Edd}}$).}. Since almost all of our targets have low Eddington ratios, we fix the cutoff energy of all the sources at 370~keV. We have confirmed that the obtained results are consistent within the 90\% confidence interval even when we fix the cutoff energy at 160~keV for the sources with the high Eddington ratios.

\item $\mathbf{const3} \times \mathbf{zcutoffpl}$. This component represents the scattered component. The const3 term is the scattering factor ($f_{\mathrm{scat}}$). We link the photon index ($\Gamma$), the cutoff energy ($E_{\mathrm{cut}}$), and the normalization ($N_{\mathrm{Dir}}$) to those of the transmitted continuum.

\item $\mathbf{atable\{xclumpy\_v01\_RC.fits\}}$. This component represents the reflection continuum from the torus based on the XClumpy model \citep{Tanimoto19}\footnote{In this study, we updated the XClumpy model using the National Astronomical Observatory of Japan's XC50 supercomputer. Please see the following link for details (\url{https://github.com/AtsushiTanimoto/XClumpy}).}. We assumed the power-law distribution in the radial direction and the Gaussian distribution in the elevation direction as the clump distribution. The clump number density function $d(r,\theta,\phi)$ in the spherical coordinate system (where $r$ is radius, $\theta$ is inclination angle, and $\phi$ is azimuth) is the following.
\begin{equation}
d(r,\theta,\phi) \propto \left(\frac{r}{r_{\mathrm{in}}}\right)^{-1/2} \exp{\left(-\frac{(\theta-\pi/2)^2}{\sigma^2}\right)}.
\end{equation}
Here $r_{\mathrm{in}}$ is the inner radius of the torus and $\sigma$ is the torus angular width. In the XClumpy model, \cite{Tanimoto19} assumed that the radius of the clump is 0.002~pc, the inner radius of the torus is 0.05~pc, and the outer radius of the torus is 1.00~pc. This model has five free parameters: (1) hydrogen column density along the equatorial plane ($\log N_{\mathrm{H}}^{\mathrm{Equ}}/\mathrm{cm}^{-2}$: 23--25), (2) torus angular width ($\sigma$: 10\degr--90\degr), (3) inclination angle ($i$: 20\degr--87\degr), (4) photon index ($\Gamma$: 1.0--3.0), and (5) normalization ($N_{\mathrm{Dir}}$). The hydrogen column density along the line of sight ($N_{\mathrm{H}}^{\mathrm{LOS}}$) is determined by the following equation.
\begin{equation}\label{Equation03}
N_{\mathrm{H}}^{\mathrm{LOS}} = N_{\mathrm{H}}^{\mathrm{Equ}} \exp{\left(-\frac{(i-\pi/2)^2}{\sigma^2}\right)}.
\end{equation}
We limit $i$ value within a range of 60\degr--87\degr. Especially, we fix its value at 87\degr for the eight objects (Mrk~0003, NGC~3079, NGC~3393, NGC~4945, NGC~5643, NGC~5728, NGC~6240, and NGC~7479) whose water maser has been detected \citep{Panessa20}. This is because, if $i$ is free parameter, the obtained torus covering factor is unphysically large, which is inconsistent with the existence of a narrow line region. We link $\Gamma$ and $N_{\mathrm{Dir}}$ to those of the transmitted continuum. We note that we fix $E_{\mathrm{cut}} = 370$ keV in the XClumpy model.

\item $\mathbf{atable\{xclumpy\_v01\_RL.fits\}+(zgauss)}$. This component represents fluorescent lines from the torus based on the XClumpy model. We link all parameters to those of the reflection continuum. If the F-test shows an improvement at a significance level of 99\% or higher, we add zgauss to the model. This is because recent studies implied the contribution from spatially extended fluorescent lines such as Fe K$\alpha$ (6.4~keV) \citep[e.g.,][]{Arevalo14, Bauer15, Fabbiano17, Kawamuro19, Kawamuro20} and the XClumpy model does not include the ionized fluorescent lines such as He-like Fe K$\alpha$ (6.7~keV) and H-like Fe K$\alpha$ (7.0~keV). In fact, it is necessary to add Fe K$\alpha$ fluorescent lines to reproduce the observed X-ray spectra for some objects.

\item $\mathbf{apec}$. This component represents emission from an optically thin thermal plasma in the host galaxy. If the F-test shows an improvement at a significance level of 99\% or higher, we add apec to the model.
\end{enumerate}
 \clearpage
\begin{figure*}\label{Figure01}
\gridline{\fig{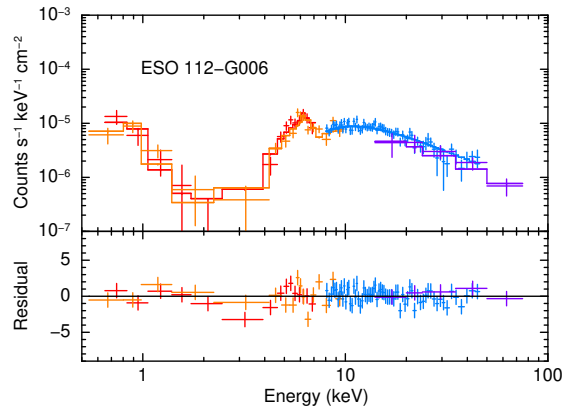}{0.45\textwidth}{}}
\caption{The folded X-ray spectra fitted with the XClumpy model. The pink crosses are Chandra/ACIS, the red crosses are XMM--Newton/MOS, the orange crosses are XMM--Newton/PN, the light green crosses are Swift/XRT, the green crosses are Suzaku/BIXIS, the light blue crosses are Suzaku/FIXIS, the blue crosses are NuSTAR/FPM, and the purple crosses are Swift/BAT. The solid curves represent best fitting model. The lower panel shows residuals. The complete figure set (52 images) is available in the online journal.}
\end{figure*}

\figsetstart
\figsetnum      {52}
\figsettitle    {The Folded X-Ray Spectra Fitted with the XClumpy Model}
\figsetgrpstart
\figsetgrpnum   {01.01}
\figsetgrptitle {ESO~112--G006}
\figsetplot     {0001}
\figsetgrpnote  {The folded X-ray spectra of ESO~112--G006.}
\figsetgrpend
\figsetgrpstart
\figsetgrpnum   {01.02}
\figsetgrptitle {MCG~--07--03--007}
\figsetplot     {0002}
\figsetgrpnote  {The folded X-ray spectra of MCG~--07--03--007.}
\figsetgrpend
\figsetgrpstart
\figsetgrpnum   {01.03}
\figsetgrptitle {NGC~0424}
\figsetplot     {0003}
\figsetgrpnote  {The folded X-ray spectra of NGC~0424.}
\figsetgrpend
\figsetgrpstart
\figsetgrpnum   {01.04}
\figsetgrptitle {MCG~+08--03--018}
\figsetplot     {0004}
\figsetgrpnote  {The folded X-ray spectra of MCG~+08--03--018.}
\figsetgrpend
\figsetgrpstart
\figsetgrpnum   {01.05}
\figsetgrptitle {2MASX~J01290761--6038423}
\figsetplot     {0005}
\figsetgrpnote  {The folded X-ray spectra of 2MASX~J01290761--6038423.}
\figsetgrpend
\figsetgrpstart
\figsetgrpnum   {01.06}
\figsetgrptitle {ESO~244--IG030}
\figsetplot     {0006}
\figsetgrpnote  {The folded X-ray spectra of ESO~244--IG030.}
\figsetgrpend
\figsetgrpstart
\figsetgrpnum   {01.07}
\figsetgrptitle {NGC~1106}
\figsetplot     {0007}
\figsetgrpnote  {The folded X-ray spectra of NGC~1106.}
\figsetgrpend
\figsetgrpstart
\figsetgrpnum   {01.08}
\figsetgrptitle {2MFGC~02280}
\figsetplot     {0008}
\figsetgrpnote  {The folded X-ray spectra of 2MFGC~02280.}
\figsetgrpend
\figsetgrpstart
\figsetgrpnum   {01.09}
\figsetgrptitle {NGC~1125}
\figsetplot     {0009}
\figsetgrpnote  {The folded X-ray spectra of NGC~1125.}
\figsetgrpend
\figsetgrpstart
\figsetgrpnum   {01.10}
\figsetgrptitle {NGC~1194}
\figsetplot     {0010}
\figsetgrpnote  {The folded X-ray spectra of NGC~1194.}
\figsetgrpend
\figsetgrpstart
\figsetgrpnum   {01.11}
\figsetgrptitle {NGC~1229}
\figsetplot     {0011}
\figsetgrpnote  {The folded X-ray spectra of NGC~1229.}
\figsetgrpend
\figsetgrpstart
\figsetgrpnum   {01.12}
\figsetgrptitle {ESO~201--IG004}
\figsetplot     {0012}
\figsetgrpnote  {The folded X-ray spectra of ESO~201--IG004.}
\figsetgrpend
\figsetgrpstart
\figsetgrpnum   {01.13}
\figsetgrptitle {2MASX~J03561995--6251391}
\figsetplot     {0013}
\figsetgrpnote  {The folded X-ray spectra of 2MASX~J03561995--6251391.}
\figsetgrpend
\figsetgrpstart
\figsetgrpnum   {01.14}
\figsetgrptitle {MCG~--02--12--017}
\figsetplot     {0014}
\figsetgrpnote  {The folded X-ray spectra of MCG~--02--12--017.}
\figsetgrpend
\figsetgrpstart
\figsetgrpnum   {01.15}
\figsetgrptitle {CGCG~420--015}
\figsetplot     {0015}
\figsetgrpnote  {The folded X-ray spectra of CGCG~420--015.}
\figsetgrpend
\figsetgrpstart
\figsetgrpnum   {01.16}
\figsetgrptitle {ESO~005--G004}
\figsetplot     {0016}
\figsetgrpnote  {The folded X-ray spectra of ESO~005--G004.}
\figsetgrpend
\figsetgrpstart
\figsetgrpnum   {01.17}
\figsetgrptitle {Mrk~0003}
\figsetplot     {0017}
\figsetgrpnote  {The folded X-ray spectra of Mrk~0003.}
\figsetgrpend
\figsetgrpstart
\figsetgrpnum   {01.18}
\figsetgrptitle {2MASX~J06561197--4919499}
\figsetplot     {0018}
\figsetgrpnote  {The folded X-ray spectra of 2MASX~J06561197--4919499.}
\figsetgrpend
\figsetgrpstart
\figsetgrpnum   {01.19}
\figsetgrptitle {MCG~+06--16--028}
\figsetplot     {0019}
\figsetgrpnote  {The folded X-ray spectra of MCG~+06--16--028.}
\figsetgrpend
\figsetgrpstart
\figsetgrpnum   {01.20}
\figsetgrptitle {Mrk~0078}
\figsetplot     {0020}
\figsetgrpnote  {The folded X-ray spectra of Mrk~0078.}
\figsetgrpend
\figsetgrpstart
\figsetgrpnum   {01.21}
\figsetgrptitle {Mrk~0622}
\figsetplot     {0021}
\figsetgrpnote  {The folded X-ray spectra of Mrk~0622.}
\figsetgrpend
\figsetgrpstart
\figsetgrpnum   {01.22}
\figsetgrptitle {NGC~2788A}
\figsetplot     {0022}
\figsetgrpnote  {The folded X-ray spectra of NGC~2788A.}
\figsetgrpend
\figsetgrpstart
\figsetgrpnum   {01.23}
\figsetgrptitle {SBS~0915+556}
\figsetplot     {0023}
\figsetgrpnote  {The folded X-ray spectra of SBS~0915+556.}
\figsetgrpend
\figsetgrpstart
\figsetgrpnum   {01.24}
\figsetgrptitle {2MASX~J09235371--3141305}
\figsetplot     {0024}
\figsetgrpnote  {The folded X-ray spectra of 2MASX~J09235371--3141305.}
\figsetgrpend
\figsetgrpstart
\figsetgrpnum   {01.25}
\figsetgrptitle {ESO~565--G019}
\figsetplot     {0025}
\figsetgrpnote  {The folded X-ray spectra of ESO~565--G019.}
\figsetgrpend
\figsetgrpstart
\figsetgrpnum   {01.26}
\figsetgrptitle {MCG~+10--14--025}
\figsetplot     {0026}
\figsetgrpnote  {The folded X-ray spectra of MCG~+10--14--025.}
\figsetgrpend
\figsetgrpstart
\figsetgrpnum   {01.27}
\figsetgrptitle {NGC~3079}
\figsetplot     {0027}
\figsetgrpnote  {The folded X-ray spectra of NGC~3079.}
\figsetgrpend
\figsetgrpstart
\figsetgrpnum   {01.28}
\figsetgrptitle {ESO~317--G041}
\figsetplot     {0028}
\figsetgrpnote  {The folded X-ray spectra of ESO~317--G041.}
\figsetgrpend
\figsetgrpstart
\figsetgrpnum   {01.29}
\figsetgrptitle {SDSS~J103315.71+525217.8}
\figsetplot     {0029}
\figsetgrpnote  {The folded X-ray spectra of SDSS~J103315.71+525217.8.}
\figsetgrpend
\figsetgrpstart
\figsetgrpnum   {01.30}
\figsetgrptitle {NGC~3393}
\figsetplot     {0030}
\figsetgrpnote  {The folded X-ray spectra of NGC~3393.}
\figsetgrpend
\figsetgrpstart
\figsetgrpnum   {01.31}
\figsetgrptitle {NGC~4102}
\figsetplot     {0031}
\figsetgrpnote  {The folded X-ray spectra of NGC~4102.}
\figsetgrpend
\figsetgrpstart
\figsetgrpnum   {01.32}
\figsetgrptitle {NGC~4180}
\figsetplot     {0032}
\figsetgrpnote  {The folded X-ray spectra of NGC~4180.}
\figsetgrpend
\figsetgrpstart
\figsetgrpnum   {01.33}
\figsetgrptitle {ESO~323--G032}
\figsetplot     {0033}
\figsetgrpnote  {The folded X-ray spectra of ESO~323--G032.}
\figsetgrpend
\figsetgrpstart
\figsetgrpnum   {01.34}
\figsetgrptitle {NGC~4945}
\figsetplot     {0034}
\figsetgrpnote  {The folded X-ray spectra of NGC~4945.}
\figsetgrpend
\figsetgrpstart
\figsetgrpnum   {01.35}
\figsetgrptitle {IGR~J14175--4641}
\figsetplot     {0035}
\figsetgrpnote  {The folded X-ray spectra of IGR~J14175--4641.}
\figsetgrpend
\figsetgrpstart
\figsetgrpnum   {01.36}
\figsetgrptitle {NGC~5643}
\figsetplot     {0036}
\figsetgrpnote  {The folded X-ray spectra of NGC~5643.}
\figsetgrpend
\figsetgrpstart
\figsetgrpnum   {01.37}
\figsetgrptitle {NGC~5728}
\figsetplot     {0037}
\figsetgrpnote  {The folded X-ray spectra of NGC~5728.}
\figsetgrpend
\figsetgrpstart
\figsetgrpnum   {01.38}
\figsetgrptitle {CGCG~164--019}
\figsetplot     {0038}
\figsetgrpnote  {The folded X-ray spectra of CGCG~164--019.}
\figsetgrpend
\figsetgrpstart
\figsetgrpnum   {01.39}
\figsetgrptitle {ESO~137--G034}
\figsetplot     {0039}
\figsetgrpnote  {The folded X-ray spectra of ESO~137--G034.}
\figsetgrpend
\figsetgrpstart
\figsetgrpnum   {01.40}
\figsetgrptitle {NGC~6232}
\figsetplot     {0040}
\figsetgrpnote  {The folded X-ray spectra of NGC~6232.}
\figsetgrpend
\figsetgrpstart
\figsetgrpnum   {01.41}
\figsetgrptitle {ESO~138--G001}
\figsetplot     {0041}
\figsetgrpnote  {The folded X-ray spectra of ESO~138--G001.}
\figsetgrpend
\figsetgrpstart
\figsetgrpnum   {01.42}
\figsetgrptitle {NGC~6240}
\figsetplot     {0042}
\figsetgrpnote  {The folded X-ray spectra of NGC~6240.}
\figsetgrpend
\figsetgrpstart
\figsetgrpnum   {01.43}
\figsetgrptitle {NGC~6552}
\figsetplot     {0043}
\figsetgrpnote  {The folded X-ray spectra of NGC~6552.}
\figsetgrpend
\figsetgrpstart
\figsetgrpnum   {01.44}
\figsetgrptitle {2MASX~J20145928+2523010}
\figsetplot     {0044}
\figsetgrpnote  {The folded X-ray spectra of 2MASX~J20145928+2523010.}
\figsetgrpend
\figsetgrpstart
\figsetgrpnum   {01.45}
\figsetgrptitle {ESO~464--G016}
\figsetplot     {0045}
\figsetgrpnote  {The folded X-ray spectra of ESO~464--G016.}
\figsetgrpend
\figsetgrpstart
\figsetgrpnum   {01.46}
\figsetgrptitle {NGC~7130}
\figsetplot     {0046}
\figsetgrpnote  {The folded X-ray spectra of NGC~7130.}
\figsetgrpend
\figsetgrpstart
\figsetgrpnum   {01.47}
\figsetgrptitle {NGC~7212}
\figsetplot     {0047}
\figsetgrpnote  {The folded X-ray spectra of NGC~7212.}
\figsetgrpend
\figsetgrpstart
\figsetgrpnum   {01.48}
\figsetgrptitle {ESO~406--G004}
\figsetplot     {0048}
\figsetgrpnote  {The folded X-ray spectra of ESO~406--G004.}
\figsetgrpend
\figsetgrpstart
\figsetgrpnum   {01.49}
\figsetgrptitle {NGC~7479}
\figsetplot     {0049}
\figsetgrpnote  {The folded X-ray spectra of NGC~7479.}
\figsetgrpend
\figsetgrpstart
\figsetgrpnum   {01.50}
\figsetgrptitle {SWIFT~J2307.9+2245}
\figsetplot     {0050}
\figsetgrpnote  {The folded X-ray spectra of SWIFT~J2307.9+2245.}
\figsetgrpend
\figsetgrpstart
\figsetgrpnum   {01.51}
\figsetgrptitle {NGC~7582}
\figsetplot     {0051}
\figsetgrpnote  {The folded X-ray spectra of NGC~7582.}
\figsetgrpend
\figsetgrpstart
\figsetgrpnum   {01.52}
\figsetgrptitle {NGC~7682}
\figsetplot     {0052}
\figsetgrpnote  {the folded X-ray spectra of NGC~7682.}
\figsetgrpend
\figsetend

\begin{figure*}\label{Figure02}
\gridline{\fig{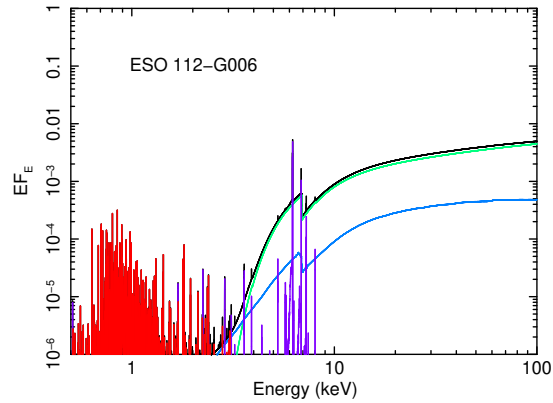}{0.45\textwidth}{}}
\caption{The best fitting model. The black line is total. the red line is thermal emission from optically thin plasma, the orange line is scattered component, the green line is direct component, the blue line is reflection continuum from the torus, and the purple line is emission lines from the torus. The complete figure set (52 images) is available in the online journal.}
\end{figure*}

\figsetstart
\figsetnum      {52}
\figsettitle    {The Best Fitting Model}
\figsetgrpstart
\figsetgrpnum   {02.01}
\figsetgrptitle {ESO~112--G006}
\figsetplot     {0053}
\figsetgrpnote  {The best fitting model of ESO~112--G006.}
\figsetgrpend
\figsetgrpstart
\figsetgrpnum   {02.02}
\figsetgrptitle {MCG~--07--03--007}
\figsetplot     {0054}
\figsetgrpnote  {The best fitting model of MCG~--07--03--007.}
\figsetgrpend
\figsetgrpstart
\figsetgrpnum   {02.03}
\figsetgrptitle {NGC~0424}
\figsetplot     {0055}
\figsetgrpnote  {The best fitting model of NGC~0424.}
\figsetgrpend
\figsetgrpstart
\figsetgrpnum   {02.04}
\figsetgrptitle {MCG~+08--03--018}
\figsetplot     {0056}
\figsetgrpnote  {The best fitting model of MCG~+08--03--018.}
\figsetgrpend
\figsetgrpstart
\figsetgrpnum   {02.05}
\figsetgrptitle {2MASX~J0129761--6038423}
\figsetplot     {0057}
\figsetgrpnote  {The best fitting model of 2MASX~J0129761--6038423.}
\figsetgrpend
\figsetgrpstart
\figsetgrpnum   {02.06}
\figsetgrptitle {ESO~244--IG030}
\figsetplot     {0058}
\figsetgrpnote  {The best fitting model of ESO~244--IG030.}
\figsetgrpend
\figsetgrpstart
\figsetgrpnum   {02.07}
\figsetgrptitle {NGC~1106}
\figsetplot     {0059}
\figsetgrpnote  {The best fitting model of NGC~1106.}
\figsetgrpend
\figsetgrpstart
\figsetgrpnum   {02.08}
\figsetgrptitle {2MFGC~02280}
\figsetplot     {0060}
\figsetgrpnote  {The best fitting model of 2MFGC~02280.}
\figsetgrpend
\figsetgrpstart
\figsetgrpnum   {02.09}
\figsetgrptitle {NGC~1125}
\figsetplot     {0061}
\figsetgrpnote  {The best fitting model of NGC~1125.}
\figsetgrpend
\figsetgrpstart
\figsetgrpnum   {02.10}
\figsetgrptitle {NGC~1194}
\figsetplot     {0062}
\figsetgrpnote  {The best fitting model of NGC~1194.}
\figsetgrpend
\figsetgrpstart
\figsetgrpnum   {02.11}
\figsetgrptitle {NGC~1229}
\figsetplot     {0063}
\figsetgrpnote  {The best fitting model of NGC~1229.}
\figsetgrpend
\figsetgrpstart
\figsetgrpnum   {02.12}
\figsetgrptitle {ESO~201--IG004}
\figsetplot     {0064}
\figsetgrpnote  {The best fitting model of ESO~201--IG004.}
\figsetgrpend
\figsetgrpstart
\figsetgrpnum   {02.13}
\figsetgrptitle {2MASX~J03561995--6251391}
\figsetplot     {0065}
\figsetgrpnote  {The best fitting model of 2MASX~J03561995--6251391.}
\figsetgrpend
\figsetgrpstart
\figsetgrpnum   {02.14}
\figsetgrptitle {MCG~--02--12--017}
\figsetplot     {0066}
\figsetgrpnote  {The best fitting model of MCG~--02--12--017.}
\figsetgrpend
\figsetgrpstart
\figsetgrpnum   {02.15}
\figsetgrptitle {CGCG~420--015}
\figsetplot     {0067}
\figsetgrpnote  {The best fitting model of CGCG~420--015.}
\figsetgrpend
\figsetgrpstart
\figsetgrpnum   {02.16}
\figsetgrptitle {ESO~005--G004}
\figsetplot     {0068}
\figsetgrpnote  {The best fitting model of ESO~005--G004.}
\figsetgrpend
\figsetgrpstart
\figsetgrpnum   {02.17}
\figsetgrptitle {Mrk~0003}
\figsetplot     {0069}
\figsetgrpnote  {The best fitting model of Mrk~0003.}
\figsetgrpend
\figsetgrpstart
\figsetgrpnum   {02.18}
\figsetgrptitle {2MASX~J06561197--4919499}
\figsetplot     {0070}
\figsetgrpnote  {The best fitting model of 2MASX~J06561197--4919499.}
\figsetgrpend
\figsetgrpstart
\figsetgrpnum   {02.19}
\figsetgrptitle {MCG~+06--16--028}
\figsetplot     {0071}
\figsetgrpnote  {The best fitting model of MCG~+06--16--028.}
\figsetgrpend
\figsetgrpstart
\figsetgrpnum   {02.20}
\figsetgrptitle {Mrk~0078}
\figsetplot     {0072}
\figsetgrpnote  {The best fitting model of Mrk~0078.}
\figsetgrpend
\figsetgrpstart
\figsetgrpnum   {02.21}
\figsetgrptitle {Mrk~0622}
\figsetplot     {0073}
\figsetgrpnote  {The best fitting model of Mrk~0622.}
\figsetgrpend
\figsetgrpstart
\figsetgrpnum   {02.22}
\figsetgrptitle {NGC~2788A}
\figsetplot     {0074}
\figsetgrpnote  {The best fitting model of NGC~2788A.}
\figsetgrpend
\figsetgrpstart
\figsetgrpnum   {02.23}
\figsetgrptitle {SBS~0915+556}
\figsetplot     {0075}
\figsetgrpnote  {The best fitting model of SBS~0915+556.}
\figsetgrpend
\figsetgrpstart
\figsetgrpnum   {02.24}
\figsetgrptitle {2MASX~J09235371--3141305}
\figsetplot     {0076}
\figsetgrpnote  {The best fitting model of 2MASX~J09235371--3141305.}
\figsetgrpend
\figsetgrpstart
\figsetgrpnum   {02.25}
\figsetgrptitle {ESO~565--G019}
\figsetplot     {0077}
\figsetgrpnote  {The best fitting model of ESO~565--G019.}
\figsetgrpend
\figsetgrpstart
\figsetgrpnum   {02.26}
\figsetgrptitle {MCG~+10--14--025}
\figsetplot     {0078}
\figsetgrpnote  {The best fitting model of MCG~+10--14--025.}
\figsetgrpend
\figsetgrpstart
\figsetgrpnum   {02.27}
\figsetgrptitle {NGC~3079}
\figsetplot     {0079}
\figsetgrpnote  {The best fitting model of NGC~3079.}
\figsetgrpend
\figsetgrpstart
\figsetgrpnum   {02.28}
\figsetgrptitle {ESO~317--G041}
\figsetplot     {0080}
\figsetgrpnote  {The best fitting model of ESO~317--G041.}
\figsetgrpend
\figsetgrpstart
\figsetgrpnum   {02.29}
\figsetgrptitle {SDSS~J103315.71+525217.8}
\figsetplot     {0081}
\figsetgrpnote  {The best fitting model of SDSS~J103315.71+525217.8.}
\figsetgrpend
\figsetgrpstart
\figsetgrpnum   {02.30}
\figsetgrptitle {NGC~3393}
\figsetplot     {0082}
\figsetgrpnote  {The best fitting model of NGC~3393.}
\figsetgrpend
\figsetgrpstart
\figsetgrpnum   {02.31}
\figsetgrptitle {NGC~4102}
\figsetplot     {0083}
\figsetgrpnote  {The best fitting model of NGC~4102.}
\figsetgrpend
\figsetgrpstart
\figsetgrpnum   {02.32}
\figsetgrptitle {NGC~4180}
\figsetplot     {0084}
\figsetgrpnote  {The best fitting model of NGC~4180.}
\figsetgrpend
\figsetgrpstart
\figsetgrpnum   {02.33}
\figsetgrptitle {ESO~323--G032}
\figsetplot     {0085}
\figsetgrpnote  {The best fitting model of ESO~323--G032.}
\figsetgrpend
\figsetgrpstart
\figsetgrpnum   {02.34}
\figsetgrptitle {NGC~4945}
\figsetplot     {0086}
\figsetgrpnote  {The best fitting model of NGC~4945.}
\figsetgrpend
\figsetgrpstart
\figsetgrpnum   {02.35}
\figsetgrptitle {IGR~J14175--4641}
\figsetplot     {0087}
\figsetgrpnote  {The best fitting model of IGR~J14175--4641.}
\figsetgrpend
\figsetgrpstart
\figsetgrpnum   {02.36}
\figsetgrptitle {NGC~5643}
\figsetplot     {0088}
\figsetgrpnote  {The best fitting model of NGC~5643.}
\figsetgrpend
\figsetgrpstart
\figsetgrpnum   {02.37}
\figsetgrptitle {NGC~5728}
\figsetplot     {0089}
\figsetgrpnote  {The best fitting model of NGC~5728.}
\figsetgrpend
\figsetgrpstart
\figsetgrpnum   {02.38}
\figsetgrptitle {CGCG~164--019}
\figsetplot     {0090}
\figsetgrpnote  {The best fitting model of CGCG~164--019.}
\figsetgrpend
\figsetgrpstart
\figsetgrpnum   {02.39}
\figsetgrptitle {ESO~137--G034}
\figsetplot     {0091}
\figsetgrpnote  {The best fitting model of ESO~137--G034.}
\figsetgrpend
\figsetgrpstart
\figsetgrpnum   {02.40}
\figsetgrptitle {NGC~6232}
\figsetplot     {0092}
\figsetgrpnote  {The best fitting model of NGC~6232.}
\figsetgrpend
\figsetgrpstart
\figsetgrpnum   {02.41}
\figsetgrptitle {ESO~138--G001}
\figsetplot     {0093}
\figsetgrpnote  {The best fitting model of ESO~138--G001.}
\figsetgrpend
\figsetgrpstart
\figsetgrpnum   {02.42}
\figsetgrptitle {NGC~6240}
\figsetplot     {0094}
\figsetgrpnote  {The best fitting model of NGC~6240.}
\figsetgrpend
\figsetgrpstart
\figsetgrpnum   {02.43}
\figsetgrptitle {NGC~6552}
\figsetplot     {0095}
\figsetgrpnote  {The best fitting model of NGC~6552.}
\figsetgrpend
\figsetgrpstart
\figsetgrpnum   {02.44}
\figsetgrptitle {2MASX~J20145928+2523010}
\figsetplot     {0096}
\figsetgrpnote  {The best fitting model of 2MASX~J20145928+2523010.}
\figsetgrpend
\figsetgrpstart
\figsetgrpnum   {02.45}
\figsetgrptitle {ESO~464--G016}
\figsetplot     {0097}
\figsetgrpnote  {The best fitting model of ESO~464--G016.}
\figsetgrpend
\figsetgrpstart
\figsetgrpnum   {02.46}
\figsetgrptitle {NGC~7130}
\figsetplot     {0098}
\figsetgrpnote  {The best fitting model of NGC~7130.}
\figsetgrpend
\figsetgrpstart
\figsetgrpnum   {02.47}
\figsetgrptitle {NGC~7212}
\figsetplot     {0099}
\figsetgrpnote  {The best fitting model of NGC~7212.}
\figsetgrpend
\figsetgrpstart
\figsetgrpnum   {02.48}
\figsetgrptitle {ESO~406--G004}
\figsetplot     {0100}
\figsetgrpnote  {The best fitting model of ESO~406--G004.}
\figsetgrpend
\figsetgrpstart
\figsetgrpnum   {02.49}
\figsetgrptitle {NGC~7479}
\figsetplot     {0101}
\figsetgrpnote  {The best fitting model of NGC~7479.}
\figsetgrpend
\figsetgrpstart
\figsetgrpnum   {02.50}
\figsetgrptitle {SWIFT~J2307.9+2245}
\figsetplot     {0102}
\figsetgrpnote  {The best fitting model of SWIFT~J2307.9+2245.}
\figsetgrpend
\figsetgrpstart
\figsetgrpnum   {02.51}
\figsetgrptitle {NGC~7582}
\figsetplot     {0103}
\figsetgrpnote  {The best fitting model of NGC~7582.}
\figsetgrpend
\figsetgrpstart
\figsetgrpnum   {02.52}
\figsetgrptitle {NGC~7682}
\figsetplot     {0104}
\figsetgrpnote  {The best fitting model of NGC~7682.}
\figsetgrpend
\figsetend    \clearpage
\begin{figure*}\label{Figure03}
\gridline{\fig{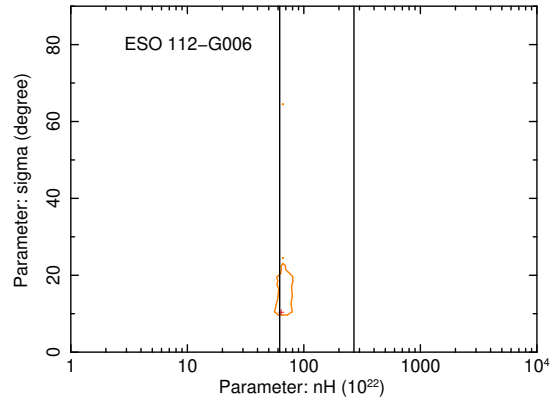}{0.45\textwidth}{}}
\caption{The two-dimensional $\Delta \chi^2$ contour between the hydrogen column density along the line of sight ($N_{\mathrm{H}}^{\mathrm{LOS}}$) and the torus angular width ($\sigma$). The orange line is the 90\% confidence. The black line represents the 90\% confidence interval of $N_{\mathrm{H}}^{\mathrm{LOS}}$ obtained from \cite{Ricci15}. The complete figure set (52 images) is available in the online journal.}
\end{figure*}

\figsetstart
\figsetnum      {52}
\figsettitle    {The two-dimensional $\Delta \chi^2$ contour between the hydrogen column density along the line of sight and the torus angular width}
\figsetgrpstart
\figsetgrpnum   {03.01}
\figsetgrptitle {ESO~112--G006}
\figsetplot     {0105}
\figsetgrpnote  {The two-dimensional $\Delta \chi^2$ contour between the hydrogen column density along the line of sight and the torus angular width of ESO~112--G006.}
\figsetgrpend
\figsetgrpstart
\figsetgrpnum   {03.02}
\figsetgrptitle {MCG~--07--03--007}
\figsetplot     {0106}
\figsetgrpnote  {The two-dimensional $\Delta \chi^2$ contour between the hydrogen column density along the line of sight and the torus angular width of MCG~--07--03--007.}
\figsetgrpend
\figsetgrpstart
\figsetgrpnum   {03.03}
\figsetgrptitle {NGC~0424}
\figsetplot     {0107}
\figsetgrpnote  {The two-dimensional $\Delta \chi^2$ contour between the hydrogen column density along the line of sight and the torus angular width of NGC~0424.}
\figsetgrpend
\figsetgrpstart
\figsetgrpnum   {03.04}
\figsetgrptitle {MCG~+08--03--018}
\figsetplot     {0108}
\figsetgrpnote  {The two-dimensional $\Delta \chi^2$ contour between the hydrogen column density along the line of sight and the torus angular width of MCG~+08--03--018.}
\figsetgrpend
\figsetgrpstart
\figsetgrpnum   {03.05}
\figsetgrptitle {2MASX~J01290761--6038423}
\figsetplot     {0109}
\figsetgrpnote  {The two-dimensional $\Delta \chi^2$ contour between the hydrogen column density along the line of sight and the torus angular width of 2MASX~J01290761--6038423.}
\figsetgrpend
\figsetgrpstart
\figsetgrpnum   {03.06}
\figsetgrptitle {ESO~244--IG030}
\figsetplot     {0110}
\figsetgrpnote  {The two-dimensional $\Delta \chi^2$ contour between the hydrogen column density along the line of sight and the torus angular width of ESO~244--IG030.}
\figsetgrpend
\figsetgrpstart
\figsetgrpnum   {03.07}
\figsetgrptitle {NGC~1106}
\figsetplot     {0111}
\figsetgrpnote  {The two-dimensional $\Delta \chi^2$ contour between the hydrogen column density along the line of sight and the torus angular width of NGC~1106.}
\figsetgrpend
\figsetgrpstart
\figsetgrpnum   {03.08}
\figsetgrptitle {2MFGC~02280}
\figsetplot     {0112}
\figsetgrpnote  {The two-dimensional $\Delta \chi^2$ contour between the hydrogen column density along the line of sight and the torus angular width of 2MFGC~02280.}
\figsetgrpend
\figsetgrpstart
\figsetgrpnum   {03.09}
\figsetgrptitle {NGC~1125}
\figsetplot     {0113}
\figsetgrpnote  {The two-dimensional $\Delta \chi^2$ contour between the hydrogen column density along the line of sight and the torus angular width of NGC~1125.}
\figsetgrpend
\figsetgrpstart
\figsetgrpnum   {03.10}
\figsetgrptitle {NGC~1194}
\figsetplot     {0114}
\figsetgrpnote  {The two-dimensional $\Delta \chi^2$ contour between the hydrogen column density along the line of sight and the torus angular width of NGC~1194.}
\figsetgrpend
\figsetgrpstart
\figsetgrpnum   {03.11}
\figsetgrptitle {NGC~1229}
\figsetplot     {0115}
\figsetgrpnote  {The two-dimensional $\Delta \chi^2$ contour between the hydrogen column density along the line of sight and the torus angular width of NGC~1229.}
\figsetgrpend
\figsetgrpstart
\figsetgrpnum   {03.12}
\figsetgrptitle {ESO~201--IG004}
\figsetplot     {0116}
\figsetgrpnote  {The two-dimensional $\Delta \chi^2$ contour between the hydrogen column density along the line of sight and the torus angular width of ESO~201--IG004.}
\figsetgrpend
\figsetgrpstart
\figsetgrpnum   {03.13}
\figsetgrptitle {2MASX~J03561995--6251391}
\figsetplot     {0117}
\figsetgrpnote  {The two-dimensional $\Delta \chi^2$ contour between the hydrogen column density along the line of sight and the torus angular width of 2MASX~J03561995--6251391.}
\figsetgrpend
\figsetgrpstart
\figsetgrpnum   {03.14}
\figsetgrptitle {MCG~--02--12--017}
\figsetplot     {0118}
\figsetgrpnote  {The two-dimensional $\Delta \chi^2$ contour between the hydrogen column density along the line of sight and the torus angular width of MCG~--02--12--017.}
\figsetgrpend
\figsetgrpstart
\figsetgrpnum   {03.15}
\figsetgrptitle {CGCG~420--015}
\figsetplot     {0119}
\figsetgrpnote  {The two-dimensional $\Delta \chi^2$ contour between the hydrogen column density along the line of sight and the torus angular width of CGCG~420--015.}
\figsetgrpend
\figsetgrpstart
\figsetgrpnum   {03.16}
\figsetgrptitle {ESO~005--G004}
\figsetplot     {0120}
\figsetgrpnote  {The two-dimensional $\Delta \chi^2$ contour between the hydrogen column density along the line of sight and the torus angular width of ESO~005--G004.}
\figsetgrpend
\figsetgrpstart
\figsetgrpnum   {03.17}
\figsetgrptitle {Mrk~0003}
\figsetplot     {0121}
\figsetgrpnote  {The two-dimensional $\Delta \chi^2$ contour between the hydrogen column density along the line of sight and the torus angular width of Mrk~0003.}
\figsetgrpend
\figsetgrpstart
\figsetgrpnum   {03.18}
\figsetgrptitle {2MASX~J06561197--4919499}
\figsetplot     {0122}
\figsetgrpnote  {The two-dimensional $\Delta \chi^2$ contour between the hydrogen column density along the line of sight and the torus angular width of 2MASX~J06561197--4919499.}
\figsetgrpend
\figsetgrpstart
\figsetgrpnum   {03.19}
\figsetgrptitle {MCG~+06--16--028}
\figsetplot     {0123}
\figsetgrpnote  {The two-dimensional $\Delta \chi^2$ contour between the hydrogen column density along the line of sight and the torus angular width of MCG~+06--16--028.}
\figsetgrpend
\figsetgrpstart
\figsetgrpnum   {03.20}
\figsetgrptitle {Mrk~0078}
\figsetplot     {0124}
\figsetgrpnote  {The two-dimensional $\Delta \chi^2$ contour between the hydrogen column density along the line of sight and the torus angular width of Mrk~0078.}
\figsetgrpend
\figsetgrpstart
\figsetgrpnum   {03.21}
\figsetgrptitle {Mrk~0622}
\figsetplot     {0125}
\figsetgrpnote  {The two-dimensional $\Delta \chi^2$ contour between the hydrogen column density along the line of sight and the torus angular width of Mrk~0622.}
\figsetgrpend
\figsetgrpstart
\figsetgrpnum   {03.22}
\figsetgrptitle {NGC~2788A}
\figsetplot     {0126}
\figsetgrpnote  {The two-dimensional $\Delta \chi^2$ contour between the hydrogen column density along the line of sight and the torus angular width of NGC~2788A.}
\figsetgrpend
\figsetgrpstart
\figsetgrpnum   {03.23}
\figsetgrptitle {SBS~0915+556}
\figsetplot     {0127}
\figsetgrpnote  {The two-dimensional $\Delta \chi^2$ contour between the hydrogen column density along the line of sight and the torus angular width of SBS~0915+556.}
\figsetgrpend
\figsetgrpstart
\figsetgrpnum   {03.24}
\figsetgrptitle {2MASX~J09235371--3141305}
\figsetplot     {0128}
\figsetgrpnote  {The two-dimensional $\Delta \chi^2$ contour between the hydrogen column density along the line of sight and the torus angular width of 2MASX~J09235371--3141305.}
\figsetgrpend
\figsetgrpstart
\figsetgrpnum   {03.25}
\figsetgrptitle {ESO~565--G019}
\figsetplot     {0129}
\figsetgrpnote  {The two-dimensional $\Delta \chi^2$ contour between the hydrogen column density along the line of sight and the torus angular width of ESO~565--G019.}
\figsetgrpend
\figsetgrpstart
\figsetgrpnum   {03.26}
\figsetgrptitle {MCG~+10--14--025}
\figsetplot     {0130}
\figsetgrpnote  {The two-dimensional $\Delta \chi^2$ contour between the hydrogen column density along the line of sight and the torus angular width of MCG~+10--14--025.}
\figsetgrpend
\figsetgrpstart
\figsetgrpnum   {03.27}
\figsetgrptitle {NGC~3079}
\figsetplot     {0131}
\figsetgrpnote  {The two-dimensional $\Delta \chi^2$ contour between the hydrogen column density along the line of sight and the torus angular width of NGC~3079.}
\figsetgrpend
\figsetgrpstart
\figsetgrpnum   {03.28}
\figsetgrptitle {ESO~317--G041}
\figsetplot     {0132}
\figsetgrpnote  {The two-dimensional $\Delta \chi^2$ contour between the hydrogen column density along the line of sight and the torus angular width of ESO~317--G041.}
\figsetgrpend
\figsetgrpstart
\figsetgrpnum   {03.29}
\figsetgrptitle {SDSS~J103315.71+525217.8}
\figsetplot     {0133}
\figsetgrpnote  {The two-dimensional $\Delta \chi^2$ contour between the hydrogen column density along the line of sight and the torus angular width of SDSS~J103315.71+525217.8.}
\figsetgrpend
\figsetgrpstart
\figsetgrpnum   {03.30}
\figsetgrptitle {NGC~3393}
\figsetplot     {0134}
\figsetgrpnote  {The two-dimensional $\Delta \chi^2$ contour between the hydrogen column density along the line of sight and the torus angular width of NGC~3393.}
\figsetgrpend
\figsetgrpstart
\figsetgrpnum   {03.31}
\figsetgrptitle {NGC~4102}
\figsetplot     {0135}
\figsetgrpnote  {The two-dimensional $\Delta \chi^2$ contour between the hydrogen column density along the line of sight and the torus angular width of NGC~4102.}
\figsetgrpend
\figsetgrpstart
\figsetgrpnum   {03.32}
\figsetgrptitle {NGC~4180}
\figsetplot     {0136}
\figsetgrpnote  {The two-dimensional $\Delta \chi^2$ contour between the hydrogen column density along the line of sight and the torus angular width of NGC~4180.}
\figsetgrpend
\figsetgrpstart
\figsetgrpnum   {03.33}
\figsetgrptitle {ESO~323--G032}
\figsetplot     {0137}
\figsetgrpnote  {The two-dimensional $\Delta \chi^2$ contour between the hydrogen column density along the line of sight and the torus angular width of ESO~323--G032.}
\figsetgrpend
\figsetgrpstart
\figsetgrpnum   {03.34}
\figsetgrptitle {NGC~4945}
\figsetplot     {0138}
\figsetgrpnote  {The two-dimensional $\Delta \chi^2$ contour between the hydrogen column density along the line of sight and the torus angular width of NGC~4945.}
\figsetgrpend
\figsetgrpstart
\figsetgrpnum   {03.35}
\figsetgrptitle {IGR~J14175--4641}
\figsetplot     {0139}
\figsetgrpnote  {The two-dimensional $\Delta \chi^2$ contour between the hydrogen column density along the line of sight and the torus angular width of IGR~J14175--4641.}
\figsetgrpend
\figsetgrpstart
\figsetgrpnum   {03.36}
\figsetgrptitle {NGC~5643}
\figsetplot     {0140}
\figsetgrpnote  {The two-dimensional $\Delta \chi^2$ contour between the hydrogen column density along the line of sight and the torus angular width of NGC~5643.}
\figsetgrpend
\figsetgrpstart
\figsetgrpnum   {03.37}
\figsetgrptitle {NGC~5728}
\figsetplot     {0141}
\figsetgrpnote  {The two-dimensional $\Delta \chi^2$ contour between the hydrogen column density along the line of sight and the torus angular width of NGC~5728.}
\figsetgrpend
\figsetgrpstart
\figsetgrpnum   {03.38}
\figsetgrptitle {CGCG~164--019}
\figsetplot     {0142}
\figsetgrpnote  {The two-dimensional $\Delta \chi^2$ contour between the hydrogen column density along the line of sight and the torus angular width of CGCG~164--019.}
\figsetgrpend
\figsetgrpstart
\figsetgrpnum   {03.39}
\figsetgrptitle {ESO~137--G034}
\figsetplot     {0143}
\figsetgrpnote  {The two-dimensional $\Delta \chi^2$ contour between the hydrogen column density along the line of sight and the torus angular width of ESO~137--G034.}
\figsetgrpend
\figsetgrpstart
\figsetgrpnum   {03.40}
\figsetgrptitle {NGC~6232}
\figsetplot     {0144}
\figsetgrpnote  {The two-dimensional $\Delta \chi^2$ contour between the hydrogen column density along the line of sight and the torus angular width of NGC~6232.}
\figsetgrpend
\figsetgrpstart
\figsetgrpnum   {03.41}
\figsetgrptitle {ESO~138--G001}
\figsetplot     {0145}
\figsetgrpnote  {The two-dimensional $\Delta \chi^2$ contour between the hydrogen column density along the line of sight and the torus angular width of ESO~138--G001.}
\figsetgrpend
\figsetgrpstart
\figsetgrpnum   {03.42}
\figsetgrptitle {NGC~6240}
\figsetplot     {0146}
\figsetgrpnote  {The two-dimensional $\Delta \chi^2$ contour between the hydrogen column density along the line of sight and the torus angular width of NGC~6240.}
\figsetgrpend
\figsetgrpstart
\figsetgrpnum   {03.43}
\figsetgrptitle {NGC~6552}
\figsetplot     {0147}
\figsetgrpnote  {The two-dimensional $\Delta \chi^2$ contour between the hydrogen column density along the line of sight and the torus angular width of NGC~6552.}
\figsetgrpend
\figsetgrpstart
\figsetgrpnum   {03.44}
\figsetgrptitle {2MASX~J20145928+2523010}
\figsetplot     {0148}
\figsetgrpnote  {The two-dimensional $\Delta \chi^2$ contour between the hydrogen column density along the line of sight and the torus angular width of 2MASX~J20145928+2523010.}
\figsetgrpend
\figsetgrpstart
\figsetgrpnum   {03.45}
\figsetgrptitle {ESO~464--G016}
\figsetplot     {0149}
\figsetgrpnote  {The two-dimensional $\Delta \chi^2$ contour between the hydrogen column density along the line of sight and the torus angular width of ESO~464--G016.}
\figsetgrpend
\figsetgrpstart
\figsetgrpnum   {03.46}
\figsetgrptitle {NGC~7130}
\figsetplot     {0150}
\figsetgrpnote  {The two-dimensional $\Delta \chi^2$ contour between the hydrogen column density along the line of sight and the torus angular width of NGC~7130.}
\figsetgrpend
\figsetgrpstart
\figsetgrpnum   {03.47}
\figsetgrptitle {NGC~7212}
\figsetplot     {0151}
\figsetgrpnote  {The two-dimensional $\Delta \chi^2$ contour between the hydrogen column density along the line of sight and the torus angular width of NGC~7212.}
\figsetgrpend
\figsetgrpstart
\figsetgrpnum   {03.48}
\figsetgrptitle {ESO~406--G004}
\figsetplot     {0152}
\figsetgrpnote  {The two-dimensional $\Delta \chi^2$ contour between the hydrogen column density along the line of sight and the torus angular width of ESO~406--G004.}
\figsetgrpend
\figsetgrpstart
\figsetgrpnum   {03.49}
\figsetgrptitle {NGC~7479}
\figsetplot     {0153}
\figsetgrpnote  {The two-dimensional $\Delta \chi^2$ contour between the hydrogen column density along the line of sight and the torus angular width of NGC~7479.}
\figsetgrpend
\figsetgrpstart
\figsetgrpnum   {03.50}
\figsetgrptitle {SWIFT~J2307.9+2245}
\figsetplot     {0154}
\figsetgrpnote  {The two-dimensional $\Delta \chi^2$ contour between the hydrogen column density along the line of sight and the torus angular width of SWIFT~J2307.9+2245.}
\figsetgrpend
\figsetgrpstart
\figsetgrpnum   {03.51}
\figsetgrptitle {NGC~7582}
\figsetplot     {0155}
\figsetgrpnote  {The two-dimensional $\Delta \chi^2$ contour between the hydrogen column density along the line of sight and the torus angular width of NGC~7582.}
\figsetgrpend
\figsetgrpstart
\figsetgrpnum   {03.52}
\figsetgrptitle {NGC~7682}
\figsetplot     {0156}
\figsetgrpnote  {The two-dimensional $\Delta \chi^2$ contour between the hydrogen column density along the line of sight and the torus angular width of NGC~7682.}
\figsetgrpend
\figsetend
    
\begin{figure*}\label{Figure04}
\gridline{\fig{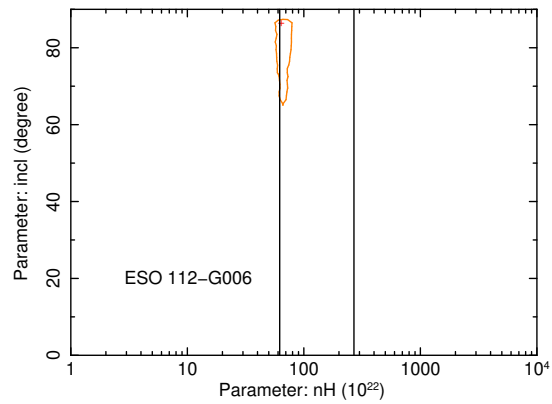}{0.45\textwidth}{}}
\caption{The two-dimensional $\Delta \chi^2$ contour between the hydrogen column density along the line of sight ($N_{\mathrm{H}}^{\mathrm{LOS}}$) and the inclination angle ($i$). The orange line is the 90\% confidence. The black line represents the 90\% confidence interval of $N_{\mathrm{H}}^{\mathrm{LOS}}$ obtained from \cite{Ricci15}. The complete figure set (44 images) is available in the online journal.}
\end{figure*}

\figsetstart
\figsetnum      {44}
\figsettitle    {The two-dimensional $\Delta \chi^2$ contour between the hydrogen column density along the line of sight and the inclination angle}
\figsetgrpstart
\figsetgrpnum   {04.01}
\figsetgrptitle {ESO~112--G006}
\figsetplot     {0157}
\figsetgrpnote  {The two-dimensional $\Delta \chi^2$ contour between the hydrogen column density along the line of sight and the inclination angle of ESO~112--G006.}
\figsetgrpend
\figsetgrpstart
\figsetgrpnum   {04.02}
\figsetgrptitle {MCG~--07--03--007}
\figsetplot     {0158}
\figsetgrpnote  {The two-dimensional $\Delta \chi^2$ contour between the hydrogen column density along the line of sight and the inclination angle of MCG~--07--03--007.}
\figsetgrpend
\figsetgrpstart
\figsetgrpnum   {04.03}
\figsetgrptitle {NGC~0424}
\figsetplot     {0159}
\figsetgrpnote  {The two-dimensional $\Delta \chi^2$ contour between the hydrogen column density along the line of sight and the inclination angle of NGC~0424.}
\figsetgrpend
\figsetgrpstart
\figsetgrpnum   {04.04}
\figsetgrptitle {MCG~+08--03--018}
\figsetplot     {0160}
\figsetgrpnote  {The two-dimensional $\Delta \chi^2$ contour between the hydrogen column density along the line of sight and the inclination angle of MCG~+08--03--018.}
\figsetgrpend
\figsetgrpstart
\figsetgrpnum   {04.05}
\figsetgrptitle {2MASX~J01290761--6038423}
\figsetplot     {0161}
\figsetgrpnote  {The two-dimensional $\Delta \chi^2$ contour between the hydrogen column density along the line of sight and the inclination angle of 2MASX~J01290761--6038423.}
\figsetgrpend
\figsetgrpstart
\figsetgrpnum   {04.06}
\figsetgrptitle {ESO~244--IG030}
\figsetplot     {0162}
\figsetgrpnote  {The two-dimensional $\Delta \chi^2$ contour between the hydrogen column density along the line of sight and the inclination angle of ESO~244--IG030.}
\figsetgrpend
\figsetgrpstart
\figsetgrpnum   {04.07}
\figsetgrptitle {NGC~1106}
\figsetplot     {0163}
\figsetgrpnote  {The two-dimensional $\Delta \chi^2$ contour between the hydrogen column density along the line of sight and the inclination angle of NGC~1106.}
\figsetgrpend
\figsetgrpstart
\figsetgrpnum   {04.08}
\figsetgrptitle {2MFGC~02280}
\figsetplot     {0164}
\figsetgrpnote  {The two-dimensional $\Delta \chi^2$ contour between the hydrogen column density along the line of sight and the inclination angle of 2MFGC~02280.}
\figsetgrpend
\figsetgrpstart
\figsetgrpnum   {04.09}
\figsetgrptitle {NGC~1125}
\figsetplot     {0165}
\figsetgrpnote  {The two-dimensional $\Delta \chi^2$ contour between the hydrogen column density along the line of sight and the inclination angle of NGC~1125.}
\figsetgrpend
\figsetgrpstart
\figsetgrpnum   {04.10}
\figsetgrptitle {NGC~1194}
\figsetplot     {0166}
\figsetgrpnote  {The two-dimensional $\Delta \chi^2$ contour between the hydrogen column density along the line of sight and the inclination angle of NGC~1194.}
\figsetgrpend
\figsetgrpstart
\figsetgrpnum   {04.11}
\figsetgrptitle {NGC~1229}
\figsetplot     {0167}
\figsetgrpnote  {The two-dimensional $\Delta \chi^2$ contour between the hydrogen column density along the line of sight and the inclination angle of NGC~1229.}
\figsetgrpend
\figsetgrpstart
\figsetgrpnum   {04.12}
\figsetgrptitle {ESO~201--IG004}
\figsetplot     {0168}
\figsetgrpnote  {The two-dimensional $\Delta \chi^2$ contour between the hydrogen column density along the line of sight and the inclination angle of ESO~201--IG004.}
\figsetgrpend
\figsetgrpstart
\figsetgrpnum   {04.13}
\figsetgrptitle {2MASX~J03561995--6251391}
\figsetplot     {0169}
\figsetgrpnote  {The two-dimensional $\Delta \chi^2$ contour between the hydrogen column density along the line of sight and the inclination angle of 2MASX~J03561995--6251391.}
\figsetgrpend
\figsetgrpstart
\figsetgrpnum   {04.14}
\figsetgrptitle {MCG~--02--12--017}
\figsetplot     {0170}
\figsetgrpnote  {The two-dimensional $\Delta \chi^2$ contour between the hydrogen column density along the line of sight and the inclination angle of MCG~--02--12--017.}
\figsetgrpend
\figsetgrpstart
\figsetgrpnum   {04.15}
\figsetgrptitle {CGCG~420--015}
\figsetplot     {0171}
\figsetgrpnote  {The two-dimensional $\Delta \chi^2$ contour between the hydrogen column density along the line of sight and the inclination angle of CGCG~420--015.}
\figsetgrpend
\figsetgrpstart
\figsetgrpnum   {04.16}
\figsetgrptitle {ESO~005--G004}
\figsetplot     {0172}
\figsetgrpnote  {The two-dimensional $\Delta \chi^2$ contour between the hydrogen column density along the line of sight and the inclination angle of ESO~005--G004.}
\figsetgrpend
\figsetgrpstart
\figsetgrpnum   {04.18}
\figsetgrptitle {2MASX~J06561197--4919499}
\figsetplot     {0174}
\figsetgrpnote  {The two-dimensional $\Delta \chi^2$ contour between the hydrogen column density along the line of sight and the inclination angle of 2MASX~J06561197--4919499.}
\figsetgrpend
\figsetgrpstart
\figsetgrpnum   {04.19}
\figsetgrptitle {MCG~+06--16--028}
\figsetplot     {0175}
\figsetgrpnote  {The two-dimensional $\Delta \chi^2$ contour between the hydrogen column density along the line of sight and the inclination angle of MCG~+06--16--028.}
\figsetgrpend
\figsetgrpstart
\figsetgrpnum   {04.20}
\figsetgrptitle {Mrk~0078}
\figsetplot     {0176}
\figsetgrpnote  {The two-dimensional $\Delta \chi^2$ contour between the hydrogen column density along the line of sight and the inclination angle of Mrk~0078.}
\figsetgrpend
\figsetgrpstart
\figsetgrpnum   {04.21}
\figsetgrptitle {Mrk~0622}
\figsetplot     {0177}
\figsetgrpnote  {The two-dimensional $\Delta \chi^2$ contour between the hydrogen column density along the line of sight and the inclination angle of Mrk~0622.}
\figsetgrpend
\figsetgrpstart
\figsetgrpnum   {04.22}
\figsetgrptitle {NGC~2788A}
\figsetplot     {0178}
\figsetgrpnote  {The two-dimensional $\Delta \chi^2$ contour between the hydrogen column density along the line of sight and the inclination angle of NGC~2788A.}
\figsetgrpend
\figsetgrpstart
\figsetgrpnum   {04.23}
\figsetgrptitle {SBS~0915+556}
\figsetplot     {0179}
\figsetgrpnote  {The two-dimensional $\Delta \chi^2$ contour between the hydrogen column density along the line of sight and the inclination angle of SBS~0915+556.}
\figsetgrpend
\figsetgrpstart
\figsetgrpnum   {04.24}
\figsetgrptitle {2MASX~J09235371--3141305}
\figsetplot     {0180}
\figsetgrpnote  {The two-dimensional $\Delta \chi^2$ contour between the hydrogen column density along the line of sight and the inclination angle of 2MASX~J09235371--3141305.}
\figsetgrpend
\figsetgrpstart
\figsetgrpnum   {04.25}
\figsetgrptitle {ESO~565--G019}
\figsetplot     {0181}
\figsetgrpnote  {The two-dimensional $\Delta \chi^2$ contour between the hydrogen column density along the line of sight and the inclination angle of ESO~565--G019.}
\figsetgrpend
\figsetgrpstart
\figsetgrpnum   {04.26}
\figsetgrptitle {MCG~+10--14--025}
\figsetplot     {0182}
\figsetgrpnote  {The two-dimensional $\Delta \chi^2$ contour between the hydrogen column density along the line of sight and the inclination angle of MCG~+10--14--025.}
\figsetgrpend
\figsetgrpstart
\figsetgrpnum   {04.28}
\figsetgrptitle {ESO~317--G041}
\figsetplot     {0184}
\figsetgrpnote  {The two-dimensional $\Delta \chi^2$ contour between the hydrogen column density along the line of sight and the inclination angle of ESO~317--G041.}
\figsetgrpend
\figsetgrpstart
\figsetgrpnum   {04.29}
\figsetgrptitle {SDSS~J103315.71+525217.8}
\figsetplot     {0185}
\figsetgrpnote  {The two-dimensional $\Delta \chi^2$ contour between the hydrogen column density along the line of sight and the inclination angle of SDSS~J103315.71+525217.8.}
\figsetgrpend
\figsetgrpstart
\figsetgrpnum   {04.31}
\figsetgrptitle {NGC~4102}
\figsetplot     {0187}
\figsetgrpnote  {The two-dimensional $\Delta \chi^2$ contour between the hydrogen column density along the line of sight and the inclination angle of NGC~4102.}
\figsetgrpend
\figsetgrpstart
\figsetgrpnum   {04.32}
\figsetgrptitle {NGC~4180}
\figsetplot     {0188}
\figsetgrpnote  {The two-dimensional $\Delta \chi^2$ contour between the hydrogen column density along the line of sight and the inclination angle of NGC~4180.}
\figsetgrpend
\figsetgrpstart
\figsetgrpnum   {04.33}
\figsetgrptitle {ESO~323--G032}
\figsetplot     {0189}
\figsetgrpnote  {The two-dimensional $\Delta \chi^2$ contour between the hydrogen column density along the line of sight and the inclination angle of ESO~323--G032.}
\figsetgrpend
\figsetgrpstart
\figsetgrpnum   {04.35}
\figsetgrptitle {IGR~J14175--4641}
\figsetplot     {0191}
\figsetgrpnote  {The two-dimensional $\Delta \chi^2$ contour between the hydrogen column density along the line of sight and the inclination angle of IGR~J14175--4641.}
\figsetgrpend
\figsetgrpstart
\figsetgrpnum   {04.38}
\figsetgrptitle {CGCG~164--019}
\figsetplot     {0194}
\figsetgrpnote  {The two-dimensional $\Delta \chi^2$ contour between the hydrogen column density along the line of sight and the inclination angle of CGCG~164--019.}
\figsetgrpend
\figsetgrpstart
\figsetgrpnum   {04.39}
\figsetgrptitle {ESO~137--G034}
\figsetplot     {0195}
\figsetgrpnote  {The two-dimensional $\Delta \chi^2$ contour between the hydrogen column density along the line of sight and the inclination angle of ESO~137--G034.}
\figsetgrpend
\figsetgrpstart
\figsetgrpnum   {04.40}
\figsetgrptitle {NGC~6232}
\figsetplot     {0196}
\figsetgrpnote  {The two-dimensional $\Delta \chi^2$ contour between the hydrogen column density along the line of sight and the inclination angle of NGC~6232.}
\figsetgrpend
\figsetgrpstart
\figsetgrpnum   {04.41}
\figsetgrptitle {ESO~138--G001}
\figsetplot     {0197}
\figsetgrpnote  {The two-dimensional $\Delta \chi^2$ contour between the hydrogen column density along the line of sight and the inclination angle of ESO~138--G001.}
\figsetgrpend
\figsetgrpstart
\figsetgrpnum   {04.43}
\figsetgrptitle {NGC~6552}
\figsetplot     {0199}
\figsetgrpnote  {The two-dimensional $\Delta \chi^2$ contour between the hydrogen column density along the line of sight and the inclination angle of NGC~6552.}
\figsetgrpend
\figsetgrpstart
\figsetgrpnum   {04.44}
\figsetgrptitle {2MASX~J20145928+2523010}
\figsetplot     {0200}
\figsetgrpnote  {The two-dimensional $\Delta \chi^2$ contour between the hydrogen column density along the line of sight and the inclination angle of 2MASX~J20145928+2523010.}
\figsetgrpend
\figsetgrpstart
\figsetgrpnum   {04.45}
\figsetgrptitle {ESO~464--G016}
\figsetplot     {0201}
\figsetgrpnote  {The two-dimensional $\Delta \chi^2$ contour between the hydrogen column density along the line of sight and the inclination angle of ESO~464--G016.}
\figsetgrpend
\figsetgrpstart
\figsetgrpnum   {04.46}
\figsetgrptitle {NGC~7130}
\figsetplot     {0202}
\figsetgrpnote  {The two-dimensional $\Delta \chi^2$ contour between the hydrogen column density along the line of sight and the inclination angle of NGC~7130.}
\figsetgrpend
\figsetgrpstart
\figsetgrpnum   {04.47}
\figsetgrptitle {NGC~7212}
\figsetplot     {0203}
\figsetgrpnote  {The two-dimensional $\Delta \chi^2$ contour between the hydrogen column density along the line of sight and the inclination angle of NGC~7212.}
\figsetgrpend
\figsetgrpstart
\figsetgrpnum   {04.48}
\figsetgrptitle {ESO~406--G004}
\figsetplot     {0204}
\figsetgrpnote  {The two-dimensional $\Delta \chi^2$ contour between the hydrogen column density along the line of sight and the inclination angle of ESO~406--G004.}
\figsetgrpend
\figsetgrpstart
\figsetgrpnum   {04.50}
\figsetgrptitle {SWIFT~J2307.9+2245}
\figsetplot     {0206}
\figsetgrpnote  {The two-dimensional $\Delta \chi^2$ contour between the hydrogen column density along the line of sight and the inclination angle of SWIFT~J2307.9+2245.}
\figsetgrpend
\figsetgrpstart
\figsetgrpnum   {04.51}
\figsetgrptitle {NGC~7582}
\figsetplot     {0207}
\figsetgrpnote  {The two-dimensional $\Delta \chi^2$ contour between the hydrogen column density along the line of sight and the inclination angle of NGC~7582.}
\figsetgrpend
\figsetgrpstart
\figsetgrpnum   {04.52}
\figsetgrptitle {NGC~7682}
\figsetplot     {0208}
\figsetgrpnote  {The two-dimensional $\Delta \chi^2$ contour between the hydrogen column density along the line of sight and the inclination angle of NGC~7682.}
\figsetgrpend
\figsetend    \clearpage 
\begin{figure*}\label{Figure05}
\gridline{\fig{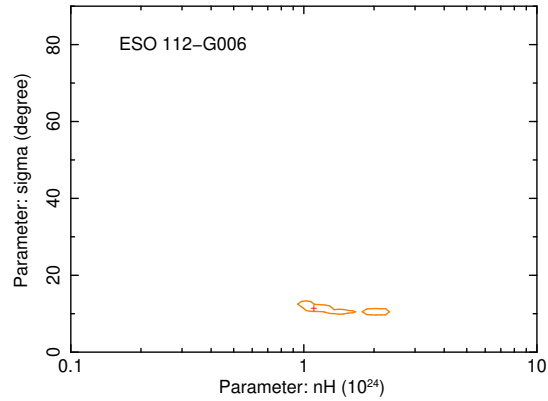}{0.45\textwidth}{}}
\caption{The two-dimensional $\Delta \chi^2$ contour between the hydrogen column density along the equatorial direction ($N_{\mathrm{H}}^{\mathrm{Equ}}$) and the torus angular width ($\sigma$). The orange line is the 90\% confidence. The complete figure set (52 images) is available in the online journal.}
\end{figure*}

\figsetstart
\figsetnum      {52}
\figsettitle    {The two-dimensional $\Delta \chi^2$ contour between the hydrogen column density along the equatorial direction and the torus angular width}
\figsetgrpstart
\figsetgrpnum   {05.01}
\figsetgrptitle {ESO~112--G006}
\figsetplot     {0209}
\figsetgrpnote  {The two-dimensional $\Delta \chi^2$ contour between the hydrogen column density along the equatorial direction and the torus angular width of ESO~112--G006.}
\figsetgrpend
\figsetgrpstart
\figsetgrpnum   {05.02}
\figsetgrptitle {MCG~--07--03--007}
\figsetplot     {0210}
\figsetgrpnote  {The two-dimensional $\Delta \chi^2$ contour between the hydrogen column density along the equatorial direction and the torus angular width of MCG~--07--03--007.}
\figsetgrpend
\figsetgrpstart
\figsetgrpnum   {05.03}
\figsetgrptitle {NGC~0424}
\figsetplot     {0211}
\figsetgrpnote  {The two-dimensional $\Delta \chi^2$ contour between the hydrogen column density along the equatorial direction and the torus angular width of NGC~0424.}
\figsetgrpend
\figsetgrpstart
\figsetgrpnum   {05.04}
\figsetgrptitle {MCG~+08--03--018}
\figsetplot     {0212}
\figsetgrpnote  {The two-dimensional $\Delta \chi^2$ contour between the hydrogen column density along the equatorial direction and the torus angular width of MCG~+08--03--018.}
\figsetgrpend
\figsetgrpstart
\figsetgrpnum   {05.05}
\figsetgrptitle {2MASX~J01290761--6038423}
\figsetplot     {0213}
\figsetgrpnote  {The two-dimensional $\Delta \chi^2$ contour between the hydrogen column density along the equatorial direction and the torus angular width of 2MASX~J01290761--6038423.}
\figsetgrpend
\figsetgrpstart
\figsetgrpnum   {05.06}
\figsetgrptitle {ESO~244--IG030}
\figsetplot     {0214}
\figsetgrpnote  {The two-dimensional $\Delta \chi^2$ contour between the hydrogen column density along the equatorial direction and the torus angular width of ESO~244--IG030.}
\figsetgrpend
\figsetgrpstart
\figsetgrpnum   {05.07}
\figsetgrptitle {NGC~1106}
\figsetplot     {0215}
\figsetgrpnote  {The two-dimensional $\Delta \chi^2$ contour between the hydrogen column density along the equatorial direction and the torus angular width of NGC~1106.}
\figsetgrpend
\figsetgrpstart
\figsetgrpnum   {05.08}
\figsetgrptitle {2MFGC~02280}
\figsetplot     {0216}
\figsetgrpnote  {The two-dimensional $\Delta \chi^2$ contour between the hydrogen column density along the equatorial direction and the torus angular width of 2MFGC~02280.}
\figsetgrpend
\figsetgrpstart
\figsetgrpnum   {05.09}
\figsetgrptitle {NGC~1125}
\figsetplot     {0217}
\figsetgrpnote  {The two-dimensional $\Delta \chi^2$ contour between the hydrogen column density along the equatorial direction and the torus angular width of NGC~1125.}
\figsetgrpend
\figsetgrpstart
\figsetgrpnum   {05.10}
\figsetgrptitle {NGC~1194}
\figsetplot     {0218}
\figsetgrpnote  {The two-dimensional $\Delta \chi^2$ contour between the hydrogen column density along the equatorial direction and the torus angular width of NGC~1194.}
\figsetgrpend
\figsetgrpstart
\figsetgrpnum   {05.11}
\figsetgrptitle {NGC~1229}
\figsetplot     {0219}
\figsetgrpnote  {The two-dimensional $\Delta \chi^2$ contour between the hydrogen column density along the equatorial direction and the torus angular width of NGC~1229.}
\figsetgrpend
\figsetgrpstart
\figsetgrpnum   {05.12}
\figsetgrptitle {ESO~201--IG004}
\figsetplot     {0220}
\figsetgrpnote  {The two-dimensional $\Delta \chi^2$ contour between the hydrogen column density along the equatorial direction and the torus angular width of ESO~201--IG004.}
\figsetgrpend
\figsetgrpstart
\figsetgrpnum   {05.13}
\figsetgrptitle {2MASX~J03561995--6251391}
\figsetplot     {0221}
\figsetgrpnote  {The two-dimensional $\Delta \chi^2$ contour between the hydrogen column density along the equatorial direction and the torus angular width of 2MASX~J03561995--6251391.}
\figsetgrpend
\figsetgrpstart
\figsetgrpnum   {05.14}
\figsetgrptitle {MCG~--02--12--017}
\figsetplot     {0222}
\figsetgrpnote  {The two-dimensional $\Delta \chi^2$ contour between the hydrogen column density along the equatorial direction and the torus angular width of MCG~--02--12--017.}
\figsetgrpend
\figsetgrpstart
\figsetgrpnum   {05.15}
\figsetgrptitle {CGCG~420--015}
\figsetplot     {0223}
\figsetgrpnote  {The two-dimensional $\Delta \chi^2$ contour between the hydrogen column density along the equatorial direction and the torus angular width of CGCG~420--015.}
\figsetgrpend
\figsetgrpstart
\figsetgrpnum   {05.16}
\figsetgrptitle {ESO~005--G004}
\figsetplot     {0224}
\figsetgrpnote  {The two-dimensional $\Delta \chi^2$ contour between the hydrogen column density along the equatorial direction and the torus angular width of ESO~005--G004.}
\figsetgrpend
\figsetgrpstart
\figsetgrpnum   {05.17}
\figsetgrptitle {Mrk~0003}
\figsetplot     {0225}
\figsetgrpnote  {The two-dimensional $\Delta \chi^2$ contour between the hydrogen column density along the equatorial direction and the torus angular width of Mrk~0003.}
\figsetgrpend
\figsetgrpstart
\figsetgrpnum   {05.18}
\figsetgrptitle {2MASX~J06561197--4919499}
\figsetplot     {0226}
\figsetgrpnote  {The two-dimensional $\Delta \chi^2$ contour between the hydrogen column density along the equatorial direction and the torus angular width of 2MASX~J06561197--4919499.}
\figsetgrpend
\figsetgrpstart
\figsetgrpnum   {05.19}
\figsetgrptitle {MCG~+06--16--028}
\figsetplot     {0227}
\figsetgrpnote  {The two-dimensional $\Delta \chi^2$ contour between the hydrogen column density along the equatorial direction and the torus angular width of MCG~+06--16--028.}
\figsetgrpend
\figsetgrpstart
\figsetgrpnum   {05.20}
\figsetgrptitle {Mrk~0078}
\figsetplot     {0228}
\figsetgrpnote  {The two-dimensional $\Delta \chi^2$ contour between the hydrogen column density along the equatorial direction and the torus angular width of Mrk~0078.}
\figsetgrpend
\figsetgrpstart
\figsetgrpnum   {05.21}
\figsetgrptitle {Mrk~0622}
\figsetplot     {0229}
\figsetgrpnote  {The two-dimensional $\Delta \chi^2$ contour between the hydrogen column density along the equatorial direction and the torus angular width of Mrk~0622.}
\figsetgrpend
\figsetgrpstart
\figsetgrpnum   {05.22}
\figsetgrptitle {NGC~2788A}
\figsetplot     {0230}
\figsetgrpnote  {The two-dimensional $\Delta \chi^2$ contour between the hydrogen column density along the equatorial direction and the torus angular width of NGC~2788A.}
\figsetgrpend
\figsetgrpstart
\figsetgrpnum   {05.23}
\figsetgrptitle {SBS~0915+556}
\figsetplot     {0231}
\figsetgrpnote  {The two-dimensional $\Delta \chi^2$ contour between the hydrogen column density along the equatorial direction and the torus angular width of SBS~0915+556.}
\figsetgrpend
\figsetgrpstart
\figsetgrpnum   {05.24}
\figsetgrptitle {2MASX~J09235371--3141305}
\figsetplot     {0232}
\figsetgrpnote  {The two-dimensional $\Delta \chi^2$ contour between the hydrogen column density along the equatorial direction and the torus angular width of 2MASX~J09235371--3141305.}
\figsetgrpend
\figsetgrpstart
\figsetgrpnum   {05.25}
\figsetgrptitle {ESO~565--G019}
\figsetplot     {0233}
\figsetgrpnote  {The two-dimensional $\Delta \chi^2$ contour between the hydrogen column density along the equatorial direction and the torus angular width of ESO~565--G019.}
\figsetgrpend
\figsetgrpstart
\figsetgrpnum   {05.26}
\figsetgrptitle {MCG~+10--14--025}
\figsetplot     {0234}
\figsetgrpnote  {The two-dimensional $\Delta \chi^2$ contour between the hydrogen column density along the equatorial direction and the torus angular width of MCG~+10--14--025.}
\figsetgrpend
\figsetgrpstart
\figsetgrpnum   {05.27}
\figsetgrptitle {NGC~3079}
\figsetplot     {0235}
\figsetgrpnote  {The two-dimensional $\Delta \chi^2$ contour between the hydrogen column density along the equatorial direction and the torus angular width of NGC~3079.}
\figsetgrpend
\figsetgrpstart
\figsetgrpnum   {05.28}
\figsetgrptitle {ESO~317--G041}
\figsetplot     {0236}
\figsetgrpnote  {The two-dimensional $\Delta \chi^2$ contour between the hydrogen column density along the equatorial direction and the torus angular width of ESO~317--G041.}
\figsetgrpend
\figsetgrpstart
\figsetgrpnum   {05.29}
\figsetgrptitle {SDSS~J103315.71+525217.8}
\figsetplot     {0237}
\figsetgrpnote  {The two-dimensional $\Delta \chi^2$ contour between the hydrogen column density along the equatorial direction and the torus angular width of SDSS~J103315.71+525217.8.}
\figsetgrpend
\figsetgrpstart
\figsetgrpnum   {05.30}
\figsetgrptitle {NGC~3393}
\figsetplot     {0238}
\figsetgrpnote  {The two-dimensional $\Delta \chi^2$ contour between the hydrogen column density along the equatorial direction and the torus angular width of NGC~3393.}
\figsetgrpend
\figsetgrpstart
\figsetgrpnum   {05.31}
\figsetgrptitle {NGC~4102}
\figsetplot     {0239}
\figsetgrpnote  {The two-dimensional $\Delta \chi^2$ contour between the hydrogen column density along the equatorial direction and the torus angular width of NGC~4102.}
\figsetgrpend
\figsetgrpstart
\figsetgrpnum   {05.32}
\figsetgrptitle {NGC~4180}
\figsetplot     {0240}
\figsetgrpnote  {The two-dimensional $\Delta \chi^2$ contour between the hydrogen column density along the equatorial direction and the torus angular width of NGC~4180.}
\figsetgrpend
\figsetgrpstart
\figsetgrpnum   {05.33}
\figsetgrptitle {ESO~323--G032}
\figsetplot     {0241}
\figsetgrpnote  {The two-dimensional $\Delta \chi^2$ contour between the hydrogen column density along the equatorial direction and the torus angular width of ESO~323--G032.}
\figsetgrpend
\figsetgrpstart
\figsetgrpnum   {05.34}
\figsetgrptitle {NGC~4945}
\figsetplot     {0242}
\figsetgrpnote  {The two-dimensional $\Delta \chi^2$ contour between the hydrogen column density along the equatorial direction and the torus angular width of NGC~4945.}
\figsetgrpend
\figsetgrpstart
\figsetgrpnum   {05.35}
\figsetgrptitle {IGR~J14175--4641}
\figsetplot     {0243}
\figsetgrpnote  {The two-dimensional $\Delta \chi^2$ contour between the hydrogen column density along the equatorial direction and the torus angular width of IGR~J14175--4641.}
\figsetgrpend
\figsetgrpstart
\figsetgrpnum   {05.36}
\figsetgrptitle {NGC~5643}
\figsetplot     {0244}
\figsetgrpnote  {The two-dimensional $\Delta \chi^2$ contour between the hydrogen column density along the equatorial direction and the torus angular width of NGC~5643.}
\figsetgrpend
\figsetgrpstart
\figsetgrpnum   {05.37}
\figsetgrptitle {NGC~5728}
\figsetplot     {0245}
\figsetgrpnote  {The two-dimensional $\Delta \chi^2$ contour between the hydrogen column density along the equatorial direction and the torus angular width of NGC~5728.}
\figsetgrpend
\figsetgrpstart
\figsetgrpnum   {05.38}
\figsetgrptitle {CGCG~164--019}
\figsetplot     {0246}
\figsetgrpnote  {The two-dimensional $\Delta \chi^2$ contour between the hydrogen column density along the equatorial direction and the torus angular width of CGCG~164--019.}
\figsetgrpend
\figsetgrpstart
\figsetgrpnum   {05.39}
\figsetgrptitle {ESO~137--G034}
\figsetplot     {0247}
\figsetgrpnote  {The two-dimensional $\Delta \chi^2$ contour between the hydrogen column density along the equatorial direction and the torus angular width of ESO~137--G034.}
\figsetgrpend
\figsetgrpstart
\figsetgrpnum   {05.40}
\figsetgrptitle {NGC~6232}
\figsetplot     {0248}
\figsetgrpnote  {The two-dimensional $\Delta \chi^2$ contour between the hydrogen column density along the equatorial direction and the torus angular width of NGC~6232.}
\figsetgrpend
\figsetgrpstart
\figsetgrpnum   {05.41}
\figsetgrptitle {ESO~138--G001}
\figsetplot     {0249}
\figsetgrpnote  {The two-dimensional $\Delta \chi^2$ contour between the hydrogen column density along the equatorial direction and the torus angular width of ESO~138--G001.}
\figsetgrpend
\figsetgrpstart
\figsetgrpnum   {05.42}
\figsetgrptitle {NGC~6240}
\figsetplot     {0250}
\figsetgrpnote  {The two-dimensional $\Delta \chi^2$ contour between the hydrogen column density along the equatorial direction and the torus angular width of NGC~6240.}
\figsetgrpend
\figsetgrpstart
\figsetgrpnum   {05.43}
\figsetgrptitle {NGC~6552}
\figsetplot     {0251}
\figsetgrpnote  {The two-dimensional $\Delta \chi^2$ contour between the hydrogen column density along the equatorial direction and the torus angular width of NGC~6552.}
\figsetgrpend
\figsetgrpstart
\figsetgrpnum   {05.44}
\figsetgrptitle {2MASX~J20145928+2523010}
\figsetplot     {0252}
\figsetgrpnote  {The two-dimensional $\Delta \chi^2$ contour between the hydrogen column density along the equatorial direction and the torus angular width of 2MASX~J20145928+2523010.}
\figsetgrpend
\figsetgrpstart
\figsetgrpnum   {05.45}
\figsetgrptitle {ESO~464--G016}
\figsetplot     {0253}
\figsetgrpnote  {The two-dimensional $\Delta \chi^2$ contour between the hydrogen column density along the equatorial direction and the torus angular width of ESO~464--G016.}
\figsetgrpend
\figsetgrpstart
\figsetgrpnum   {05.46}
\figsetgrptitle {NGC~7130}
\figsetplot     {0254}
\figsetgrpnote  {The two-dimensional $\Delta \chi^2$ contour between the hydrogen column density along the equatorial direction and the torus angular width of NGC~7130.}
\figsetgrpend
\figsetgrpstart
\figsetgrpnum   {05.47}
\figsetgrptitle {NGC~7212}
\figsetplot     {0255}
\figsetgrpnote  {The two-dimensional $\Delta \chi^2$ contour between the hydrogen column density along the equatorial direction and the torus angular width of NGC~7212.}
\figsetgrpend
\figsetgrpstart
\figsetgrpnum   {05.48}
\figsetgrptitle {ESO~406--G004}
\figsetplot     {0256}
\figsetgrpnote  {The two-dimensional $\Delta \chi^2$ contour between the hydrogen column density along the equatorial direction and the torus angular width of ESO~406--G004.}
\figsetgrpend
\figsetgrpstart
\figsetgrpnum   {05.49}
\figsetgrptitle {NGC~7479}
\figsetplot     {0257}
\figsetgrpnote  {The two-dimensional $\Delta \chi^2$ contour between the hydrogen column density along the equatorial direction and the torus angular width of NGC~7479.}
\figsetgrpend
\figsetgrpstart
\figsetgrpnum   {05.50}
\figsetgrptitle {SWIFT~J2307.9+2245}
\figsetplot     {0258}
\figsetgrpnote  {The two-dimensional $\Delta \chi^2$ contour between the hydrogen column density along the equatorial direction and the torus angular width of SWIFT~J2307.9+2245.}
\figsetgrpend
\figsetgrpstart
\figsetgrpnum   {05.51}
\figsetgrptitle {NGC~7582}
\figsetplot     {0259}
\figsetgrpnote  {The two-dimensional $\Delta \chi^2$ contour between the hydrogen column density along the equatorial direction and the torus angular width of NGC~7582.}
\figsetgrpend
\figsetgrpstart
\figsetgrpnum   {05.52}
\figsetgrptitle {NGC~7682}
\figsetplot     {0260}
\figsetgrpnote  {The two-dimensional $\Delta \chi^2$ contour between the hydrogen column density along the equatorial direction and the torus angular width of NGC~7682.}
\figsetgrpend
\figsetend
\begin{figure*}\label{Figure06}
\gridline{\fig{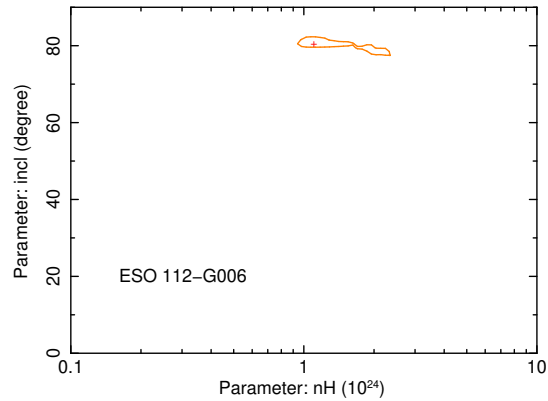}{0.45\textwidth}{}}
\caption{The two-dimensional $\Delta \chi^2$ contour between the hydrogen column density along the equatorial direction ($N_{\mathrm{H}}^{\mathrm{Equ}}$) and the inclination angle ($i$). The orange line is the 90\% confidence. The complete figure set (44 images) is available in the online journal.}
\end{figure*}

\figsetstart
\figsetnum      {44}
\figsettitle    {The two-dimensional $\Delta \chi^2$ contour between the hydrogen column density along the equatorial direction and the inclination angle}
\figsetgrpstart
\figsetgrpnum   {06.01}
\figsetgrptitle {ESO~112--G006}
\figsetplot     {0261}
\figsetgrpnote  {The two-dimensional $\Delta \chi^2$ contour between the hydrogen column density along the equatorial direction and the inclination angle of ESO~112--G006.}
\figsetgrpend
\figsetgrpstart
\figsetgrpnum   {06.02}
\figsetgrptitle {MCG~--07--03--007}
\figsetplot     {0262}
\figsetgrpnote  {The two-dimensional $\Delta \chi^2$ contour between the hydrogen column density along the equatorial direction and the inclination angle of MCG~--07--03--007.}
\figsetgrpend
\figsetgrpstart
\figsetgrpnum   {06.03}
\figsetgrptitle {NGC~0424}
\figsetplot     {0263}
\figsetgrpnote  {The two-dimensional $\Delta \chi^2$ contour between the hydrogen column density along the equatorial direction and the inclination angle of NGC~0424.}
\figsetgrpend
\figsetgrpstart
\figsetgrpnum   {06.04}
\figsetgrptitle {MCG~+08--03--018}
\figsetplot     {0264}
\figsetgrpnote  {The two-dimensional $\Delta \chi^2$ contour between the hydrogen column density along the equatorial direction and the inclination angle of MCG~+08--03--018.}
\figsetgrpend
\figsetgrpstart
\figsetgrpnum   {06.05}
\figsetgrptitle {2MASX~J01290761--6038423}
\figsetplot     {0265}
\figsetgrpnote  {The two-dimensional $\Delta \chi^2$ contour between the hydrogen column density along the equatorial direction and the inclination angle of 2MASX~J01290761--6038423.}
\figsetgrpend
\figsetgrpstart
\figsetgrpnum   {06.06}
\figsetgrptitle {ESO~244--IG030}
\figsetplot     {0266}
\figsetgrpnote  {The two-dimensional $\Delta \chi^2$ contour between the hydrogen column density along the equatorial direction and the inclination angle of ESO~244--IG030.}
\figsetgrpend
\figsetgrpstart
\figsetgrpnum   {06.07}
\figsetgrptitle {NGC~1106}
\figsetplot     {0267}
\figsetgrpnote  {The two-dimensional $\Delta \chi^2$ contour between the hydrogen column density along the equatorial direction and the inclination angle of NGC~1106.}
\figsetgrpend
\figsetgrpstart
\figsetgrpnum   {06.08}
\figsetgrptitle {2MFGC~02280}
\figsetplot     {0268}
\figsetgrpnote  {The two-dimensional $\Delta \chi^2$ contour between the hydrogen column density along the equatorial direction and the inclination angle of 2MFGC~02280.}
\figsetgrpend
\figsetgrpstart
\figsetgrpnum   {06.09}
\figsetgrptitle {NGC~1125}
\figsetplot     {0269}
\figsetgrpnote  {The two-dimensional $\Delta \chi^2$ contour between the hydrogen column density along the equatorial direction and the inclination angle of NGC~1125.}
\figsetgrpend
\figsetgrpstart
\figsetgrpnum   {06.10}
\figsetgrptitle {NGC~1194}
\figsetplot     {0270}
\figsetgrpnote  {The two-dimensional $\Delta \chi^2$ contour between the hydrogen column density along the equatorial direction and the inclination angle of NGC~1194.}
\figsetgrpend
\figsetgrpstart
\figsetgrpnum   {06.11}
\figsetgrptitle {NGC~1229}
\figsetplot     {0271}
\figsetgrpnote  {The two-dimensional $\Delta \chi^2$ contour between the hydrogen column density along the equatorial direction and the inclination angle of NGC~1229.}
\figsetgrpend
\figsetgrpstart
\figsetgrpnum   {06.12}
\figsetgrptitle {ESO~201--IG004}
\figsetplot     {0272}
\figsetgrpnote  {The two-dimensional $\Delta \chi^2$ contour between the hydrogen column density along the equatorial direction and the inclination angle of ESO~201--IG004.}
\figsetgrpend
\figsetgrpstart
\figsetgrpnum   {06.13}
\figsetgrptitle {2MASX~J03561995--6251391}
\figsetplot     {0273}
\figsetgrpnote  {The two-dimensional $\Delta \chi^2$ contour between the hydrogen column density along the equatorial direction and the inclination angle of 2MASX~J03561995--6251391.}
\figsetgrpend
\figsetgrpstart
\figsetgrpnum   {06.14}
\figsetgrptitle {MCG~--02--12--017}
\figsetplot     {0274}
\figsetgrpnote  {The two-dimensional $\Delta \chi^2$ contour between the hydrogen column density along the equatorial direction and the inclination angle of MCG~--02--12--017.}
\figsetgrpend
\figsetgrpstart
\figsetgrpnum   {06.15}
\figsetgrptitle {CGCG~420--015}
\figsetplot     {0275}
\figsetgrpnote  {The two-dimensional $\Delta \chi^2$ contour between the hydrogen column density along the equatorial direction and the inclination angle of CGCG~420--015.}
\figsetgrpend
\figsetgrpstart
\figsetgrpnum   {06.16}
\figsetgrptitle {ESO~005--G004}
\figsetplot     {0276}
\figsetgrpnote  {The two-dimensional $\Delta \chi^2$ contour between the hydrogen column density along the equatorial direction and the inclination angle of ESO~005--G004.}
\figsetgrpend
\figsetgrpstart
\figsetgrpnum   {06.18}
\figsetgrptitle {2MASX~J06561197--4919499}
\figsetplot     {0278}
\figsetgrpnote  {The two-dimensional $\Delta \chi^2$ contour between the hydrogen column density along the equatorial direction and the inclination angle of 2MASX~J06561197--4919499.}
\figsetgrpend
\figsetgrpstart
\figsetgrpnum   {06.19}
\figsetgrptitle {MCG~+06--16--028}
\figsetplot     {0279}
\figsetgrpnote  {The two-dimensional $\Delta \chi^2$ contour between the hydrogen column density along the equatorial direction and the inclination angle of MCG~+06--16--028.}
\figsetgrpend
\figsetgrpstart
\figsetgrpnum   {06.20}
\figsetgrptitle {Mrk~0078}
\figsetplot     {0280}
\figsetgrpnote  {The two-dimensional $\Delta \chi^2$ contour between the hydrogen column density along the equatorial direction and the inclination angle of Mrk~0078.}
\figsetgrpend
\figsetgrpstart
\figsetgrpnum   {06.21}
\figsetgrptitle {Mrk~0622}
\figsetplot     {0281}
\figsetgrpnote  {The two-dimensional $\Delta \chi^2$ contour between the hydrogen column density along the equatorial direction and the inclination angle of Mrk~0622.}
\figsetgrpend
\figsetgrpstart
\figsetgrpnum   {06.22}
\figsetgrptitle {NGC~2788A}
\figsetplot     {0282}
\figsetgrpnote  {The two-dimensional $\Delta \chi^2$ contour between the hydrogen column density along the equatorial direction and the inclination angle of NGC~2788A.}
\figsetgrpend
\figsetgrpstart
\figsetgrpnum   {06.23}
\figsetgrptitle {SBS~0915+556}
\figsetplot     {0283}
\figsetgrpnote  {The two-dimensional $\Delta \chi^2$ contour between the hydrogen column density along the equatorial direction and the inclination angle of SBS~0915+556.}
\figsetgrpend
\figsetgrpstart
\figsetgrpnum   {06.24}
\figsetgrptitle {2MASX~J09235371--3141305}
\figsetplot     {0284}
\figsetgrpnote  {The two-dimensional $\Delta \chi^2$ contour between the hydrogen column density along the equatorial direction and the inclination angle of 2MASX~J09235371--3141305.}
\figsetgrpend
\figsetgrpstart
\figsetgrpnum   {06.25}
\figsetgrptitle {ESO~565--G019}
\figsetplot     {0285}
\figsetgrpnote  {The two-dimensional $\Delta \chi^2$ contour between the hydrogen column density along the equatorial direction and the inclination angle of ESO~565--G019.}
\figsetgrpend
\figsetgrpstart
\figsetgrpnum   {06.26}
\figsetgrptitle {MCG~+10--14--025}
\figsetplot     {0286}
\figsetgrpnote  {The two-dimensional $\Delta \chi^2$ contour between the hydrogen column density along the equatorial direction and the inclination angle of MCG~+10--14--025.}
\figsetgrpend
\figsetgrpstart
\figsetgrpnum   {06.28}
\figsetgrptitle {ESO~317--G041}
\figsetplot     {0288}
\figsetgrpnote  {The two-dimensional $\Delta \chi^2$ contour between the hydrogen column density along the equatorial direction and the inclination angle of ESO~317--G041.}
\figsetgrpend
\figsetgrpstart
\figsetgrpnum   {06.29}
\figsetgrptitle {SDSS~J103315.71+525217.8}
\figsetplot     {0289}
\figsetgrpnote  {The two-dimensional $\Delta \chi^2$ contour between the hydrogen column density along the equatorial direction and the inclination angle of SDSS~J103315.71+525217.8.}
\figsetgrpend
\figsetgrpstart
\figsetgrpnum   {06.31}
\figsetgrptitle {NGC~4102}
\figsetplot     {0291}
\figsetgrpnote  {The two-dimensional $\Delta \chi^2$ contour between the hydrogen column density along the equatorial direction and the inclination angle of NGC~4102.}
\figsetgrpend
\figsetgrpstart
\figsetgrpnum   {06.32}
\figsetgrptitle {NGC~4180}
\figsetplot     {0292}
\figsetgrpnote  {The two-dimensional $\Delta \chi^2$ contour between the hydrogen column density along the equatorial direction and the inclination angle of NGC~4180.}
\figsetgrpend
\figsetgrpstart
\figsetgrpnum   {06.33}
\figsetgrptitle {ESO~323--G032}
\figsetplot     {0293}
\figsetgrpnote  {The two-dimensional $\Delta \chi^2$ contour between the hydrogen column density along the equatorial direction and the inclination angle of ESO~323--G032.}
\figsetgrpend
\figsetgrpstart
\figsetgrpnum   {06.35}
\figsetgrptitle {IGR~J14175--4641}
\figsetplot     {0295}
\figsetgrpnote  {The two-dimensional $\Delta \chi^2$ contour between the hydrogen column density along the equatorial direction and the inclination angle of IGR~J14175--4641.}
\figsetgrpend
\figsetgrpstart
\figsetgrpnum   {06.38}
\figsetgrptitle {CGCG~164--019}
\figsetplot     {0298}
\figsetgrpnote  {The two-dimensional $\Delta \chi^2$ contour between the hydrogen column density along the equatorial direction and the inclination angle of CGCG~164--019.}
\figsetgrpend
\figsetgrpstart
\figsetgrpnum   {06.39}
\figsetgrptitle {ESO~137--G034}
\figsetplot     {0299}
\figsetgrpnote  {The two-dimensional $\Delta \chi^2$ contour between the hydrogen column density along the equatorial direction and the inclination angle of ESO~137--G034.}
\figsetgrpend
\figsetgrpstart
\figsetgrpnum   {06.40}
\figsetgrptitle {NGC~6232}
\figsetplot     {0300}
\figsetgrpnote  {The two-dimensional $\Delta \chi^2$ contour between the hydrogen column density along the equatorial direction and the inclination angle of NGC~6232.}
\figsetgrpend
\figsetgrpstart
\figsetgrpnum   {06.41}
\figsetgrptitle {ESO~138--G001}
\figsetplot     {0301}
\figsetgrpnote  {The two-dimensional $\Delta \chi^2$ contour between the hydrogen column density along the equatorial direction and the inclination angle of ESO~138--G001.}
\figsetgrpend
\figsetgrpstart
\figsetgrpnum   {06.43}
\figsetgrptitle {NGC~6552}
\figsetplot     {0303}
\figsetgrpnote  {The two-dimensional $\Delta \chi^2$ contour between the hydrogen column density along the equatorial direction and the inclination angle of NGC~6552.}
\figsetgrpend
\figsetgrpstart
\figsetgrpnum   {06.44}
\figsetgrptitle {2MASX~J20145928+2523010}
\figsetplot     {0304}
\figsetgrpnote  {The two-dimensional $\Delta \chi^2$ contour between the hydrogen column density along the equatorial direction and the inclination angle of 2MASX~J20145928+2523010.}
\figsetgrpend
\figsetgrpstart
\figsetgrpnum   {06.45}
\figsetgrptitle {ESO~464--G016}
\figsetplot     {0305}
\figsetgrpnote  {The two-dimensional $\Delta \chi^2$ contour between the hydrogen column density along the equatorial direction and the inclination angle of ESO~464--G016.}
\figsetgrpend
\figsetgrpstart
\figsetgrpnum   {06.46}
\figsetgrptitle {NGC 7130}
\figsetplot     {0306}
\figsetgrpnote  {The two-dimensional $\Delta \chi^2$ contour between the hydrogen column density along the equatorial direction and the inclination angle of NGC~7130.}
\figsetgrpend
\figsetgrpstart
\figsetgrpnum   {06.47}
\figsetgrptitle {NGC~7212}
\figsetplot     {0307}
\figsetgrpnote  {The two-dimensional $\Delta \chi^2$ contour between the hydrogen column density along the equatorial direction and the inclination angle of NGC~7212.}
\figsetgrpend
\figsetgrpstart
\figsetgrpnum   {06.48}
\figsetgrptitle {ESO~406--G004}
\figsetplot     {0308}
\figsetgrpnote  {The two-dimensional $\Delta \chi^2$ contour between the hydrogen column density along the equatorial direction and the inclination angle of ESO~406--G004.}
\figsetgrpend
\figsetgrpstart
\figsetgrpnum   {06.50}
\figsetgrptitle {SWIFT~J2307.9+2245}
\figsetplot     {0310}
\figsetgrpnote  {The two-dimensional $\Delta \chi^2$ contour between the hydrogen column density along the equatorial direction and the inclination angle of SWIFT~J2307.9+2245.}
\figsetgrpend
\figsetgrpstart
\figsetgrpnum   {06.51}
\figsetgrptitle {NGC~7582}
\figsetplot     {0311}
\figsetgrpnote  {The two-dimensional $\Delta \chi^2$ contour between the hydrogen column density along the equatorial direction and the inclination angle of NGC~7582.}
\figsetgrpend
\figsetend
\figsetgrpstart
\figsetgrpnum   {06.52}
\figsetgrptitle {NGC~7682}
\figsetplot     {0312}
\figsetgrpnote  {The two-dimensional $\Delta \chi^2$ contour between the hydrogen column density along the equatorial direction and the inclination angle of NGC~7682.}
\figsetgrpend
\figsetend    \clearpage
\begin{figure*}\label{Figure07}
\gridline{\fig{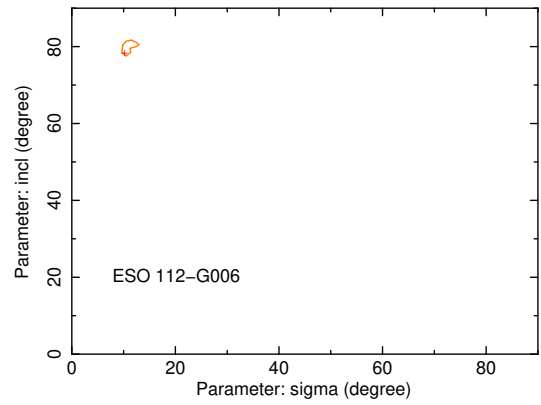}{0.45\textwidth}{}}
\caption{The two-dimensional $\Delta \chi^2$ contour between the torus angular width ($\sigma$) and the inclination angle ($i$). The orange line is the 90\% confidence. The complete figure set (44 images) is available in the online journal.}
\end{figure*}

\figsetstart
\figsetnum      {44}
\figsettitle    {The two-dimensional $\Delta \chi^2$ contour between the torus angular width and the inclination angle.}
\figsetgrpstart
\figsetgrpnum   {07.01}
\figsetgrptitle {ESO~112--G006}
\figsetplot     {0313}
\figsetgrpnote  {The two-dimensional $\Delta \chi^2$ contour between the torus angular width and the inclination angle of ESO~112--G006.}
\figsetgrpend
\figsetgrpstart
\figsetgrpnum   {07.02}
\figsetgrptitle {MCG~--07--03--007}
\figsetplot     {0314}
\figsetgrpnote  {The two-dimensional $\Delta \chi^2$ contour between the torus angular width and the inclination angle of MCG~--07--03--007.}
\figsetgrpend
\figsetgrpstart
\figsetgrpnum   {07.03}
\figsetgrptitle {NGC~0424}
\figsetplot     {0315}
\figsetgrpnote  {The two-dimensional $\Delta \chi^2$ contour between the torus angular width and the inclination angle of NGC~0424.}
\figsetgrpend
\figsetgrpstart
\figsetgrpnum   {07.04}
\figsetgrptitle {MCG~+08--03--018}
\figsetplot     {0316}
\figsetgrpnote  {The two-dimensional $\Delta \chi^2$ contour between the torus angular width and the inclination angle of MCG~+08--03--018.}
\figsetgrpend
\figsetgrpstart
\figsetgrpnum   {07.05}
\figsetgrptitle {2MASX~J01290761--6038423}
\figsetplot     {0317}
\figsetgrpnote  {The two-dimensional $\Delta \chi^2$ contour between the torus angular width and the inclination angle of 2MASX~J01290761--6038423.}
\figsetgrpend
\figsetgrpstart
\figsetgrpnum   {07.06}
\figsetgrptitle {ESO~244--IG030}
\figsetplot     {0318}
\figsetgrpnote  {The two-dimensional $\Delta \chi^2$ contour between the torus angular width and the inclination angle of ESO~244--IG030.}
\figsetgrpend
\figsetgrpstart
\figsetgrpnum   {07.07}
\figsetgrptitle {NGC~1106}
\figsetplot     {0319}
\figsetgrpnote  {The two-dimensional $\Delta \chi^2$ contour between the torus angular width and the inclination angle of NGC~1106.}
\figsetgrpend
\figsetgrpstart
\figsetgrpnum   {07.08}
\figsetgrptitle {2MFGC~02280}
\figsetplot     {0320}
\figsetgrpnote  {The two-dimensional $\Delta \chi^2$ contour between the torus angular width and the inclination angle of 2MFGC~02280.}
\figsetgrpend
\figsetgrpstart
\figsetgrpnum   {07.09}
\figsetgrptitle {NGC~1125}
\figsetplot     {0321}
\figsetgrpnote  {The two-dimensional $\Delta \chi^2$ contour between the torus angular width and the inclination angle of NGC~1125.}
\figsetgrpend
\figsetgrpstart
\figsetgrpnum   {07.10}
\figsetgrptitle {NGC~1194}
\figsetplot     {0322}
\figsetgrpnote  {The two-dimensional $\Delta \chi^2$ contour between the torus angular width and the inclination angle of NGC~1194.}
\figsetgrpend
\figsetgrpstart
\figsetgrpnum   {07.11}
\figsetgrptitle {NGC~1229}
\figsetplot     {0323}
\figsetgrpnote  {The two-dimensional $\Delta \chi^2$ contour between the torus angular width and the inclination angle of NGC~1229.}
\figsetgrpend
\figsetgrpstart
\figsetgrpnum   {07.12}
\figsetgrptitle {ESO~201--IG004}
\figsetplot     {0324}
\figsetgrpnote  {The two-dimensional $\Delta \chi^2$ contour between the torus angular width and the inclination angle of ESO~201--IG004.}
\figsetgrpend
\figsetgrpstart
\figsetgrpnum   {07.13}
\figsetgrptitle {2MASX~J03561995--6251391}
\figsetplot     {0325}
\figsetgrpnote  {The two-dimensional $\Delta \chi^2$ contour between the torus angular width and the inclination angle of 2MASX~J03561995--6251391.}
\figsetgrpend
\figsetgrpstart
\figsetgrpnum   {07.14}
\figsetgrptitle {MCG~--02--12--017}
\figsetplot     {0326}
\figsetgrpnote  {The two-dimensional $\Delta \chi^2$ contour between the torus angular width and the inclination angle of MCG~--02--12--017.}
\figsetgrpend
\figsetgrpstart
\figsetgrpnum   {07.15}
\figsetgrptitle {CGCG~420--015}
\figsetplot     {0327}
\figsetgrpnote  {The two-dimensional $\Delta \chi^2$ contour between the torus angular width and the inclination angle of CGCG~420--015.}
\figsetgrpend
\figsetgrpstart
\figsetgrpnum   {07.16}
\figsetgrptitle {ESO~005--G004}
\figsetplot     {0328}
\figsetgrpnote  {The two-dimensional $\Delta \chi^2$ contour between the torus angular width and the inclination angle of ESO~005--G004.}
\figsetgrpend
\figsetgrpstart
\figsetgrpnum   {07.18}
\figsetgrptitle {2MASX~J06561197--4919499}
\figsetplot     {0330}
\figsetgrpnote  {The two-dimensional $\Delta \chi^2$ contour between the torus angular width and the inclination angle of 2MASX~J06561197--4919499.}
\figsetgrpend
\figsetgrpstart
\figsetgrpnum   {07.19}
\figsetgrptitle {MCG~+06--16--028}
\figsetplot     {0331}
\figsetgrpnote  {The two-dimensional $\Delta \chi^2$ contour between the torus angular width and the inclination angle of MCG~+06--16--028.}
\figsetgrpend
\figsetgrpstart
\figsetgrpnum   {07.20}
\figsetgrptitle {Mrk~0078}
\figsetplot     {0332}
\figsetgrpnote  {The two-dimensional $\Delta \chi^2$ contour between the torus angular width and the inclination angle of Mrk~0078.}
\figsetgrpend
\figsetgrpstart
\figsetgrpnum   {07.21}
\figsetgrptitle {Mrk~0622}
\figsetplot     {0333}
\figsetgrpnote  {The two-dimensional $\Delta \chi^2$ contour between the torus angular width and the inclination angle of Mrk~0622.}
\figsetgrpend
\figsetgrpstart
\figsetgrpnum   {07.22}
\figsetgrptitle {NGC~2788A}
\figsetplot     {0334}
\figsetgrpnote  {The two-dimensional $\Delta \chi^2$ contour between the torus angular width and the inclination angle of NGC~2788A.}
\figsetgrpend
\figsetgrpstart
\figsetgrpnum   {07.23}
\figsetgrptitle {SBS~0915+556}
\figsetplot     {0335}
\figsetgrpnote  {The two-dimensional $\Delta \chi^2$ contour between the torus angular width and the inclination angle of SBS~0915+556.}
\figsetgrpend
\figsetgrpstart
\figsetgrpnum   {07.24}
\figsetgrptitle {2MASX~J09235371--3141305}
\figsetplot     {0336}
\figsetgrpnote  {The two-dimensional $\Delta \chi^2$ contour between the torus angular width and the inclination angle of 2MASX~J09235371--3141305.}
\figsetgrpend
\figsetgrpstart
\figsetgrpnum   {07.25}
\figsetgrptitle {ESO~565--G019}
\figsetplot     {0337}
\figsetgrpnote  {The two-dimensional $\Delta \chi^2$ contour between the torus angular width and the inclination angle of ESO~565--G019.}
\figsetgrpend
\figsetgrpstart
\figsetgrpnum   {07.26}
\figsetgrptitle {MCG~+10--14--025}
\figsetplot     {0338}
\figsetgrpnote  {The two-dimensional $\Delta \chi^2$ contour between the torus angular width and the inclination angle of MCG~+10--14--025.}
\figsetgrpend
\figsetgrpstart
\figsetgrpnum   {07.28}
\figsetgrptitle {ESO~317--G041}
\figsetplot     {0340}
\figsetgrpnote  {The two-dimensional $\Delta \chi^2$ contour between the torus angular width and the inclination angle of ESO~317--G041.}
\figsetgrpend
\figsetgrpstart
\figsetgrpnum   {07.29}
\figsetgrptitle {SDSS~J103315.71+525217.8}
\figsetplot     {0341}
\figsetgrpnote  {The two-dimensional $\Delta \chi^2$ contour between the torus angular width and the inclination angle of SDSS~J103315.71+525217.8.}
\figsetgrpend
\figsetgrpstart
\figsetgrpnum   {07.31}
\figsetgrptitle {NGC~4102}
\figsetplot     {0343}
\figsetgrpnote  {The two-dimensional $\Delta \chi^2$ contour between the torus angular width and the inclination angle of NGC~4102.}
\figsetgrpend
\figsetgrpstart
\figsetgrpnum   {07.32}
\figsetgrptitle {NGC~4180}
\figsetplot     {0344}
\figsetgrpnote  {The two-dimensional $\Delta \chi^2$ contour between the torus angular width and the inclination angle of NGC~4180.}
\figsetgrpend
\figsetgrpstart
\figsetgrpnum   {07.33}
\figsetgrptitle {ESO~323--G032}
\figsetplot     {0345}
\figsetgrpnote  {The two-dimensional $\Delta \chi^2$ contour between the torus angular width and the inclination angle of ESO~323--G032.}
\figsetgrpend
\figsetgrpstart
\figsetgrpnum   {07.35}
\figsetgrptitle {IGR~J14175--4641}
\figsetplot     {0347}
\figsetgrpnote  {The two-dimensional $\Delta \chi^2$ contour between the torus angular width and the inclination angle of IGR~J14175--4641.}
\figsetgrpend
\figsetgrpstart
\figsetgrpnum   {07.38}
\figsetgrptitle {CGCG~164--019}
\figsetplot     {0350}
\figsetgrpnote  {The two-dimensional $\Delta \chi^2$ contour between the torus angular width and the inclination angle of CGCG~164--019.}
\figsetgrpend
\figsetgrpstart
\figsetgrpnum   {07.39}
\figsetgrptitle {ESO~137--G034}
\figsetplot     {0351}
\figsetgrpnote  {The two-dimensional $\Delta \chi^2$ contour between the torus angular width and the inclination angle of ESO~137--G034.}
\figsetgrpend
\figsetgrpstart
\figsetgrpnum   {07.40}
\figsetgrptitle {NGC~6232}
\figsetplot     {0352}
\figsetgrpnote  {The two-dimensional $\Delta \chi^2$ contour between the torus angular width and the inclination angle of NGC~6232.}
\figsetgrpend
\figsetgrpstart
\figsetgrpnum   {07.41}
\figsetgrptitle {ESO~138--G001}
\figsetplot     {0353}
\figsetgrpnote  {The two-dimensional $\Delta \chi^2$ contour between the torus angular width and the inclination angle of ESO~138--G001.}
\figsetgrpend
\figsetgrpstart
\figsetgrpnum   {07.43}
\figsetgrptitle {NGC~6552}
\figsetplot     {0355}
\figsetgrpnote  {The two-dimensional $\Delta \chi^2$ contour between the torus angular width and the inclination angle of NGC~6552.}
\figsetgrpend
\figsetgrpstart
\figsetgrpnum   {07.44}
\figsetgrptitle {2MASX~J20145928+2523010}
\figsetplot     {0356}
\figsetgrpnote  {The two-dimensional $\Delta \chi^2$ contour between the torus angular width and the inclination angle of 2MASX~J20145928+2523010.}
\figsetgrpend
\figsetgrpstart
\figsetgrpnum   {07.45}
\figsetgrptitle {ESO~464--G016}
\figsetplot     {0357}
\figsetgrpnote  {The two-dimensional $\Delta \chi^2$ contour between the torus angular width and the inclination angle of ESO~464--G016.}
\figsetgrpend
\figsetgrpstart
\figsetgrpnum   {07.46}
\figsetgrptitle {NGC 7130}
\figsetplot     {0358}
\figsetgrpnote  {The two-dimensional $\Delta \chi^2$ contour between the torus angular width and the inclination angle of NGC~7130.}
\figsetgrpend
\figsetgrpstart
\figsetgrpnum   {07.47}
\figsetgrptitle {NGC~7212}
\figsetplot     {0359}
\figsetgrpnote  {The two-dimensional $\Delta \chi^2$ contour between the torus angular width and the inclination angle of NGC~7212.}
\figsetgrpend
\figsetgrpstart
\figsetgrpnum   {07.48}
\figsetgrptitle {ESO~406--G004}
\figsetplot     {0360}
\figsetgrpnote  {The two-dimensional $\Delta \chi^2$ contour between the torus angular width and the inclination angle of ESO~406--G004.}
\figsetgrpend
\figsetgrpstart
\figsetgrpnum   {07.50}
\figsetgrptitle {SWIFT~J2307.9+2245}
\figsetplot     {0362}
\figsetgrpnote  {The two-dimensional $\Delta \chi^2$ contour between the torus angular width and the inclination angle of SWIFT~J2307.9+2245.}
\figsetgrpend
\figsetgrpstart
\figsetgrpnum   {07.51}
\figsetgrptitle {NGC~7582}
\figsetplot     {0363}
\figsetgrpnote  {The two-dimensional $\Delta \chi^2$ contour between the torus angular width and the inclination angle of NGC~7582.}
\figsetgrpend
\figsetgrpstart
\figsetgrpnum   {07.52}
\figsetgrptitle {NGC~7682}
\figsetplot     {0364}
\figsetgrpnote  {The two-dimensional $\Delta \chi^2$ contour between the torus angular width and the inclination angle of NGC~7682.}
\figsetgrpend
\figsetend    \clearpage
\startlongtable
\begin{deluxetable*}{llllllll}\label{Table03}
\tablecaption{Best Fitting Parameters}
\tablehead
{
Galaxy Name                                                 & $T_{\mathrm{Satellite1}}$                                 & 
$T_{\mathrm{Satellite2}}$                                   & $\Gamma$                                                  & 
$N_{\mathrm{Dir}}$                                          & $f_{\mathrm{scat}}$                                       &       
$N_{\mathrm{H}}^{\mathrm{LOS}}$                             & $N_{\mathrm{H}}^{\mathrm{Equ}}$                           \\
(01)                                                        & (02)							                            &
(03)                                                        & (04)							                            &
(05)                                                        & (06)							                            &
(07)	                                                    & (08)                                                      \\
                                                            & $\sigma$                                                  &
$i$                                                         & $E_{\mathrm{line}}$                                       &
$N_{\mathrm{line}}$                                         & $k_{\mathrm{B}}T$			                                &
$N_{k_B T}$	                                                & $\chi^2_{\mathrm{red}}$                                   \\
                                                            & (09)                                                      &
(10)                                                        & (11)                                                      &
(12)                                                        & (13)                                                      &
(14)                                                        & (15)                                                      }
\startdata
ESO 112--G006                                               & $1.25_{-0.15}^{+0.63}$                                    &
$1.25$ (linked)                                             & $1.47_{-0.07}^{+0.17}$                                    &
$0.07_{-0.02}^{+0.02}$                                      & $0.00^{+0.08}$                                            &
$0.62_{-0.04}^{+0.06}$                                      & $1.55_{-0.60}^{+1.76}$                                    \\
                                                            & $10.1^{+9.53}$                                            &
$80.4_{-6.01}^{+1.47}$                                      & \nodata                                                   &
\nodata                                                     & $0.82_{-0.09}^{+0.03}$                                    &
$0.03_{-0.01}^{+0.01}$                                      & $1.21$                                                    \\
MCG --07--03--007                                           & $0.84_{-0.10}^{+0.43}$                                    &
$0.84$ (linked)                                             & $1.61_{-0.20}^{+0.05}$                                    &
$0.13_{-0.08}^{+0.01}$                                      & $0.45_{-0.08}^{+0.54}$                                    &
$0.90_{-0.09}^{+0.11}$                                      & $1.49_{-0.69}$                                            \\
                                                            & $38.9_{-3.68}^{+46.2}$                                    &
$62.3^{+24.1}$                                              & \nodata                                                   &
\nodata                                                     & $0.86_{-0.19}^{+0.10}$                                    &
$0.03_{-0.01}^{+0.00}$                                      & $1.24$                                                    \\
\enddata
\tablecomments{Column~(01): galaxy name. Column~(02): time variability constant of the soft X-ray observations (Chandra, XMM--Newton, Swift/XRT, and Suzaku) relative to the Swift/BAT observation. Column~(03): time variability constant of the hard X-ray observations (NuSTAR) relative to the Swift/BAT observation. Column~(04): photon index. Column~(05): normalization of the direct component in units of 10$^{-2}$ photons keV$^{-1}$ cm$^{-2}$ s$^{-1}$. Column~(06): scattering fraction in units of percent. Column~(07): hydrogen column density along the line of sight in units of $10^{24} \ \mathrm{cm}^{-2}$. Column~(08): hydrogen column density along the equatorial direction in units of $10^{24} \ \mathrm{cm}^{-2}$. Column~(09): torus angular width in units of degree. Column~(10): inclination angle in units of degree. Column~(11): energy of the zgauss model in units of keV. Column~(12): normalization of the zgauss in units of. Column~(13): temperature of the apec model in units of keV. Column~(14): normalization of the apec model in units of 10$^{-19}/4\pi [D_{\mathrm{A}}(1+z)]^2 \int n_{\mathrm{e}} n_{\mathrm{H}}dV$, where $D_{\mathrm{A}}$ is the angular diameter distance to the source in units of cm, $n_{\mathrm{e}}$ and $n_{\mathrm{H}}$ are the electron and hydrogen densities in units of cm$^{-3}$. Column~(15): reduced $\chi^2$. Note that the error on a spectral parameter corresponds to the 90\% confidence interval. We estimated the error by using the Markov Chain Monte Carlo chain with a length of 100,000.}
\end{deluxetable*}
     \clearpage
\startlongtable
\begin{deluxetable*}{lllllllll}\label{Table04}
\tablecaption{Fluxes, Luminosities and Covering Factors}
\tablehead
{
Galaxy Name                                                                                                             &
$\log F_{2-10}$ 				                            & $\log F_{10-50}$				                            &
$\log L_{2-10}$                                             & $\log L_{10-50}$                                          &
$C_{22}$                                                    & $C_{24}$                                                  &
$\mathrm{EW}$                                               & $\log R_{\mathrm{Edd}}$ 						            \\
(1)                                                                                                                     &
(2)                                                         & (3)                                                       &
(4)                                                         & (5)                                                       &
(6)                                                         & (7)                                                       &
(8)                                                         & (9)                                                       }
\startdata
ESO 112--G006                                                                                                           &
$-12.3$                                                     & $-11.2$                                                   &
$ 42.9$                                                     & $ 43.2$                                                   &
$ 0.39_{-0.39}^{+0.35}$                                     & $ 0.00^{+0.01}$                                           &
$ 0.07$                                                     & \nodata                                                   \\
MCG --07--03--007                                                                                                       &
$-12.3$                                                     & $-11.2$                                                   &
$ 43.1$                                                     & $ 43.3$                                                   &
$ 1.00_{-0.01}$                                             & $ 0.42_{-0.23}^{+0.58}$                                   &
$ 0.65$                                                     & $-2.26$                                                   \\
NGC 0424                                                                                                                &
$-12.0$                                                     & $-11.1$                                                   &
$ 42.5$                                                     & $ 42.7$                                                   &
$ 1.00_{-0.02}$                                             & $ 0.43_{-0.19}^{+0.18}$                                   &
$ 0.62$                                                     & $-2.08$                                                   \\
\enddata
\tablecomments{Column~(1): galaxy name. Column~(2): logarithmic observed flux in the 2--10 keV based on Chandra/ACIS, XMM--Newton/PN, Swift/XRT, Suzaku/FIXIS. Column~(3): logarithmic observed flux in the 10--50 keV based on NuSTAR/FPM. Column~(4): logarithmic intrinsic luminosity in the 2--10 keV. We estimate the X-ray luminosities by using the redshifts except for three very nearby ($D \leq 20$ Mpc) objects (NGC~3079, NGC~4102, and NGC~4945). In the case of these objects, we utilize the following distances: (NGC~3079: 16.5~Mpc, NGC~4102: 9.1~Mpc, and NGC~4945: 3.55~Mpc) \citep{Tully09}. Column~(5): logarithmic intrinsic luminosity in the 10--50 keV. Column~(6): Compton-thin torus covering factor. Column~(7): Compton-thick torus covering factor. Column~(8): equivalent width of the Fe K$\alpha$ line in units of keV. Column~(9): logarithmic Eddington ratio ($R_{\mathrm{Edd}} = L_{\mathrm{Bol}}/L_{\mathrm{Edd}}$). Here we obtain the bolometric luminosity as $L_{\mathrm{bol}} = 20 L_{2-10}$ and define the Eddington luminosity as $L_{\mathrm{Edd}} = 1.25 \times 10^{38} M_{\mathrm{BH}}/M_{\sun}$.}
\end{deluxetable*}
     \clearpage
\begin{figure*}\label{Figure08}
\gridline{\fig{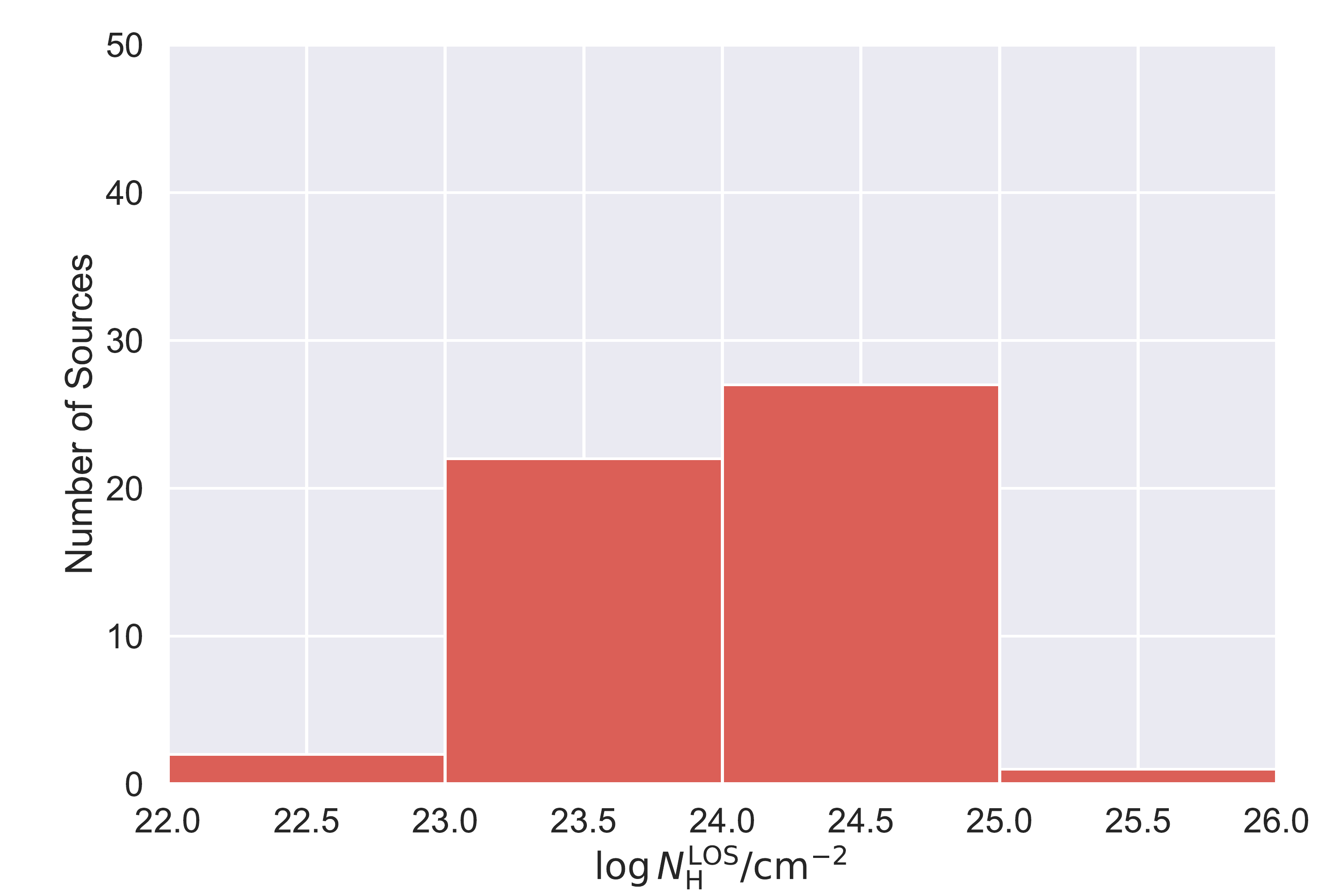}{0.45\textwidth}{}}
\caption{The distribution of the hydrogen column density along the line of sight ($\log N_{\mathrm{H}}^{\mathrm{LOS}}/\mathrm{cm}^{-2}$). Here we plot the 90\% confidence upper bound of $\log N_{\mathrm{H}}^{\mathrm{LOS}}/\mathrm{cm}^{-2}$.}
\end{figure*}

\section{Results}\label{Section05}
We applied the XClumpy model to the broadband X-ray spectra of 52 CTAGN candidates observed with Chandra, XMM--Newton, Swift, Suzaku, and NuSTAR. As a result, our model can reproduce the X-ray spectra of all objects. \hyperref[Figure01]{Figure~1} plots the folded X-ray spectra fitted with the XClumpy model and \hyperref[Figure02]{Figure~2} shows the best fitting model. \hyperref[Table03]{Table~3} summarizes the best fitting parameters. \hyperref[Figure03]{Figure~3--7} plot the two-dimensional $\Delta \chi^2$ contours between the hydrogen column density along the line of sight ($\log N_{\mathrm{H}}^{\mathrm{LOS}}/\mathrm{cm}^{-2}$) and the torus angular width ($\sigma$), between $\log N_{\mathrm{H}}^{\mathrm{LOS}}/\mathrm{cm}^{-2}$ and the inclination angle ($i$), between the hydrogen column density along the equatorial direction ($\log N_{\mathrm{H}}^{\mathrm{Equ}}/\mathrm{cm}^{-2}$) and $\sigma$, between $\log N_{\mathrm{H}}^{\mathrm{Equ}}/\mathrm{cm}^{-2}$ and $i$, and between $\sigma$ and $i$, respectively.
\section{Discussion}\label{Section06}
The structure of this Section is the following: \hyperref[Section0601]{Section~6.1} investigates the distribution of the hydrogen column density along the lien of sight and estimates the fraction of the Compton-thick AGN. \hyperref[Section0602]{Section~6.2} introduces the torus covering factor and the radiation-regulated AGN unification model. \hyperref[Section0603]{Section~6.3} examines the distributions of best fitting parameters. \hyperref[Section0604]{Section~6.4} validates the radiation-regulated AGN unification model.
\subsection{Distribution of Hydrogen Column Density and Compton-thick AGN Fraction}\label{Section0601}
We investigate the fraction of the Compton-thick AGN. \hyperref[Figure08]{Figure~8} shows the distribution of the logarithmic hydrogen column density along the line of sight ($\log N_{\mathrm{H}}^{\mathrm{LOS}}/\mathrm{cm}^{-2}$). \hyperref[Figure08]{Figure~8} indicates that 24 objects are Compton-thin AGNs and 28 objects are Compton-thick AGNs within the 90\% confidence interval. Here we classify the object whose 90\% confidence upper bound of $\log N_{\mathrm{H}}^{\mathrm{LOS}}/\mathrm{cm}^{-2}$ is larger than 24 as Compton-thick AGN.

To clarify why $\log N_{\mathrm{H}}^{\mathrm{LOS}}/\mathrm{cm}^{-2}$ of this study differ from those of \cite{Ricci15}, \hyperref[Figure03]{Figure~3} plots the two-dimensional $\Delta \chi^2$ contour between $\log N_{\mathrm{H}}^{\mathrm{LOS}}/\mathrm{cm}^{-2}$ and the torus angular width ($\sigma$). \hyperref[Figure03]{Figure~3} also plots the 90\% confidence interval of $\log N_{\mathrm{H}}^{\mathrm{LOS}}/\mathrm{cm}^{-2}$ of \cite{Ricci15}. \hyperref[Figure03]{Figure~3} shows that $\log N_{\mathrm{H}}^{\mathrm{LOS}}/\mathrm{cm}^{-2}$ of 34 objects obtained from this study are not consistent with those inferred from \cite{Ricci15}. The first reason is the difference in the torus model. In the torus model, we use the XClumpy model \citep{Tanimoto19}, while \cite{Ricci15} used the Torus model \citep{Brightman11}. It is, however, known that the Torus model had some calculation errors \citep[][Figure~4]{Liu15}. \cite{Liu15} indicated that the Torus model overestimated the flux in the soft X-rays in the case of the high hydrogen column density in the torus and the edge-on view. Due to this effect, $\log N_{\mathrm{H}}^{\mathrm{LOS}}/\mathrm{cm}^{-2}$ obtained from the XClumpy model are different from those inferred from the Torus model. The second reason is the difference in the observational data. In the observational data, \cite{Ricci15} did not analyze the NuSTAR observational data, whereas we analyze the NuSTAR observational data. \hyperref[Appendix0100]{Appendix~A} discusses why $\log N_{\mathrm{H}}^{\mathrm{LOS}}/\mathrm{cm}^{-2}$ of this study differ from those of \cite{Ricci15} in detail.

We note that \hyperref[Figure08]{Figure~8(a)} does not mean that \cite{Ricci15} overestimated the CTAGN fraction. Our parent sample consists only of sources classified as CTAGN from the best-fitting $\log N_{\mathrm{H}}^{\mathrm{LOS}}/\mathrm{cm}^{-2}$ obtained from \cite{Ricci15}. We find that $N_{\mathrm{H}}^{\mathrm{LOS}}$ varies depending on the observational data and the torus model. In other words, there is a possibility that an object that was Compton-thin AGN in their analysis may become a Compton-thick AGN as a result of the application of the XClumpy model. In fact, the largest number of AGN detected by Swift/BAT are $23 \leq \log N_{\mathrm{H}}^{\mathrm{LOS}}/\mathrm{cm}^{-2} \leq 24$, so even a few such sources can have a significant impact on the CTAGN fraction. To obtain the true CTAGN fraction, we need to analyze all object that were classified as a Compton-thin AGN by them with XClumpy model in the future.

\begin{figure*}\label{Figure09}
\gridline{\fig{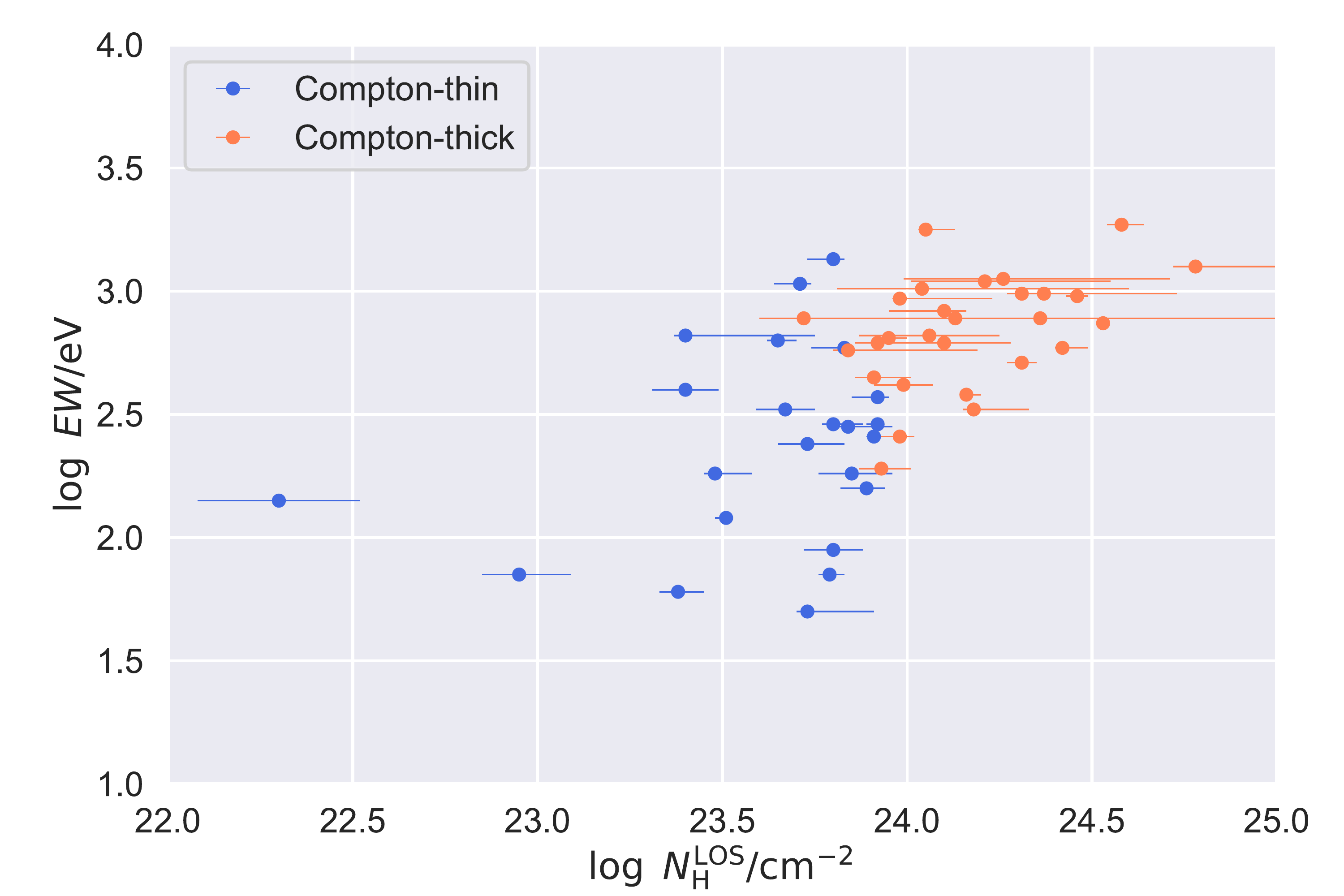}{0.45\textwidth}{}}
\caption{The scatter plot of the logarithmic hydrogen column density along the line of sight ($\log N_{\mathrm{H}}^{\mathrm{LOS}}/\mathrm{cm}^{-2}$) and the logarithmic equivalent of the Fe K$\alpha$ line ($\log \mathrm{EW}/\mathrm{eV}$).}
\end{figure*}

We study the correlation between $\log N_{\mathrm{H}}^{\mathrm{LOS}}/\mathrm{cm}^{-2}$ and the equivalent width of the Fe K$\alpha$ line ($\log \mathrm{EW}/\mathrm{eV}$). This is because $\log \mathrm{EW}/\mathrm{eV}$ have been commonly used for CTAGN diagnosis. For instance, \cite{Murphy09} investigated the correlation between $\log N_{\mathrm{H}}^{\mathrm{Equ}}/\mathrm{cm}^{-2}$ and $\log \mathrm{EW}/\mathrm{eV}$ by using the MYTorus model \citep[][Figure~8]{Murphy09}. They suggested that $\log \mathrm{EW}/\mathrm{eV}$ would be larger than 2.5 in the case of CTAGN. \hyperref[Figure09]{Figure~9} plots that the correlation between $\log N_{\mathrm{H}}^{\mathrm{LOS}}/\mathrm{cm}^{-2}$ and $\log \mathrm{EW}/\mathrm{eV}$. \hyperref[Figure09]{Figure~9} shows that almost all the objects are Compton-thin AGNs in $\log \mathrm{EW}/\mathrm{eV} \leq 2.5$, and almost all the objects are Compton-thick AGNs in $3.0 \leq \log \mathrm{EW}/\mathrm{eV}$. However, in the case of $2.5 < \log \mathrm{EW}/\mathrm{eV} < 3.0$, it is difficult to classify whether the object is a Compton-thin AGN or a Compton-thick AGN. This result is slightly different from \cite{Murphy09}. This is because $\log \mathrm{EW}/\mathrm{eV}$ depends on the geometric structure of the torus. In the case of the XClumpy model, $\log \mathrm{EW}/\mathrm{eV}$ is larger than that of the MYTorus model since the Fe K$\alpha$ emission lines can pass through the clumpy torus \citep[][Figure~4]{Tanimoto19}.

\begin{figure*}[t]\label{Figure10}
\gridline{\fig{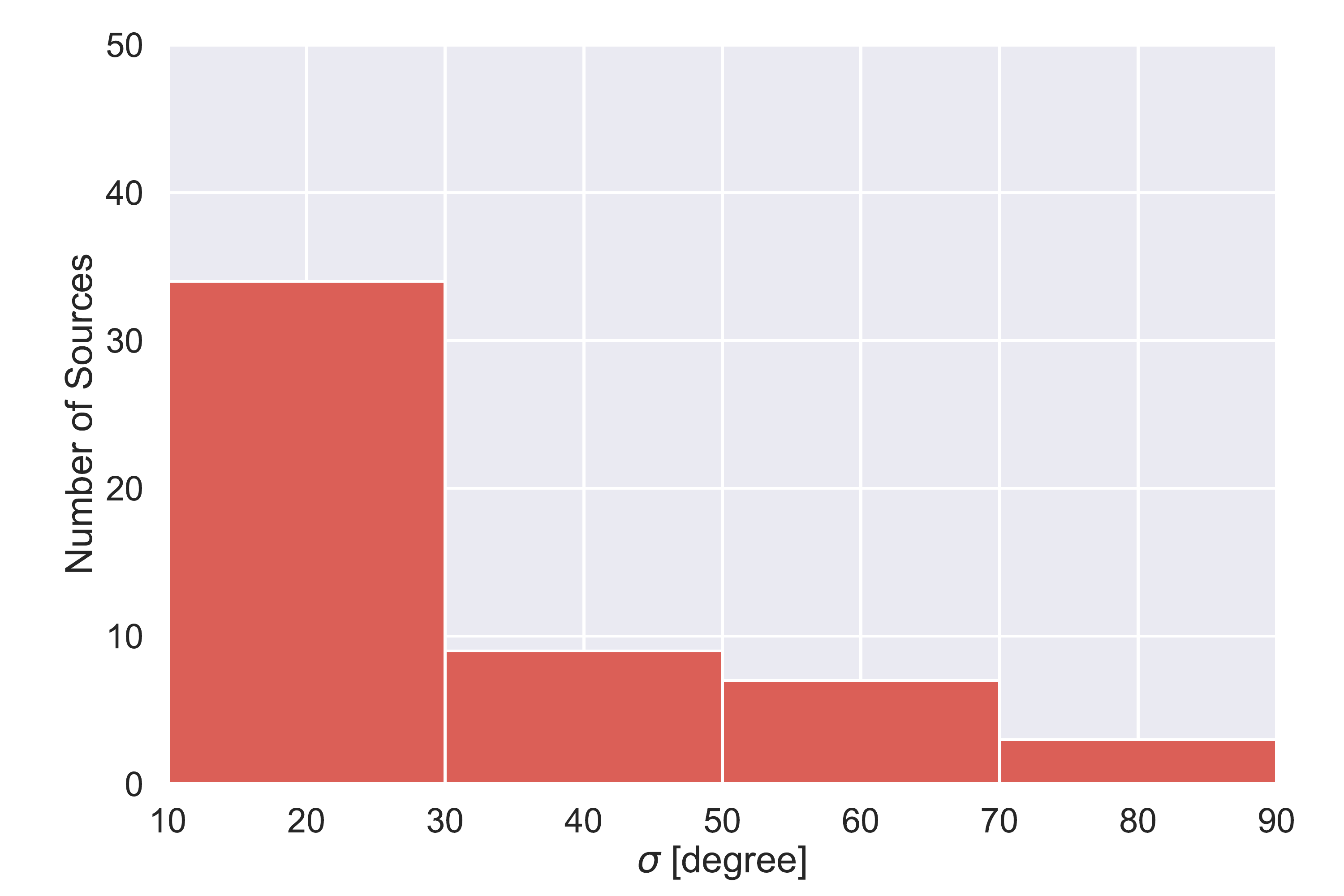}{0.45\textwidth}{(a)}\fig{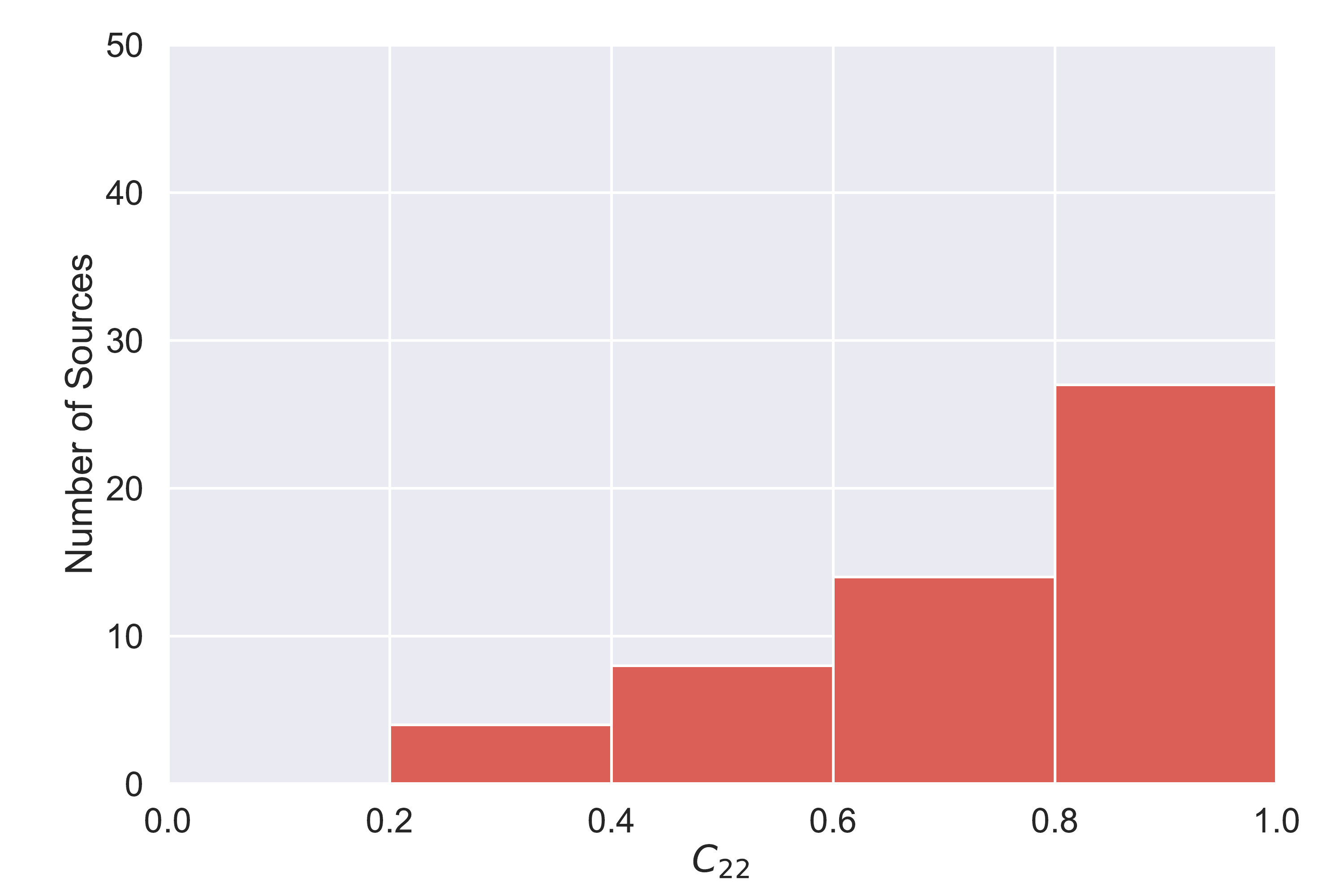}{0.45\textwidth}{(b)}}
\gridline{\fig{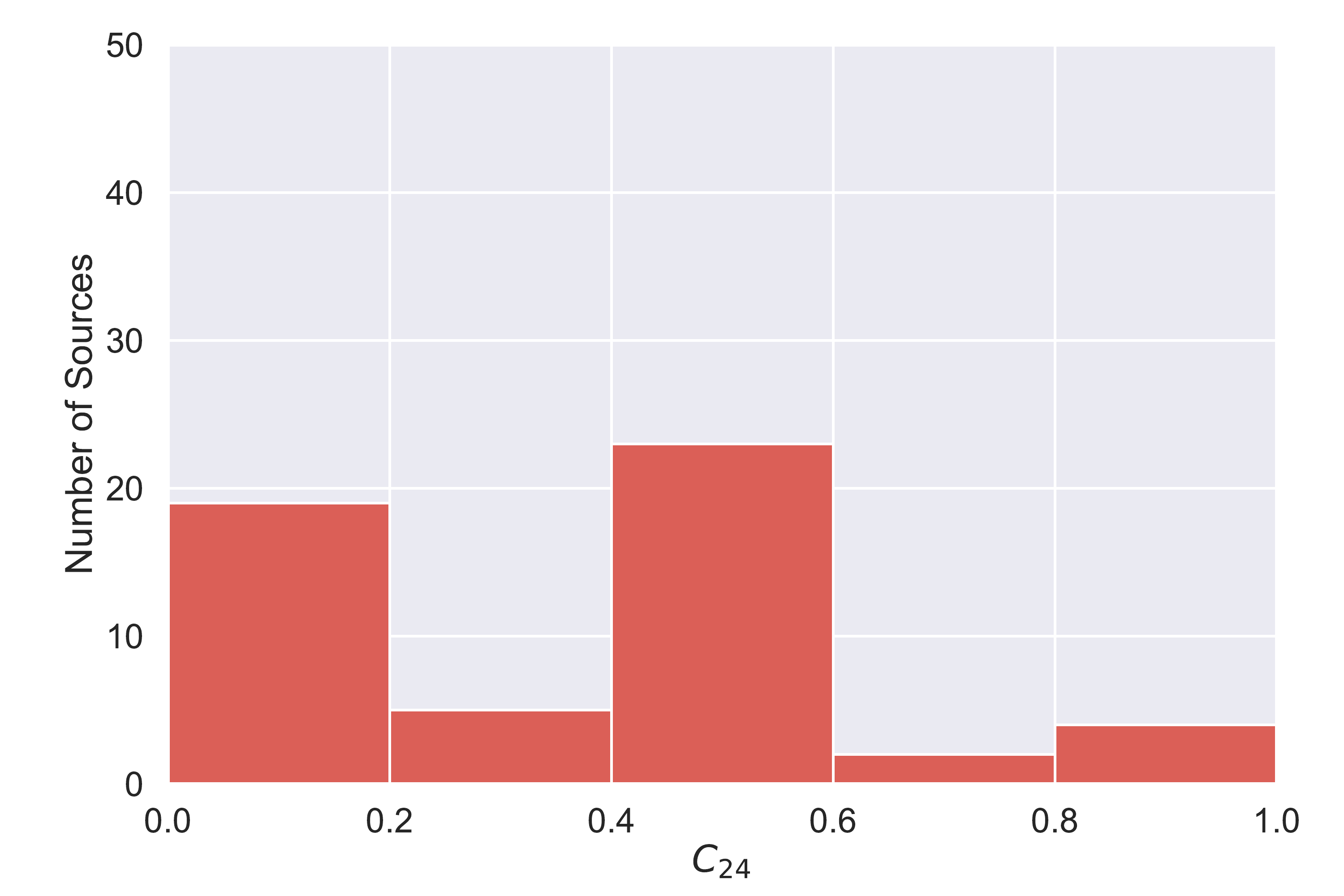}{0.45\textwidth}{(c)}\fig{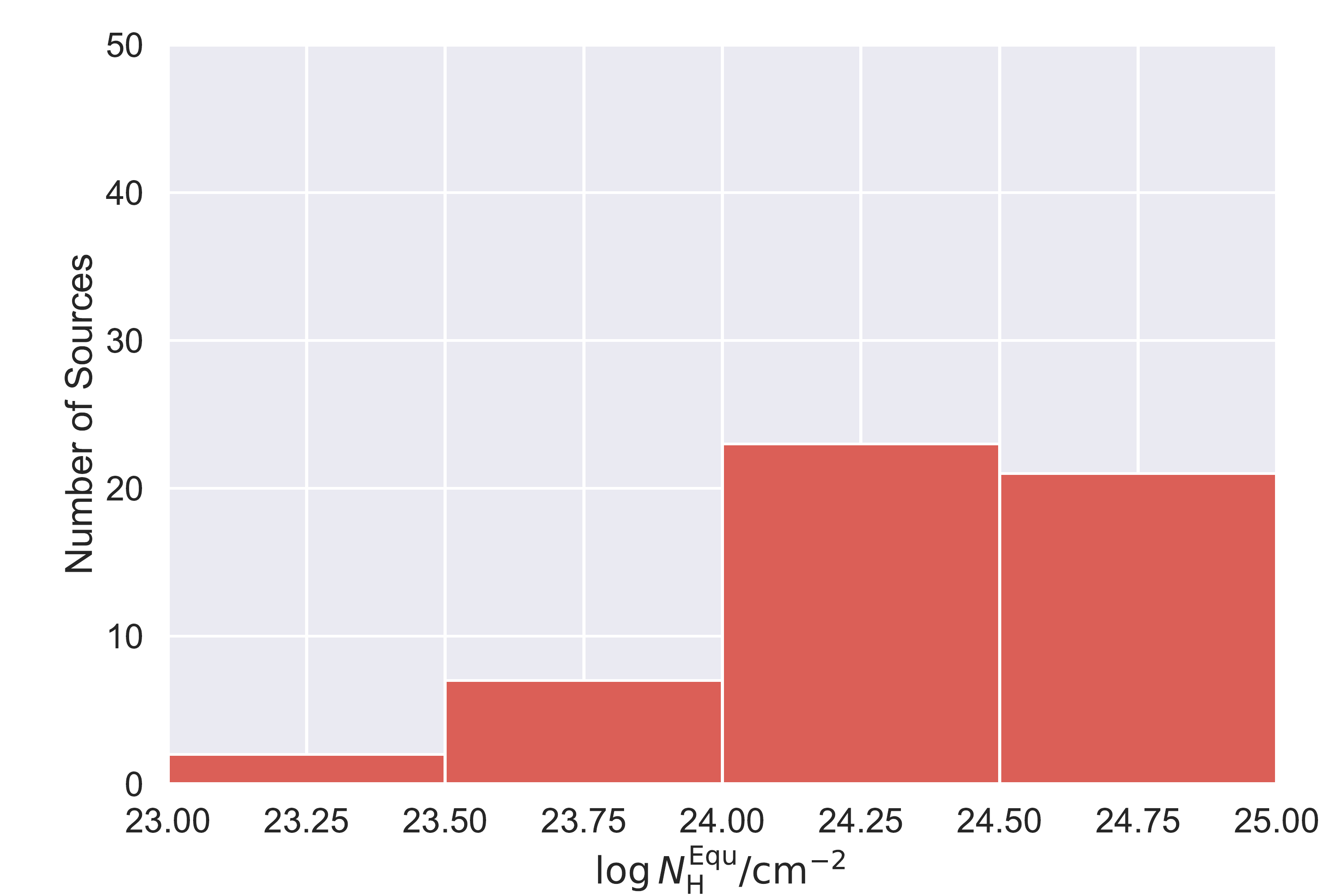}{0.45\textwidth}{(d)}}
\caption{(a) The distribution of the torus angular width ($\sigma$). (b) The distribution of the Compton-thin torus covering factor ($C_{22}$). (c) The distribution of the Compton-thick torus covering factor ($C_{24}$). (d) The distribution of the logarithmic hydrogen column density along the equatorial direction ($\log N_{\mathrm{H}}^{\mathrm{Equ}}/\mathrm{cm}^{-2}$)}
\end{figure*}
\begin{figure*}[t]\label{Figure11}
\gridline{\fig{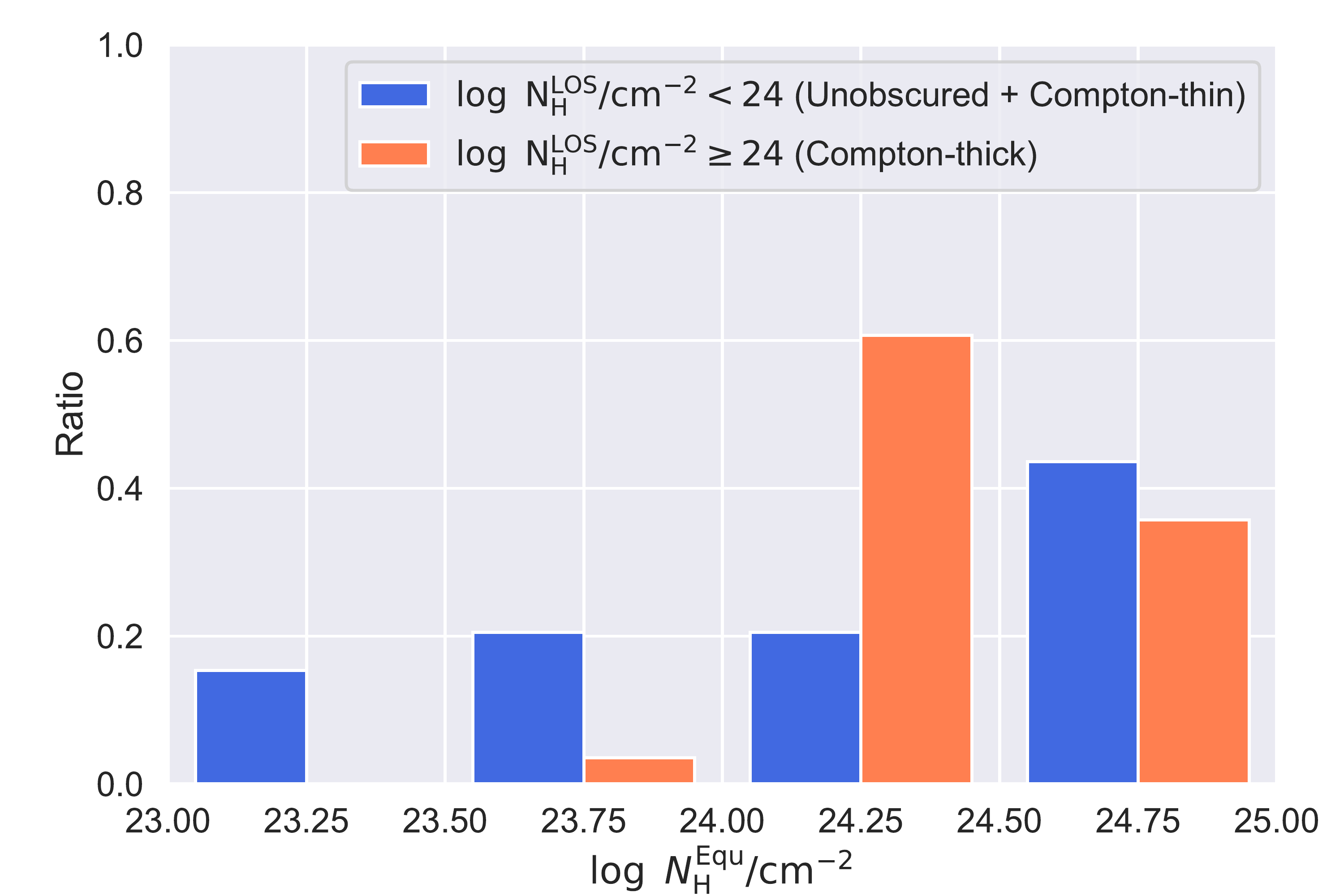}{0.45\textwidth}{}}
\caption{The normalized distribution of $\log N_{\mathrm{H}}^{\mathrm{Equ}}/\mathrm{cm}^{-2}$. The blue histogram represents $\log N_{\mathrm{H}}^{\mathrm{Equ}}/\mathrm{cm}^{-2}$ of unobscured AGNs and Compton-thin AGNs, and the orange histogram shows $\log N_{\mathrm{H}}^{\mathrm{Equ}}/\mathrm{cm}^{-2}$ of Compton-thick AGNs. Here we consider 29 CTAGNs (28 objects in this work and the Circinus Galaxy) and 40 less obscured AGNs (24 objects in this work and 16 objects in \cite{Ogawa21}). We note that Compton--thin AGNs represent the object whose logarithmic hydrogen column density along the line of sight ($\log N_{\mathrm{H}}^{\mathrm{LOS}}/\mathrm{cm}^{-2}$) is smaller than 24 and unobscured AGNs mean the object whose $\log N_{\mathrm{H}}^{\mathrm{Equ}}/\mathrm{cm}^{-2}$ is smaller than 22.}
\end{figure*}

\subsection{Torus Covering Factor and Radiation-Regulated AGN Unification Model}\label{Section0602}
The torus covering factor plays an important role in determining the ratio of type--1 and type--2 AGNs. Here we define the Compton-thin torus covering factor ($C_{22}$) as the percentage of the AGN that is covered by a material whose logarithmic hydrogen column density along the line of sight ($\log N_{\mathrm{H}}^{\mathrm{LOS}}/\mathrm{cm}^{-2}$) is larger than $22$.

There are two ways to determine $C_{22}$. The first way to determine $C_{22}$ is to examine the $\log N_{\mathrm{H}}^{\mathrm{LOS}}/\mathrm{cm}^{-2}$ distribution by investigating $\log N_{\mathrm{H}}^{\mathrm{LOS}}/\mathrm{cm}^{-2}$ for each source. \cite{Ricci17a} systematically analyzed X-ray spectra of 838 AGNs detected by the Swift/BAT 70-month catalog \citep{Baumgartner13} and estimated $\log N_{\mathrm{H}}^{\mathrm{LOS}}/\mathrm{cm}^{-2}$ for all objects. They suggested that the Eddington ratio ($R_{\mathrm{Edd}}$) is a key parameter to determine $C_{22}$. If $R_{\mathrm{Edd}}$ is larger than effective $R_{\mathrm{Edd}}$ of the dusty gas ($\log R_{\mathrm{Edd}} \simeq -2.0$), the radiation pressure blows the dusty gas away. As a result, $C_{22}$ is large in the case of low Eddington ratio ($\log R_{\mathrm{Edd}} \leq -2.0$), whereas $C_{22}$ becomes small in the case of high Eddington ratio ($-2.0 \leq \log R_{\mathrm{Edd}}$) \citep[][Figure~4]{Ricci17c}. This model is called the radiation-regulated AGN unification model. In fact, \cite{Ananna22} indicated the clear difference in the Eddington ratio distribution functions of type--1 and type--2 AGNs.

The second way to determine $C_{22}$ is to analyze the X-ray spectrum by using a torus model that can change the geometric structure of the torus, such as the UXCLUMPY model \citep{Buchner19} and the XClumpy model \citep{Tanimoto19}. In this method, we can examine ``individual" $C_{22}$ rather than ``statistical" $C_{22}$ that have been mainly examined so far. Here we use the XClumpy model. In the case of the XClumpy model, we can define $C_{22}$ in the following way. We express Equation~3 using the elevation angle ($\theta = i-\pi/2$)
\begin{equation}
N_{\mathrm{H}}^{\mathrm{LOS}}(\theta) = N_{\mathrm{H}}^{\mathrm{Equ}} \exp{\left(-\frac{\theta^2}{\sigma^2}\right)}.
\end{equation}
Here we define $\theta_{22}$ corresponding to $\log N_{\mathrm{H}}^{\mathrm{LOS}}/\mathrm{cm}^{-2} = 22$ as follows.
\begin{equation}
\theta_{22} = \sigma \left[\ln{\left(\frac{N_{\mathrm{H}}^{\mathrm{Equ}}}{10^{22} \ \mathrm{cm}^{-2}}\right)}\right]^{\frac{1}{2}}.
\end{equation}
Using $\theta_{22}$, we can obtain individual $C_{22}$ based on the following equation:
\begin{align}\label{Equation06}
C_{22} & = \frac{1}{4\pi} \int_{\frac{\pi}{2}-\theta_{22}}^{\frac{\pi}{2}+\theta_{22}} \int_{0}^{2\pi} \sin\theta d\theta d\phi. \nonumber\\
                    & = \sin \left(\sigma \left[\ln{\left(\frac{N_{\mathrm{H}}^{\mathrm{Equ}}}{10^{22} \ \mathrm{cm}^{-2}}\right)}\right]^{\frac{1}{2}}\right).
\end{align}
Similarly, we can define the Compton-thick torus covering factor ($C_{24}$) as follows.
\begin{equation}\label{Equation07}
C_{24} = \sin \left(\sigma \left[\ln{\left(\frac{N_{\mathrm{H}}^{\mathrm{Equ}}}{10^{24} \ \mathrm{cm}^{-2}}\right)}\right]^{\frac{1}{2}}\right).
\end{equation}
\hyperref[Table04]{Table~4} shows the logarithmic observed fluxes, the logarithmic intrinsic luminosities, the Compton-thin torus covering factor, the Compton-thick torus covering factor, and the Eddington ratio.
\subsection{Distribution of Torus Covering Factor}\label{Section0603}
We study the distribution of the torus covering factors. Here we add the Circinus Galaxy fitted with the XClumpy model \citep{Tanimoto19} to our sample. Hence our sample consists of 28 Compton-thin AGNs and 25 Compton-thick AGNs. \hyperref[Figure10]{Figure~10} plots the distributions of (a) the torus angular width ($\sigma$), (b) the Compton-thin torus covering factor ($C_{22}$), (c) the Compton-thick torus covering factor ($C_{24}$), and (d) the logarithmic hydrogen column density along the equatorial direction ($\log N_{\mathrm{H}}^{\mathrm{Equ}}/\mathrm{cm}^{-2}$). \hyperref[Figure10]{Figure~10(a)} shows that 34 objects have small torus angular width ($\sigma \leq 30\degr$) and 19 objects have large torus angular width ($30\degr \leq \sigma$). This could be because we limit the inclination angle within a range of 60\degr--87\degr. In this case, we note that we may underestimate $C_{22}$ and $C_{24}$. Nevertheless, \hyperref[Figure10]{Figure~10(b)} suggests that about 40 objects show large Compton-thin torus covering factor ($0.6 \leq C_{22} \leq 1.0$). \hyperref[Figure10]{Figure~10(d)} shows that nine objects have Compton-thin $\log N_{\mathrm{H}}^{\mathrm{Equ}}/\mathrm{cm}^{-2}$ and 43 objects have Compton-thick $\log N_{\mathrm{H}}^{\mathrm{Equ}}/\mathrm{cm}^{-2}$. Especially, 20 objects have very large $\log N_{\mathrm{H}}^{\mathrm{Equ}}/\mathrm{cm}^{-2}$ ($24.5 \leq \log N_{\mathrm{H}}^{\mathrm{Equ}}/\mathrm{cm}^{-2}$). In the radiative regulated AGN unification model \citep{Ricci17c}, $\log N_{\mathrm{H}}^{\mathrm{Equ}}/\mathrm{cm}^{-2}$ has been considered to be basically constant. Our results suggest, however, that $\log N_{\mathrm{H}}^{\mathrm{Equ}}/\mathrm{cm}^{-2}$ may vary from object to object.

To study whether $\log N_{\mathrm{H}}^{\mathrm{Equ}}/\mathrm{cm}^{-2}$ is different between CTAGNs and less obscured AGNs (unobscured AGNs and Compton-thin AGNs), we compare the $\log N_{\mathrm{H}}^{\mathrm{Equ}}/\mathrm{cm}^{-2}$ of CTAGNs with $\log N_{\mathrm{H}}^{\mathrm{Equ}}/\mathrm{cm}^{-2}$ of less obscured AGNs. Here unobscured AGNs mean the object whose logarithmic hydrogen column density along the line of sight ($\log N_{\mathrm{H}}^{\mathrm{LOS}}/\mathrm{cm}^{-2}$) is smaller than 22 and Compton-thin AGNs represent the object whose $\log N_{\mathrm{H}}^{\mathrm{LOS}}/\mathrm{cm}^{-2}$ is smaller than 24. \hyperref[Figure11]{Figure~11} compares normalized $\log N_{\mathrm{H}}^{\mathrm{Equ}}/\mathrm{cm}^{-2}$ distribution of CTAGNs and that of less obscured AGNs. Here we consider 29 CTAGNs (28 objects in this work and the Circinus Galaxy) and 40 less obscured AGNs (24 objects in this work and 16 objects in previous work)\footnote{\cite{Ogawa21} applied the XClumpy model to the X-ray spectra of 16 less obscured AGNs.}. \hyperref[Figure11]{Figure~11} suggests that some of less obscured AGNs have Compton-thin $\log N_{\mathrm{H}}^{\mathrm{Equ}}/\mathrm{cm}^{-2}$. To investigate whether these distributions are essentially different, we perform the Kolmogorov--Smirnov test and obtain $p = 0.02$. Hence CTAGNs have a larger $\log N_{\mathrm{H}}^{\mathrm{Equ}}/\mathrm{cm}^{-2}$ than that of less obscured AGNs, and may be at a different evolutionary stage from these populations. \clearpage
\begin{figure*}[t]\label{Figure12}
\gridline{\fig{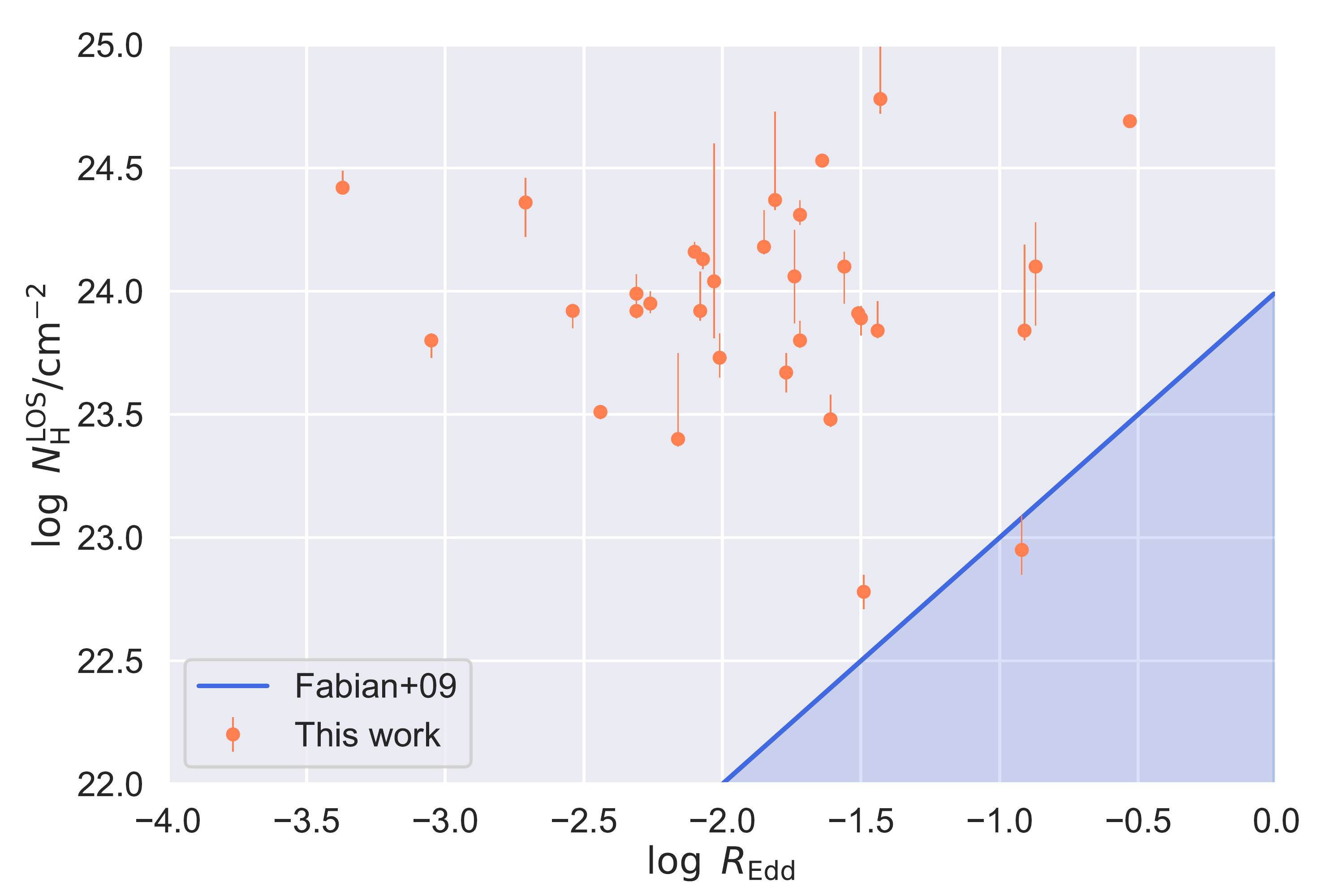}{0.45\textwidth}{}}
\caption{The scatter plot of the logarithmic Eddington ratio ($\log R_{\mathrm{Edd}}$) and the logarithmic hydrogen column density along the line of sight ($\log N_{\mathrm{H}}^{\mathrm{LOS}}$). Blue region represents that the radiation pressure push away the obscuring material (forbidden region: \citealt{Fabian09}).}
\end{figure*}

\subsection{Validation of Radiation-Regulated AGN Unification Model}\label{Section0604}
To validate the radiation-regulated AGN unification model, we investigate the correlation between the logarithmic Eddington ratio ($\log R_{\mathrm{Edd}}$) and the logarithmic hydrogen column density along the line of sight ($\log N_{\mathrm{H}}^{\mathrm{LOS}}/\mathrm{cm}^{-2}$). \hyperref[Figure12]{Figure~12} plots the correlation between $\log R_{\mathrm{Edd}}$ and $\log N_{\mathrm{H}}^{\mathrm{LOS}}/\mathrm{cm}^{-2}$. The blue region represents the area where the radiation pressure blows away the obscuring material (forbidden region: \citealt{Fabian09}). \hyperref[Figure12]{Figure~12} shows that 31 out of 32 objects have a higher $\log N_{\mathrm{H}}^{\mathrm{LOS}}/\mathrm{cm}^{-2}$ than this forbidden region. Hence our sample is expected to have a large Compton-thin torus covering factor ($C_{22}$) because the radiation pressure cannot blow away the obscuring material.

\begin{figure*}[t]\label{Figure13}
\gridline{\fig{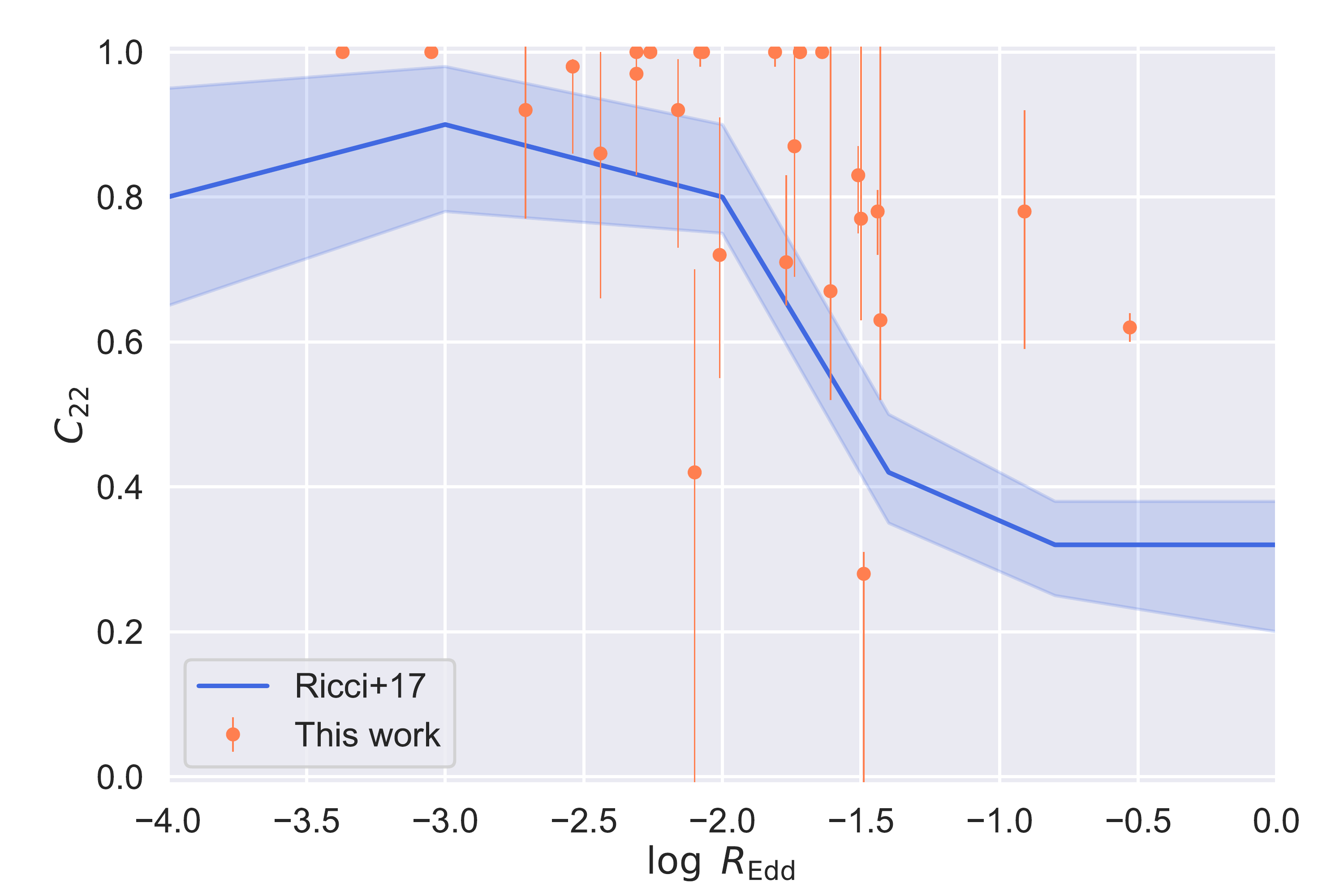}{0.45\textwidth}{(a)}\fig{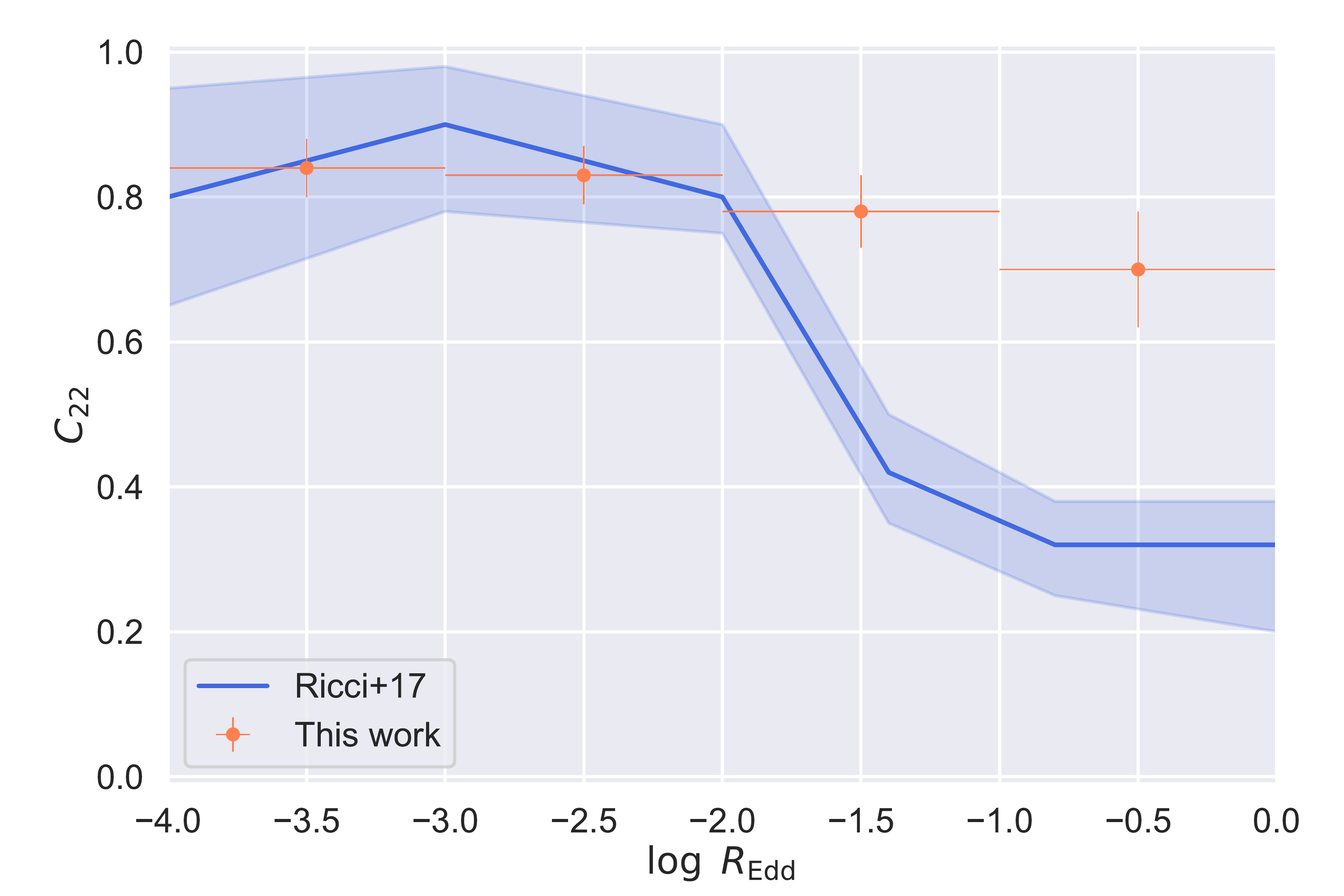}{0.45\textwidth}{(b)}}
\caption{(a) Correlation between the logarithmic Eddington ratio ($\log R_{\mathrm{Edd}}$) and the Compton-thin torus covering factor ($C_{22}$). Blue line represents the radiation regulated AGN unification model \citep{Ricci17b}. (b) Correlation between $\log R_{\mathrm{Edd}}$ and mean $C_{22}$ in the four $\log R_{\mathrm{Edd}}$ bins. Note that we apply the standard error as the error of $C_{22}$.}
\end{figure*}

\hyperref[Figure13]{Figure~13(a)} plots the correlation between the logarithmic Eddington ratio ($\log R_{\mathrm{Edd}}$) and the Compton-thin torus covering factor ($C_{22}$). Here we plot 28 objects whose black hole masses are known and whose $C_{22}$ are determined. \hyperref[Figure13]{Figure~13(b)} plots the correlation between $\log R_{\mathrm{Edd}}$ and mean $C_{22}$ in the four $\log R_{\mathrm{Edd}}$ bins. \hyperref[Figure13]{Figure~13(b)} indicates that $C_{22}$ of this work is consistent with that of \cite{Ricci17c} in the low Eddington ratio ($\log R_{\mathrm{Edd}} \leq -1.0$), while $C_{22}$ of this work is larger than that of \cite{Ricci17c} in the high Eddington ratio ($-1.0 \leq \log R_{\mathrm{Edd}}$). This is because our sample has a larger $\log N_{\mathrm{H}}^{\mathrm{LOS}}/\mathrm{cm}^{-2}$ than that of forbidden region. In fact, \cite{Kawakatu20} studied the obscuring structure of circumnuclear disks by considering supernova feedbacks and suggested that if $\log N_{\mathrm{H}}^{\mathrm{LOS}}/\mathrm{cm}^{-2}$ is sufficiently large, the feedback is less effective and the torus covering factor is approximately equal to 0.80 \citep[][Figure~2]{Kawakatu20}.

\begin{figure*}[t]\label{Figure14}
\gridline{\fig{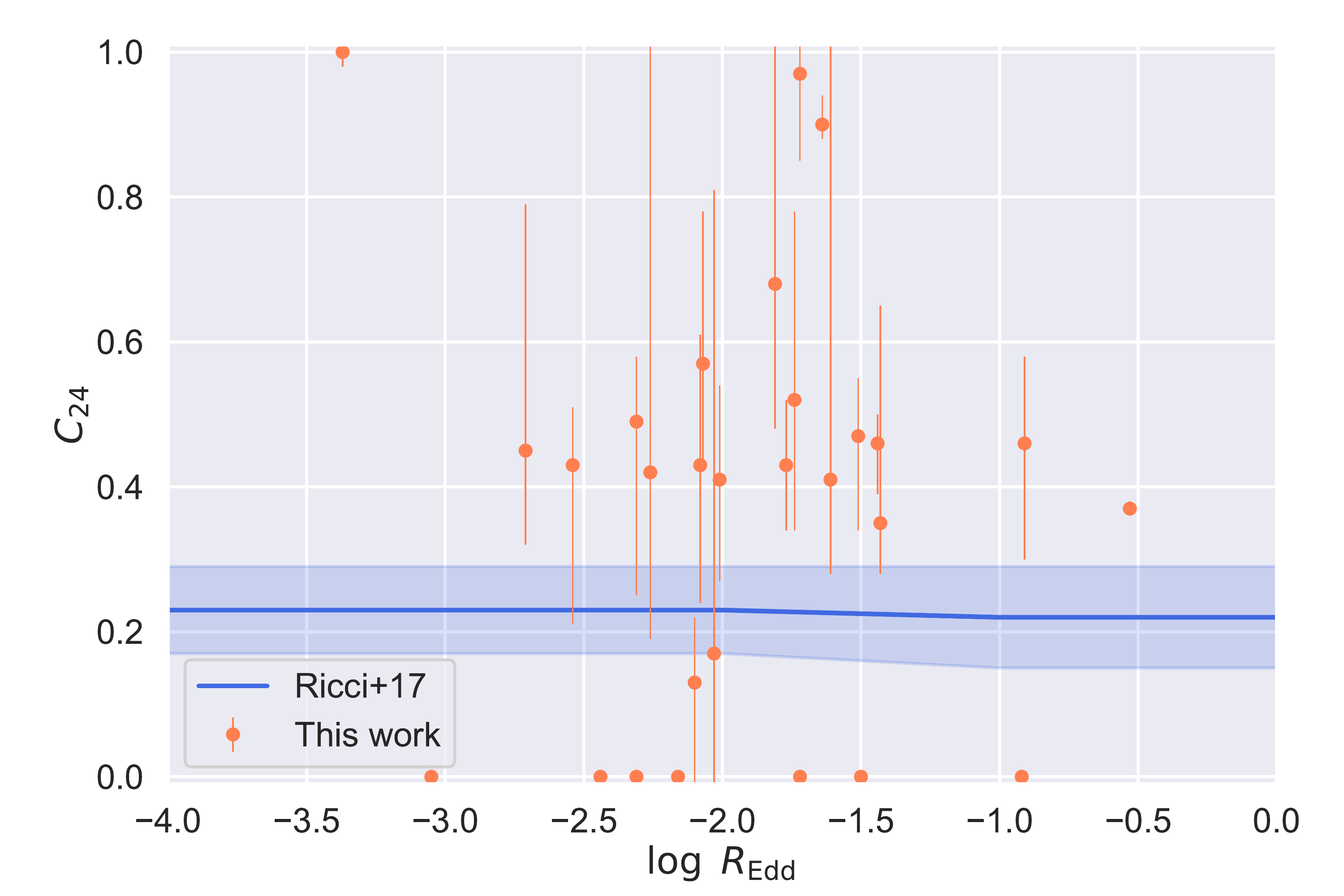}{0.45\textwidth}{(a)}\fig{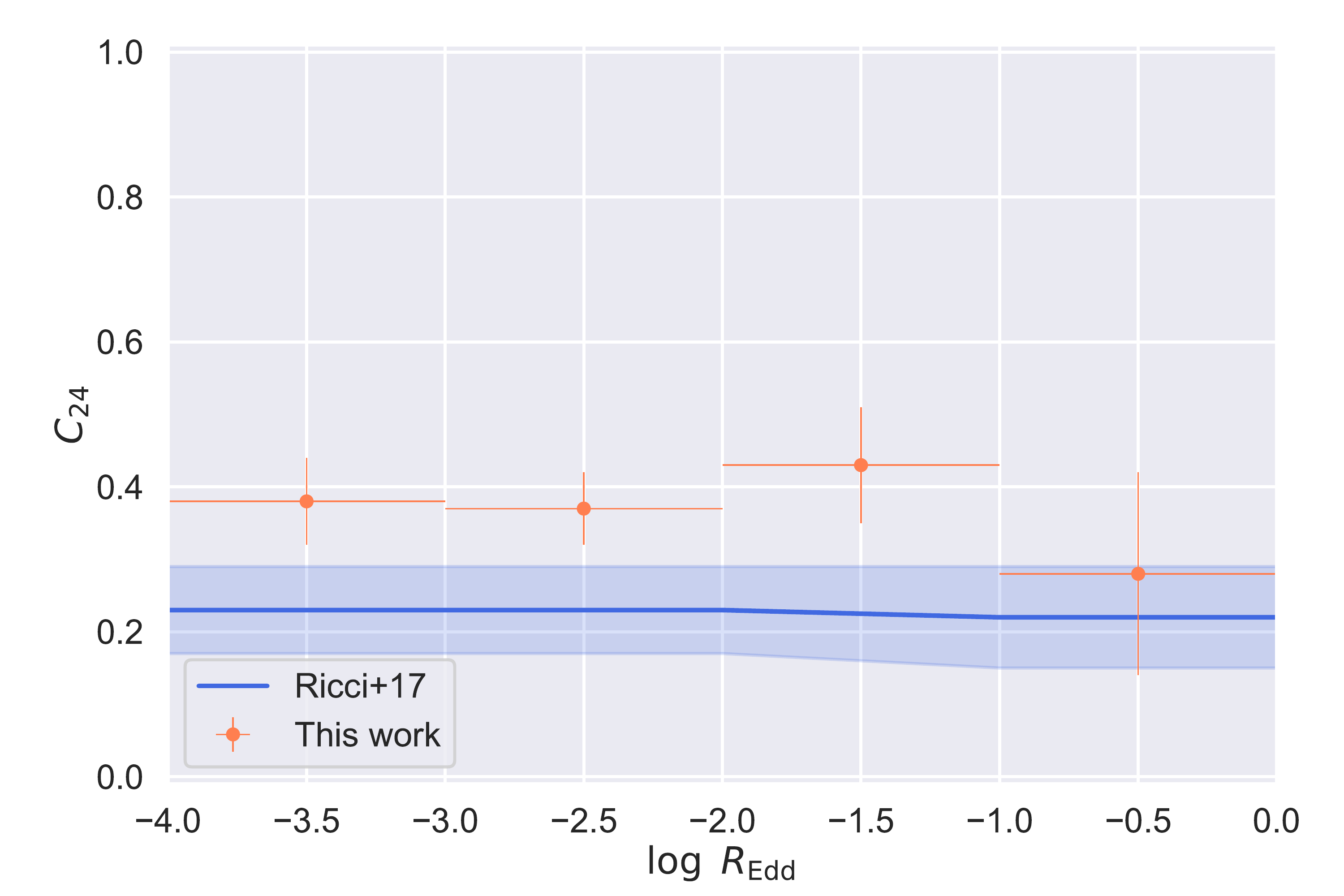}{0.45\textwidth}{(b)}}
\caption{(a) Correlation between the logarithmic Eddington ratio ($\log R_{\mathrm{Edd}}$) and the Compton-thick torus covering factor ($C_{24}$). Blue line represents the radiation regulated AGN unification model \citep{Ricci17b}. (b) Correlation between $\log R_{\mathrm{Edd}}$ and mean $C_{24}$ in the four $\log R_{\mathrm{Edd}}$ bins. Note that we apply the standard error as the error of $C_{24}$.}
\end{figure*}

\hyperref[Figure14]{Figure~14(a)} plots the correlation between the logarithmic Eddington ratio ($\log R_{\mathrm{Edd}}$) and the Compton-thick torus covering factor ($C_{24}$). Here we plot 29 objects whose black hole masses are known and whose $C_{24}$ are determined. \hyperref[Figure14]{Figure~14(b)} plots the correlation between $\log R_{\mathrm{Edd}}$ and mean $C_{24}$ in the four $\log R_{\mathrm{Edd}}$ bins. \hyperref[Figure14]{Figure~14(b)} shows that the average value of $C_{24}$ ($C_{24} = 36_{-4}^{+4}\%$). Here we apply the standard error as the error of $C_{24}$. The average $C_{24}$ of this work is larger than that of \cite{Ricci17c} ($C_{24} = 27_{-4}^{+4}\%$). In the case of \cite{Ricci17c}, they utilized AGNs detected by the Swift/BAT to calculate the statistical $C_{24}$. Here they assumed that if $\log R_{\mathrm{Edd}}$ was the same, then $\log N_{\mathrm{H}}^{\mathrm{Equ}}/\mathrm{cm}^{-2}$ would also be the same. As we mentioned in \hyperref[Section0603]{Section~6.3}, however, $\log N_{\mathrm{H}}^{\mathrm{Equ}}/\mathrm{cm}^{-2}$ of CTAGNs are larger than those of less obscured AGNs. In other words, $C_{24}$ value will be larger than that of \cite{Ricci17c}. This implies that some CTAGNs cannot be detected by Swift/BAT all-sky hard X-ray survey even in the local Universe (e.g., \citealt{Boorman16, Boorman18, LaMassa19, Zhao19b, Zhao19a, Zhao20}).
    \clearpage
\section{Conclusion}
\begin{enumerate}
\item We apply the XClumpy model to the broadband X-ray spectra of 52 CTAGN candidates observed with Chandra, XMM--Newton, Swift, Suzaku, and NuSTAR. Our model can reproduce the X-ray spectra of all objects.
\item The hydrogen column density along the line of sight ($N_{\mathrm{H}}^{\mathrm{LOS}}$) obtained from this work indicates that 24 objects are Compton-thin AGNs and 28 objects are Compton-thick AGNs within the 90\% confidence interval. The main reason is the difference in the torus model applied.
\item We compare the normalized histogram of the hydrogen column density along the equatorial direction ($N_{\mathrm{H}}^{\mathrm{Equ}}$) of CTAGNs and that of less obscured AGNs. $N_{\mathrm{H}}^{\mathrm{Equ}}$ of CTAGNs are larger than that of less obscured AGNs. This suggests that the structure of CTAGN may be intrinsically different from that of less obscured AGN.
\item We study the correlation between the logarithmic Eddington ratio ($\log R_{\mathrm{Edd}}$) and the Compton-thin torus covering factor ($C_{22}$). $C_{22}$ obtained from the XClumpy model is consistent with that inferred from \cite{Ricci17a} in the low Eddington ratio ($\log R_{\mathrm{Edd}} \leq -1.0$), whereas $C_{22}$ obtained from the XClumpy model is larger than that inferred from \cite{Ricci17a} in the high Eddington ratio ($-1.0 \leq \log R_{\mathrm{Edd}}$). This suggests that if the hydrogen column density is sufficiently large, it is difficult to blow it away by radiation pressure. In other words, there are objects with large $C_{22}$ even in the high Eddington ratio.
\item We examine the correlation between the logarithmic Eddington ratio ($\log R_{\mathrm{Edd}}$) and the Compton-thick torus covering factor ($C_{24}$). The average value of $C_{24}$ is equal to $36_{-4}^{+4}\%$. This value is larger than that of \cite{Ricci15} ($C_{24} = 27_{-4}^{+4}\%$) based on the assumption that all AGNs have intrinsically the same torus structure.
\item The above results (3--5) suggest that the structure of CTAGN may be intrinsically different from that of less obscured AGN.
\end{enumerate}

\acknowledgements
We acknowledge the anonymous referee for the helpful suggestions that incorporated the Article. This work was supported by the Grant-in-Aid for JSPS Fellows 20J00119 (A.T.) and 19J22216 (S.Y.), and by the Grants-in-Aid for Scientific Research 20H01946 (Y.U.) and 19H01906 (H.O.). Numerical computations were carried out on Cray XC50 at Center for Computational Astrophysics, National Astronomical Observatory of Japan. This research has made use of data and/or software provided by the High Energy Astrophysics Science Archive Research Center (HEASARC), which is a service of the Astrophysics Science Division at NASA/GSFC and the High Energy Astrophysics Division of the Smithsonian Astrophysical Observatory. This research has also made use of the NASA/IPAC Extragalactic Database (NED), which is operated by the Jet Propulsion Laboratory, California Institute of Technology, under contract with the National Aeronautics and Space Administration.
\facilities{Chandra, NuSTAR, Suzaku, Swift, XMM--Newton.}
\software{HEASoft 6.28 (HEASARC 2020), XSPEC \citep{Arnaud96}.}

\clearpage
\startlongtable
\begin{deluxetable*}{llllll}\label{Table05}
\tablecaption{Comparison of Hydrogen Column Density along the Line of Sight}
\tablehead
{
Galaxy Name                                                 & This work                                                 &
                                                            & \cite{Ricci15}                                            &
                                                            &                                                           \\
                                                            & XClumpy                                                   &
Borus02                                                     & Torus                                                     &
Observatory                                                 & Consistency                                               \\
(1)                                                         & (2)                                                       &
(3)                                                         & (4)                                                       &
(5)                                                         & (6)                                                       }
\startdata
ESO 112--G006                                               & $23.79_{-0.03}^{+0.04}$                                   &
$23.81_{-0.04}^{+0.04}$                                     & $24.03_{-0.24}^{+0.40}$                                   &
Swift/XRT                                                   & Yes                                                       \\
MCG --07--03--007                                           & $23.95_{-0.04}^{+0.05}$                                   &
$23.85_{-0.03}^{+0.05}$                                     & $24.18_{-0.23}^{+0.12}$                                   &
Swift/XRT                                                   & Yes                                                       \\
NGC 0424                                                    & $23.92_{-0.04}^{+0.16}$                                   &
$23.86_{-0.05}^{+0.05}$                                     & $24.33_{-0.01}^{+0.01}$                                   &
XMM--Newton                                                 & No                                                        \\
MCG +08--03--018                                            & $23.73_{-0.08}^{+0.10}$                                   &
$23.87_{-0.15}^{+0.04}$                                     & $24.24_{-0.15}^{+0.34}$                                   &
Swift/XRT                                                   & No                                                        \\
2MASX J01290761--6038423                                    & $23.98_{-0.08}^{+0.04}$                                   &
$23.81_{-0.10}^{+0.07}$                                     & $24.13_{-0.23}^{+0.19}$                                   &
Swift/XRT                                                   & Yes                                                       \\
\enddata
\tablecomments{Column~(1): galaxy name. Column~(2): logarithmic hydrogen column density along the line of sight ($\log N_{\mathrm{H}}^{\mathrm{LOS}}/\mathrm{cm}^{-2}$) obtained from the XClumpy model in this work. Column~(3): $\log N_{\mathrm{H}}^{\mathrm{LOS}}/\mathrm{cm}^{-2}$ inferred from the Borus02 model in this work. Column~(4): $\log N_{\mathrm{H}}^{\mathrm{LOS}}/\mathrm{cm}^{-2}$ obtained from the Torus model in \cite{Ricci15}. Column~(5): observatory in \cite{Ricci15}. Column~(6): whether $\log N_{\mathrm{H}}^{\mathrm{LOS}}/\mathrm{cm}^{-2}$ obtained from the XClumpy model is consistent with that inferred from the Torus model within the 90\% confidence interval.}
\end{deluxetable*}
     \clearpage
\appendix
\section{Comparison of Hydrogen Column Density along the Line of Sight with that of Parent Sample}\label{Appendix0100}
To distinguish whether the differences in the logarithmic hydrogen column density along the line of sight ($\log N_{\mathrm{H}}^{\mathrm{LOS}}/\mathrm{cm}^{-2}$), we applied the Borus02 model \citep{Balokovic18} to the X-ray spectra of all objects. \hyperref[Appendix0101]{Appendix~A.1} introduces the X-ray spectral model of \cite{Ricci15} and that of our analysis. \hyperref[Appendix0102]{Appendix~A.2} compares $\log N_{\mathrm{H}}^{\mathrm{LOS}}/\mathrm{cm}^{-2}$ obtained from the XClumpy model with those inferred from the Borus02 model and those obtained from \cite{Ricci15}.
\subsection{X-Ray Spectral Analysis}\label{Appendix0101}
We introduce the X-ray spectral model of \cite{Ricci15} and that of our analysis. First we present the X-ray spectral model of \cite{Ricci15}. The X-ray spectral model is the following:

\begin{equation}
\mathrm{const1 \times phabs} \times (\mathrm{atable\{torus1006.fits\}+ const2 \times zpowerlaw + (zgauss)+(apec))}               
\end{equation}

This model consists of five components.
\begin{enumerate}
\item $\mathrm{const1} \times \mathrm{phabs}$. These components mean the cross calibration constant and the Galactic absorption. These are basically same as the XClumpy model. \cite{Ricci15} considered the cross calibration constant if statistically required.
\item $\mathrm{atable\{torus1006.fits\}}$. This component represents the transmitted continuum absorbed by the torus, the reflection continuum, and the fluorescent lines from the torus based on the Torus model \citep{Brightman11}. They assumed that the reprocessing medium is a sphere with conical cutouts at both poles. This model have four free parameters. (1) the photon index ($\Gamma$), (2) the hydrogen column density along the equatorial direction ($N_{\mathrm{H}}^{\mathrm{Equ}}$), (3) the opening angle of the torus ($\theta_{\mathrm{open}}$), and (4) the inclination angle ($\theta_{\mathrm{incl}}$). (2) In the case of the Torus model, $N_{\mathrm{H}}^{\mathrm{Equ}}$ is equal to $N_{\mathrm{H}}^{\mathrm{LOS}}$. (4) They fixed $\theta_{\mathrm{incl}} = 87\degr$ to reduce the degree of complexity of the model.
\item $\mathrm{const2} \times \mathrm{zpowerlaw}$. This component means the scattered component. This component is the same as the XClumpy model.
\item $\mathrm{zgauss}$. This component represents the fluorescent lines. This component is the same as the XClumpy model.
\item $\mathrm{apec}$. This component means the emission from an optically thin thermal plasma in the host galaxy. This component is the same as the XClumpy model.
\end{enumerate}
We note that the Torus model had some calculation errors \citep[][Figure~4]{Liu15}. \cite{Liu15} indicated that the Torus model overestimated the flux in the soft X-rays in the case of the high hydrogen column density in the torus and the edge-on view. Due to this effect, $\log N_{\mathrm{H}}^{\mathrm{LOS}}/\mathrm{cm}^{-2}$ obtained from the XClumpy model are different from those inferred from the Torus model.

To compare $\log N_{\mathrm{H}}^{\mathrm{LOS}}/\mathrm{cm}^{-2}$ obtained from the XClumpy model with those inferred from other torus models, we applied the Borus02 model \citep{Balokovic18} to the X-ray spectra of 52 objects. \cite{Balokovic18} fixed the bug in the Torus model and released the Borus02 model whose geometric structure is the same as the Torus model. The X-ray spectral model is the following:

\begin{align}
& \mathrm{const1 \times phabs}                                                                      \nonumber\\
& \times (\mathrm{const2 \times zphabs \times cabs \times zcutoffpl}                                \nonumber\\
& +       \mathrm{const3 \times zcutoffpl+atable\{borus02\_v170323a.fits\}+(zgauss)+(apec)}).
\end{align}

This model consists of five components.

\begin{enumerate}
\item $\mathrm{const1} \times \mathrm{phabs}$. These components mean the cross calibration constant and the Galactic absorption. These components are the same as the XClumpy model.
\item $\mathrm{const2} \times \mathrm{zphabs} \times \mathrm{cabs} \times \mathrm{zcutoffpl}$. This component represents the transmitted continuum absorbed by the torus. This component is the basically same as the XClumpy model.
\item $\mathrm{const3} \times \mathrm{zcutoffpl}$. This component means the scattered component. This component is the same as the XClumpy model.
\item $\mathrm{atable\{borus02\_v170323a.fits\}}$. This component represent the reflection continuum and the fluorescent lines from the torus based on the Borus02 model \citep{Balokovic18}. This model have six free parameters. (1) the photon index ($\Gamma$: 1.4--2.6), (2) the cutoff energy ($E_{\mathrm{cut}}$: 20--2000 keV), (3) the logarithmic hydrogen column density ($\log N_{\mathrm{H}}^{\mathrm{Torus}}/\mathrm{cm}^{-2}$: 22.0--25.5), (4) the torus covering factor ($C_{\mathrm{Torus}}$: 0.1--1.0), (5) the cosine of the inclination angle ($\cos{\theta_{\mathrm{inc}}}$: 0.05--0.95), and (6) the relative abundance of iron ($A_{\mathrm{Fe}}$: 0.01--10.0). (2) We fixed $E_{\mathrm{cut}} = 370 \ \mathrm{keV}$. (3) In the case of the Borus02 model, $\log N_{\mathrm{H}}^{\mathrm{Torus}}/\mathrm{cm}^{-2}$ is equal to $\log N_{\mathrm{H}}^{\mathrm{LOS}}/\mathrm{cm}^{-2}$. (5) We limited $\cos{\theta_{\mathrm{inc}}}$ value within a range of 0.05--0.50. Especially, we fix its value at 0.05 for the eight objects (Mrk~0003, NGC~3079, NGC~3393, NGC~4945, NGC~5643, NGC~5728, NGC~6240, and NGC~7479). (6) We fixed $A_{\mathrm{Fe}} = 1.00$.
\item $\mathrm{zgauss}$. This component means the fluorescent lines. This component is the same as the XClumpy model.
\item $\mathrm{apec}$. This component represents the emission from an optically thin thermal plasma in the host galaxy. This component is the same as the XClumpy model.
\end{enumerate}
\begin{figure*}[t]\label{Figure15}
\gridline{\fig{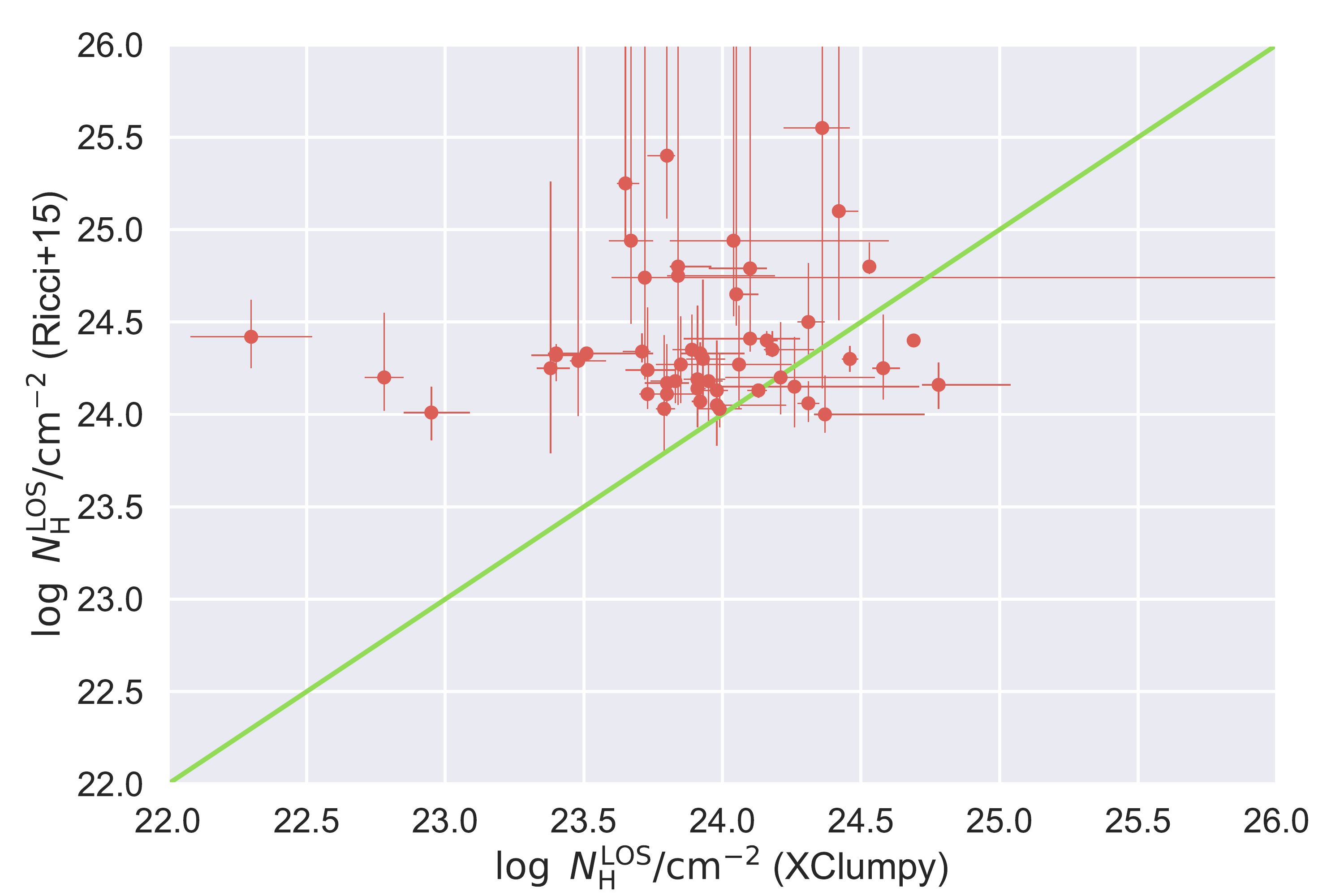}{0.45\textwidth}{(a)}\fig{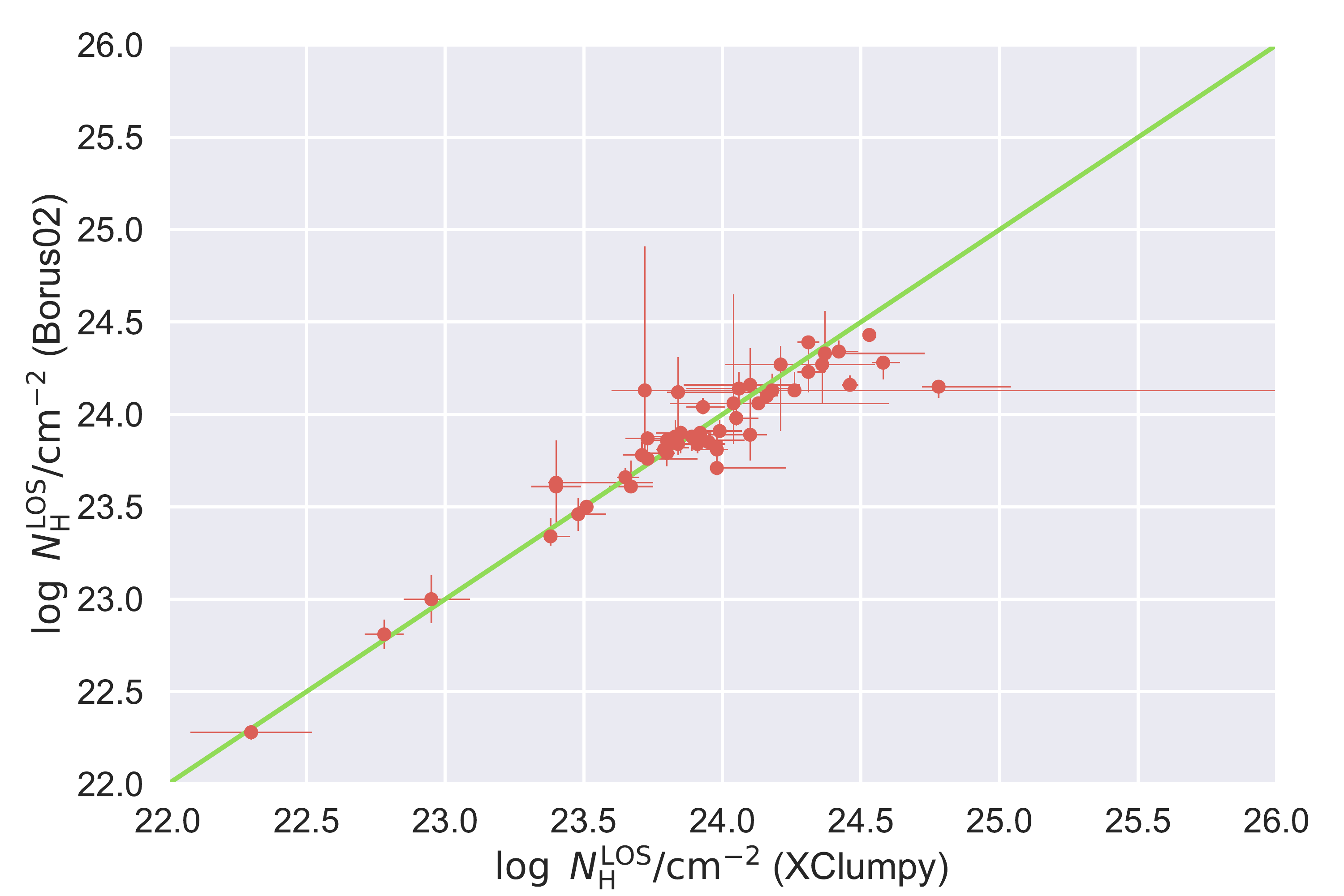}{0.45\textwidth}{(b)}}
\caption{(a) The correlation between the logarithmic hydrogen column density along the line of sight ($\log N_{\mathrm{H}}^{\mathrm{LOS}}/\mathrm{cm}^{-2}$) obtained from the XClumpy model and those inferred from the Torus model \citep{Ricci15}. (b) The correlation between $\log N_{\mathrm{H}}^{\mathrm{LOS}}/\mathrm{cm}^{-2}$ obtained from the XClumpy model and those inferred from the Borus02 model.}
\end{figure*}
\subsection{Comparison of Hydrogen Column Density along the Line of Sight among the XClumpy, Borus02, and Torus models}\label{Appendix0102}
We compare the logarithmic hydrogen column density along the line of sight ($\log N_{\mathrm{H}}^{\mathrm{LOS}}/\mathrm{cm}^{-2}$) obtained from the XClumpy model with that inferred from the Borus02 model and that obtained from the Torus model \citep{Ricci15}. \hyperref[Table05]{Table~5} summarizes $\log N_{\mathrm{H}}^{\mathrm{LOS}}/\mathrm{cm}^{-2}$ inferred from the XClumpy model, the Borus02 model, and the Torus model \citep{Ricci15}. \hyperref[Table05]{Table~5} shows that $\log N_{\mathrm{H}}^{\mathrm{LOS}}/\mathrm{cm}^{-2}$ of 34 objects obtained from the XClumpy model are not consistent with those inferred from the Torus model \citep{Ricci15} within the 90\% confidence interval. As mentioned in \hyperref[Appendix0101]{Appendix~A.1}, the main reason seems to be a calculation error in the Torus model \citep{Liu15}. In fact, we applied the Torus model to the X-ray spectra of several objects and confirmed that it can reproduce $\log N_{\mathrm{H}}^{\mathrm{LOS}}/\mathrm{cm}^{-2}$ of \cite{Ricci15}. We note that in the case of 14 objects that \cite{Ricci15} have analyzed the Swift/XRT data for, the reason could also be the observational data.

To study how much $\log N_{\mathrm{H}}^{\mathrm{LOS}}/\mathrm{cm}^{-2}$ varies with the applied torus model, we investigate the correlation between $\log N_{\mathrm{H}}^{\mathrm{LOS}}/\mathrm{cm}^{-2}$ obtained from the XClumpy model and those inferred from the Borus02 model. \hyperref[Figure15]{Figure~15} plots (a) the correlation between $\log N_{\mathrm{H}}^{\mathrm{LOS}}/\mathrm{cm}^{-2}$ obtained from the XClumpy model and those inferred from the Torus model \citep{Ricci15} and (b) the correlation between $\log N_{\mathrm{H}}^{\mathrm{LOS}}/\mathrm{cm}^{-2}$ obtained from the XClumpy model and those inferred from the Borus02 model. \hyperref[Figure15]{Figure~15(a)} shows that $\log N_{\mathrm{H}}^{\mathrm{LOS}}/\mathrm{cm}^{-2}$ obtained from the XClumpy model are different from those inferred from the Torus model \citep{Ricci15}, while \hyperref[Figure15]{Figure~15(b)} indicates that $\log N_{\mathrm{H}}^{\mathrm{LOS}}/\mathrm{cm}^{-2}$ obtained from the XClumpy model are consistent with those inferred from the Borus02 model. This result supports that the reason why $\log N_{\mathrm{H}}^{\mathrm{LOS}}/\mathrm{cm}^{-2}$ obtained from the XClumpy model are different from those inferred from the Torus model \citep{Ricci15} is due to a calculation error in the Torus model \citep{Liu15}.
\startlongtable
\begin{deluxetable*}{llllllll}\label{Table06}
\tablecaption{Best Fitting Parameters of Previous Studies}
\tablehead
{
Galaxy Name                                                 & Model                                                     &
Satellite1                                                  & Satellite2                                                &
$N_{\mathrm{H}}^{\mathrm{LOS}}$                             & $\Gamma$                                                  &
Reference                                                                                                               \\
(1)                                                         & (2)                                                       &
(3)					                                        & (4)							                            &
(5)						                                    & (6)                                                       &
(7)                                                                                                                     }
\startdata
ESO 112--G006                                               & Borus02                                                   &
XMM--Newton                                                 & NuSTAR                                                    &
$0.63_{-0.06}^{+0.04}$                                      & $1.60_{-0.13}^{+0.23}$                                    &
(06)                                                                                                                    \\
MCG --07--03--007                                           & Borus02                                                   &
XMM--Newton                                                 & NuSTAR                                                    &
$0.90_{-0.08}^{+0.07}$                                      & $1.84_{-0.15}^{+0.12}$                                    &
(06)                                                                                                                    \\
NGC 0424                                                    & Borus02                                                   &
XMM--Newton                                                 & NuSTAR                                                    &
$2.65_{-0.36}^{+0.44}$                                      & $1.81_{-0.05}^{+0.07}$                                    &
(05)                                                                                                                    \\
MCG +08--03--018                                            & Borus02                                                   &
Swift                                                       & NuSTAR                                                    &
$1.07_{-0.22}^{+0.24}$                                      & $2.17_{-0.19}^{+0.26}$                                    &
(05)                                                                                                                    \\
\enddata
\tablecomments{Column~(1): galaxy name. Column~(2): applied torus models. For more information, please see the following papers (Ehime torus model: \citealt{Ikeda09}, MYTorus model : \citealt{Murphy09}, and Borus02 model: \citealt{Balokovic18}). Column~(3): Satellite1. Column~(4): Satellite2. Column~(5): hydrogen column density along the line of sight in units of $10^{24} \ \mathrm{cm}^{-2}$. Column~(6): photon index. Column~(7): reference.}
\tablerefs{(01)~\cite{Georgantopoulos19}. (02)~\cite{Koss16}. (03)~\cite{Oda17}. (04)~\cite{Marchesi19a}. (05)~\cite{Marchesi19b}. (06)~\cite{Torres21}. (07)~\cite{Zhao20}.}
\end{deluxetable*}
     
\begin{figure*}[t]\label{Figure16}
\gridline{\fig{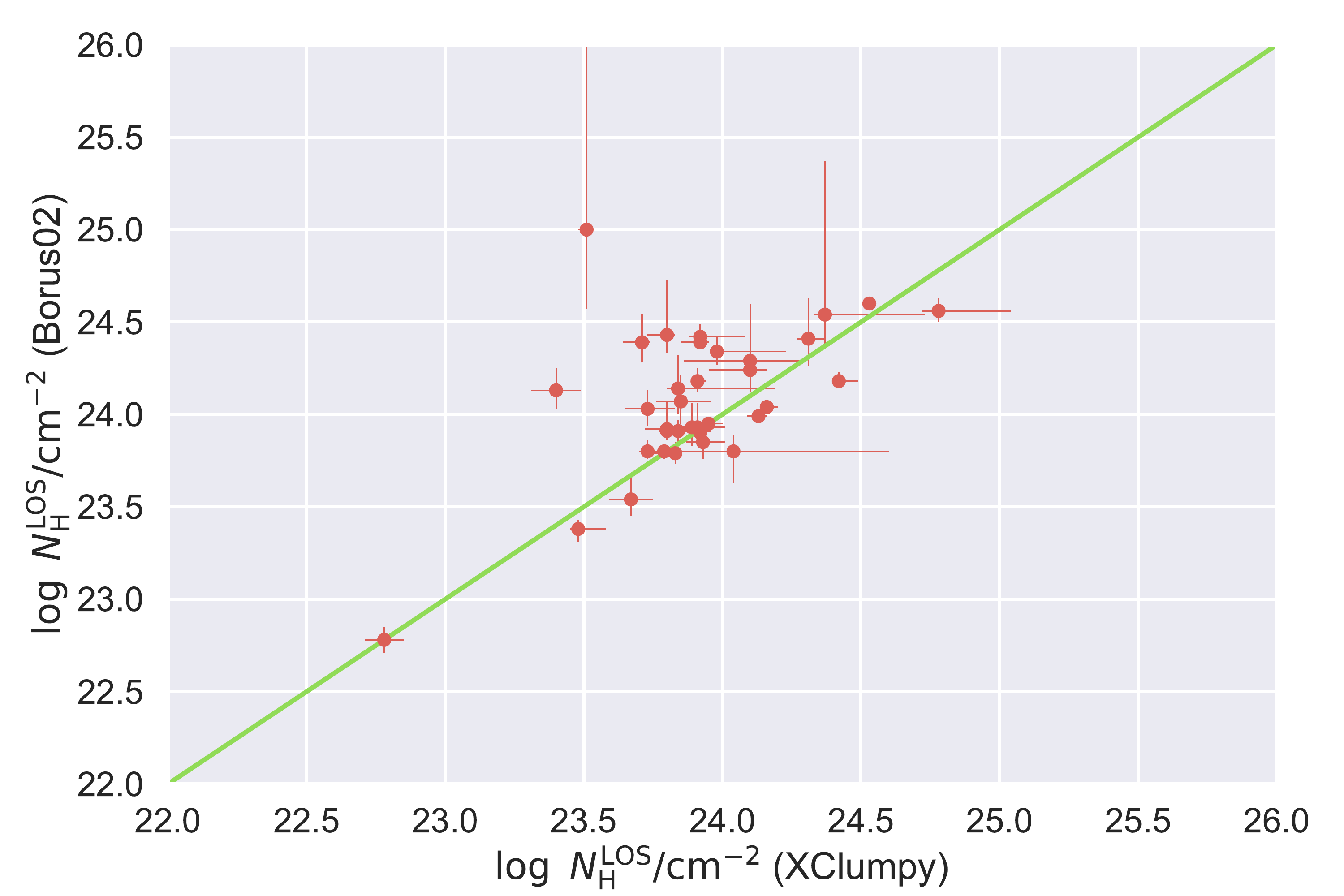}{0.45\textwidth}{(a)}\fig{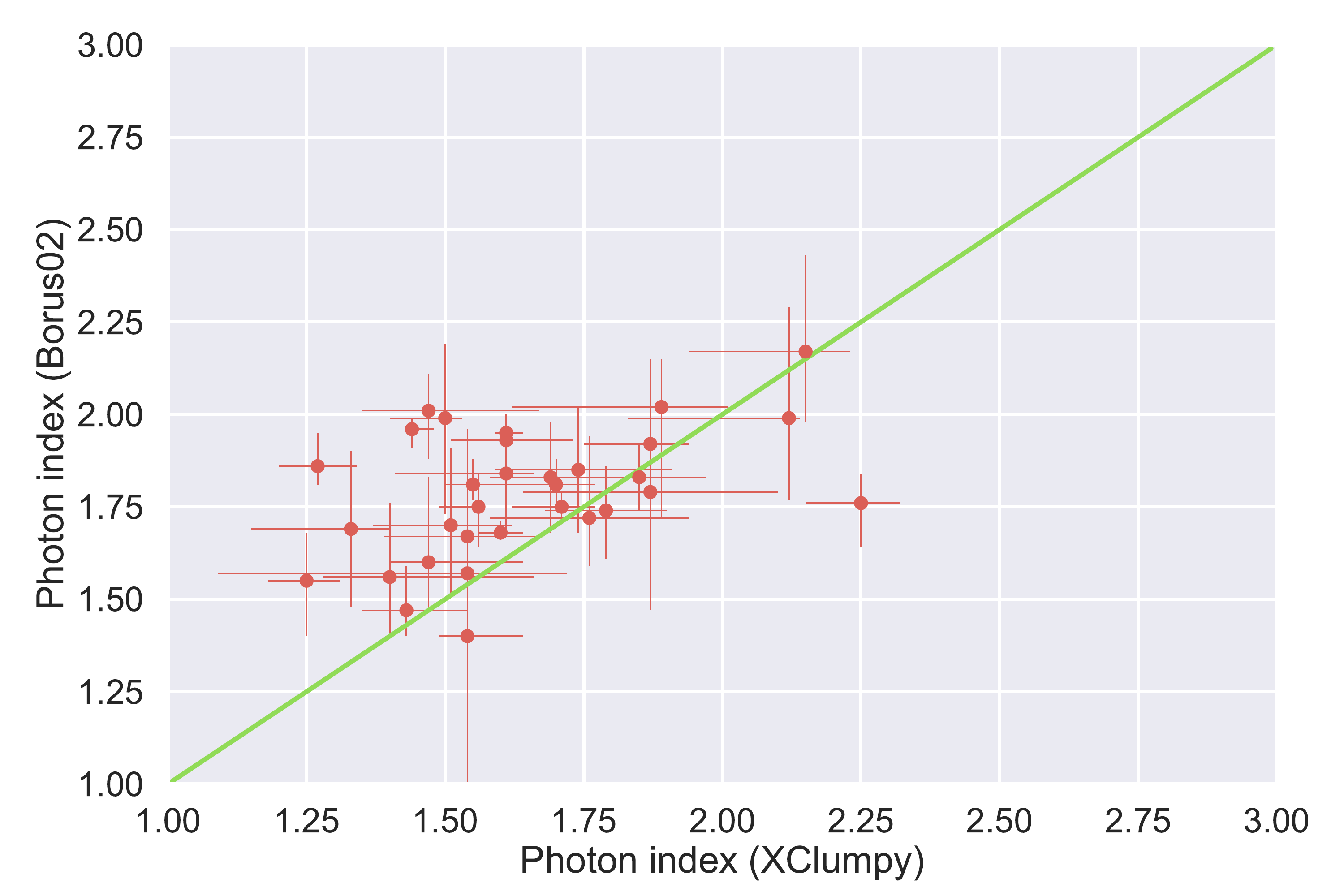}{0.45\textwidth}{(b)}}
\caption{(a) The correlation between the logarithmic hydrogen column density along the line of sight ($\log N_{\mathrm{H}}^{\mathrm{LOS}}/\mathrm{cm}^{-2}$) obtained from the XClumpy model and those inferred from the Borus02 model of previous works. (b) The correlation between the photon index ($\Gamma$) obtained from the XClumpy model and those inferred from the Borus02 model of previous works.}
\end{figure*}
\section{Comparison of Best-fitting Parameters with Previous Works}\label{Appendix0200}
\hyperref[Appendix0200]{Appexdix~B} compares our best-fitting parameters with those of previous works. To focus on differences in the spectral models, we only refer to previous works that used the NuSTAR observational data. 
\subsection{Comparison between XClumpy Model and Borus02 Model}\label{Appendix0201}
\hyperref[Table06]{Table~6} summarizes the logarithmic hydrogen column density along the line of sight ($\log N_{\mathrm{H}}^{\mathrm{LOS}}/\mathrm{cm}^{-2}$) and the photon index ($\Gamma$) of previous works. \hyperref[Appendix0201]{Appendix~B.1} mainly compares $\log N_{\mathrm{H}}^{\mathrm{LOS}}/\mathrm{cm}^{-2}$ and $\Gamma$ obtained from the XClumpy model with those inferred from the Borus02 model. \hyperref[Figure16]{Figure~16} plots (a) the correlation between $\log N_{\mathrm{H}}^{\mathrm{LOS}}/\mathrm{cm}^{-2}$ obtained from the XClumpy model and those inferred from the Borus02 model and (b) the correlation between $\Gamma$ obtained from the XClumpy model and those inferred from the Borus02 model. Here we plot 35 objects. We note that in the cases of these studies, $\log N_{\mathrm{H}}^{\mathrm{Equ}}/\mathrm{cm}^{-2}$ is different from $\log N_{\mathrm{H}}^{\mathrm{LOS}}/\mathrm{cm}^{-2}$. As a result, $\log N_{\mathrm{H}}^{\mathrm{LOS}}/\mathrm{cm}^{-2}$ obtained from the Borus02 model in their work are different from those inferred from the Borus02 model in this study (\hyperref[Appendix0102]{Appendix~A.2}).

\hyperref[Figure16]{Figure~16(a)} shows that $\log N_{\mathrm{H}}^{\mathrm{LOS}}/\mathrm{cm}^{-2}$ obtained from the XClumpy model of eight objects are consistent with those inferred from their works within the 90\% confidence interval and $\log N_{\mathrm{H}}/\mathrm{cm}^{-2}$ obtained from the XClumpy model of 15 objects are smaller than those inferred from their studies. This is because in the case of the XClumpy model, the flux of Fe K$\alpha$ emission lines and soft reflection continuum is larger than in the Borus02 model. As a result, the XClumpy model can explain the X-ray spectrum even if $\log N_{\mathrm{H}}^{\mathrm{LOS}}/\mathrm{cm}^{-2}$ and $\log N_{\mathrm{H}}^{\mathrm{Equ}}/\mathrm{cm}^{-2}$ are small. In this case, the XClumpy model is expected to have a smaller $\gamma$. As we expected, \hyperref[Figure16]{Figure~16(b)} shows that $\Gamma$ obtained from the XClumpy model of 14 objects agree with those inferred from their works within the 90\% confidence interval and $\Gamma$ obtained from the XClumpy model of 11 objects are smaller than those inferred from their studies.
\subsection{Individual Object}
\begin{enumerate}
\item \textbf{ESO 112--G006}. The model with one apec component gives an adequate fit to the X-ray spectra observed with the XMM--Newton and NuSTAR ($\chi_{\mathrm{red}}^{2} = 1.21$). We obtain $N_{\mathrm{H}}^{\mathrm{LOS}} = 0.62_{-0.04}^{+0.06} \times 10^{24} \ \mathrm{cm}^{-2}$ and $\Gamma = 1.47_{-0.07}^{+0.17}$ with the XClumpy model. The best-fitting parameters of this work are consistent with those of \cite{Torres21}. \cite{Torres21} analyzed the XMM--Newton and NuSTAR observational data, and estimated $N_{\mathrm{H}}^{\mathrm{LOS}} = 0.63_{-0.06}^{+0.04} \times 10^{24} \ \mathrm{cm}^{-2}$ and $\Gamma = 1.60_{-0.13}^{+0.23}$ with the Borus02 model.

\item \textbf{MCG --07--03--007}. The model with one apec component provides an adequate fit to the X-ray spectra observed with the XMM--Newton and NuSTAR ($\chi_{\mathrm{red}}^{2} = 1.24$). We obtain $N_{\mathrm{H}}^{\mathrm{LOS}} = 0.90_{-0.09}^{+0.11} \times 10^{24} \ \mathrm{cm}^{-2}$ and $\Gamma = 1.61_{-0.20}^{+0.05}$ with the XClumpy model. $N_{\mathrm{H}}^{\mathrm{LOS}}$ of this work agrees with that of \cite{Torres21}, while $\Gamma$ of this work is smaller than that of \cite{Torres21}. \cite{Torres21} used the XMM--Newton and NuSTAR observational data, and estimated $N_{\mathrm{H}}^{\mathrm{LOS}} = 0.90_{-0.08}^{+0.07} \times 10^{24} \ \mathrm{cm}^{-2}$ and $\Gamma = 1.84_{-0.15}^{+0.12}$ with the Borus02 model.

\item \textbf{NGC 0424}. The model with one zgauss component gives an adequate fit to the X-ray spectra observed with the XMM--Newton and NuSTAR ($\chi_{\mathrm{red}}^{2} = 1.40$). We obtain $N_{\mathrm{H}}^{\mathrm{LOS}} = 0.84_{-0.07}^{+0.31} \times 10^{24} \ \mathrm{cm}^{-2}$ and $\Gamma = 1.70_{-0.12}^{+0.07}$ with the XClumpy model. $\Gamma$ of this work is consistent with that of \cite{Marchesi19b}, whereas $N_{\mathrm{H}}^{\mathrm{LOS}}$ is smaller than that of \cite{Marchesi19b}. \cite{Marchesi19b} analyzed the XMM--Newton and NuSTAR observational data, and estimated $N_{\mathrm{H}}^{\mathrm{LOS}} = 2.65_{-0.36}^{+0.44} \times 10^{24} \ \mathrm{cm}^{-2}$ and $\Gamma = 1.81_{-0.05}^{+0.07}$ with the Borus02 model. As mentioned in \hyperref[Appendix0201]{Appendix~B.1}, such a difference can be attributed to the difference in the reflection continuum between the XClumpy model and Borus02 model.

\item \textbf{MCG +08--03--018}. The model can reproduces the X-ray spectra observed with the Swift/XRT and NuSTAR ($\chi_{\mathrm{red}}^{2} = 1.04$). We obtain $N_{\mathrm{H}}^{\mathrm{LOS}} = 0.54_{-0.10}^{+0.13} \times 10^{24} \ \mathrm{cm}^{-2}$ and $\Gamma = 2.15_{-0.21}^{+0.08}$ with the XClumpy model. $\Gamma$ of this work agrees with that of \cite{Marchesi19b}, while $N_{\mathrm{H}}^{\mathrm{LOS}}$ of this work is smaller than \cite{Marchesi19b}. \cite{Marchesi19b} used the Swift/XRT and NuSTAR observational data, and estimated $N_{\mathrm{H}}^{\mathrm{LOS}} = 1.07_{-0.22}^{+0.24} \times 10^{24} \ \mathrm{cm}^{-2}$ and $\Gamma = 2.17_{-0.19}^{+0.26}$ with the Borus02 model.

\item \textbf{2MASX J01290761--6038423}. We report the XMM--Newton and NuSTAR observational data for the first time. The model with one apec component can replicate the X-ray spectra observed with the XMM--Newton and NuSTAR ($\chi_{\mathrm{red}}^{2} = 1.14$). We obtain $N_{\mathrm{H}}^{\mathrm{LOS}} = 0.95_{-0.17}^{+0.09} \times 10^{24} \ \mathrm{cm}^{-2}$ and $\Gamma = 1.86_{-0.10}^{+0.22}$ with the XClumpy model.

\item \textbf{ESO 244--IG030}. The model with one apec component well reproduce the X-ray spectra observed with the XMM--Newton and NuSTAR ($\chi_{\mathrm{red}}^{2} = 0.94$). We obtain $N_{\mathrm{H}}^{\mathrm{LOS}} = 0.06_{-0.01}^{+0.01} \times 10^{24} \ \mathrm{cm}^{-2}$ and $\Gamma = 1.74_{-0.15}^{+0.17}$ with the XClumpy model. $N_{\mathrm{H}}^{\mathrm{LOS}}$ and $\Gamma$ of this work are consistent with those of \cite{Marchesi19b}. \cite{Marchesi19a} analyzed the XMM--Newton and NuSTAR observational data, and estimated $N_{\mathrm{H}}^{\mathrm{LOS}} = 0.06_{-0.01}^{+0.01} \times 10^{24} \ \mathrm{cm}^{-2}$ and $\Gamma = 1.85_{-0.17}^{+0.17}$ with the Borus02 model.

\item \textbf{NGC 1106}. We report the XMM--Newton and NuSTAR observational data for the first time. The model with one apec component can reproduce the X-ray spectra observed with the XMM--Newton and NuSTAR ($\chi_{\mathrm{red}}^{2} = 1.15$). We obtain $N_{\mathrm{H}}^{\mathrm{LOS}} = 3.77_{-0.32}^{+0.52} \times 10^{24} \ \mathrm{cm}^{-2}$ and $\Gamma = 1.67_{-0.30}^{+0.16}$ with the XClumpy model.

\item \textbf{2MFGC 02280}. The model well replicate the X-ray spectra observed with the Swift/XRT and NuSTAR ($\chi_{\mathrm{red}}^{2} = 0.96$). We obtain $N_{\mathrm{H}}^{\mathrm{LOS}} = 2.03_{-0.20}^{+2.17} \times 10^{24} \ \mathrm{cm}^{-2}$ and $\Gamma = 1.71_{-0.39}^{+0.18}$ with the XClumpy model. $N_{\mathrm{H}}^{\mathrm{LOS}}$ of this work agrees with that of \cite{Koss16}, whereas $\Gamma$ of this work is smaller than that of \cite{Koss16}. \cite{Koss16} used the Swift/XRT and NuSTAR observational data, and estimated $N_{\mathrm{H}}^{\mathrm{LOS}} = 3.16_{-0.30}^{+0.35} \times 10^{24} \ \mathrm{cm}^{-2}$ and $\Gamma = 2.10_{-0.10}^{+0.10}$ with the MYTorus model.

\item \textbf{NGC 1125}. We report the NuSTAR observational data for the first time. The model can reproduce the X-ray spectra observed with the Chandra and NuSTAR ($\chi_{\mathrm{red}}^{2} = 1.16$). We obtain $N_{\mathrm{H}}^{\mathrm{LOS}} = 1.16_{-0.50}^{+0.92} \times 10^{24} \ \mathrm{cm}^{-2}$ and $\Gamma = 1.82_{-0.24}^{+0.11}$ with the XClumpy model.

\item \textbf{NGC 1194}. The model can replicate the X-ray spectra observed with the Suzaku and NuSTAR ($\chi_{\mathrm{red}}^{2} = 1.20$). We obtain $N_{\mathrm{H}}^{\mathrm{LOS}} = 0.83_{-0.13}^{+0.06} \times 10^{24} \ \mathrm{cm}^{-2}$ and $\Gamma = 1.27_{-0.07}^{+0.07}$. $N_{\mathrm{H}}^{\mathrm{LOS}}$ and $\Gamma$ of this work are smaller than those of \cite{Marchesi19b}. \cite{Marchesi19b} analyzed the XMM--Newton and NuSTAR observational data, and estimated $N_{\mathrm{H}}^{\mathrm{LOS}} = 2.47_{-0.08}^{+0.41} \times 10^{24} \ \mathrm{cm}^{-2}$ and $\Gamma = 1.86_{-0.05}^{+0.09}$ with the Borus02 model.

\item \textbf{NGC 1229}. The model provides an adequate fit to the X-ray spectra observed with the Swift/XRT and NuSTAR ($\chi_{\mathrm{red}}^{2} = 1.23$). We obtain $N_{\mathrm{H}}^{\mathrm{LOS}} = 0.47_{-0.09}^{+0.26} \times 10^{24} \ \mathrm{cm}^{-2}$ and $\Gamma = 1.54_{-0.45}^{+0.18}$ with the XClumpy model. $N_{\mathrm{H}}^{\mathrm{LOS}}$ and $\Gamma$ of this work agree with those of \cite{Marchesi19b}. \cite{Marchesi19b} used the Swift/XRT and NuSTAR observational data, and estimated $N_{\mathrm{H}}^{\mathrm{LOS}} = 0.35_{-0.07}^{+0.10} \times 10^{24} \ \mathrm{cm}^{-2}$ and $\Gamma = 1.57_{-0.16}^{+0.14}$ with the Borus02 model.

\item \textbf{ESO 201--IG004}. The model gives an adequate fit to the X-ray spectra observed with the Suzaku and NuSTAR ($\chi_{\mathrm{red}}^{2} = 1.31$). We obtain $N_{\mathrm{H}}^{\mathrm{LOS}} = 0.25_{-0.05}^{+0.31} \times 10^{24} \ \mathrm{cm}^{-2}$ and $\Gamma = 1.33_{-0.18}^{+0.07}$ with the XClumpy model. $N_{\mathrm{H}}^{\mathrm{LOS}}$ and $\Gamma$ are smaller than those of \cite{Marchesi19b}. \cite{Marchesi19b} analyzed the XMM--Newton and NuSTAR observational data, and estimated $N_{\mathrm{H}}^{\mathrm{LOS}} = 1.34_{-0.32}^{+0.36} \times 10^{24} \ \mathrm{cm}^{-2}$ and $\Gamma = 1.69_{-0.21}^{+0.21}$ with the Borus02 model.

\item \textbf{2MASX J03561995--6251391}. The model with one apec component can reproduce the X-ray spectra observed with the XMM--Newton and NuSTAR ($\chi_{\mathrm{red}}^{2} = 1.12$). We obtain $N_{\mathrm{H}}^{\mathrm{LOS}} = 0.63_{-0.12}^{+0.08} \times 10^{24} \ \mathrm{cm}^{-2}$ and $\Gamma = 1.50_{-0.10}^{+0.03}$ with the XClumpy model. $N_{\mathrm{H}}^{\mathrm{LOS}}$ and $\Gamma$ of this work are smaller those of \cite{Marchesi19b}. \cite{Marchesi19b} used the Swift/XRT and NuSTAR observational data, and estimated $N_{\mathrm{H}}^{\mathrm{LOS}} = 0.84_{-0.11}^{+0.11} \times 10^{24} \ \mathrm{cm}^{-2}$ and $\Gamma = 1.99_{-0.26}^{+0.20}$ with the Borus02 model.

\item \textbf{MCG --02--12--017}. We report the XMM--Newton and NuSTAR data for the first time. The model with one apec component can replicate the X-ray spectra observed with the XMM--Newton and NuSTAR ($\chi_{\mathrm{red}}^{2} = 1.07$). We obtain $N_{\mathrm{H}}^{\mathrm{LOS}} = 0.24_{-0.03}^{+0.04} \times 10^{24} \ \mathrm{cm}^{-2}$ and $\Gamma = 1.79_{-0.26}^{+0.09}$ with the XClumpy model.

\item \textbf{CGCG 420--015}. The model with one apec component gives an adequate fit to the X-ray spectra observed with the XMM--Newton and NuSTAR ($\chi_{\mathrm{red}}^{2} = 1.22$). We obtain $N_{\mathrm{H}}^{\mathrm{LOS}} = 0.81_{-0.04}^{+0.06} \times 10^{24} \ \mathrm{cm}^{-2}$ and $\Gamma = 1.87_{-0.12}^{+0.07}$ with the XClumpy model. $\Gamma$ of this work are consistent with that of \cite{Marchesi19b}, while $N_{\mathrm{H}}^{\mathrm{LOS}}$ of this work is smaller than that of  \cite{Marchesi19b}. \cite{Marchesi19b} analyzed the XMM--Newton and NuSTAR observational data, and estimated $N_{\mathrm{H}}^{\mathrm{LOS}} = 1.50_{-0.22}^{+0.25} \times 10^{24} \ \mathrm{cm}^{-2}$ and $\Gamma = 1.92_{-0.13}^{+0.14}$.

\item \textbf{ESO 005--G004}. The model with one zgauss component provides an adequate fit to the X-ray spectra observed with the Suzaku and NuSTAR ($\chi_{\mathrm{red}}^{2} = 1.25$). We $N_{\mathrm{H}}^{\mathrm{LOS}} = 0.51_{-0.08}^{+0.04} \times 10^{24} \ \mathrm{cm}^{-2}$ and $\Gamma = 1.43_{-0.08}^{+0.11}$ with the XClumpy model. $\Gamma$ of this work agrees with that of \cite{Marchesi19b}, whereas $N_{\mathrm{H}}^{\mathrm{LOS}}$ is smaller than that of \cite{Marchesi19b}. \cite{Marchesi19b} used the Swift/XRT and NuSTAR observational data, and estimated $N_{\mathrm{H}}^{\mathrm{LOS}} = 2.48_{-0.62}^{+0.86} \times 10^{24} \ \mathrm{cm}^{-2}$ and $\Gamma = 1.47_{-0.07}^{+0.12}$ with the Borus02 model. This is because of the difference in the quality of the data in the soft X-ray band. The Suzaku/XIS has a larger effective area than that of the Swift/XRT. The scattered fraction we obtain, $f_{\mathrm{scat}} = 1.93_{-0.59}^{+1.16}$, is larger than those found in the previous studies based on the Suzaku and Swift/BAT first 9--month data \citep{Ueda07, Eguchi09}. The main reason for the difference is that the photon index, which is constrained mainly with the NuSTAR data in our analysis, is much smaller than the previous estimates ($\Gamma \simeq 2.0$). This may be related to the large flux variability between the Suzaku and NuSTAR observations (\hyperref[Table03]{Table~3}).

\item \textbf{Mrk 0003}. The model can replicate the X-ray spectra observed with the XMM--Newton and NuSTAR ($\chi_{\mathrm{red}}^{2} = 1.18$). We fixed $i$ at 87\degr because the maser has been detected in this object. We obtain $N_{\mathrm{H}}^{\mathrm{LOS}} = 0.84_{-0.06}^{+0.02} \times 10^{24} \ \mathrm{cm}^{-2}$ and $\Gamma = 1.60_{-0.04}^{+0.04}$ with the XClumpy model. $N_{\mathrm{H}}^{\mathrm{LOS}}$ of this work is consistent with that of \cite{Marchesi19b}, while $\Gamma$ of this work is smaller than that of \cite{Marchesi19b}. \cite{Marchesi19b} analyzed the XMM--Newton and NuSTAR observational data, and estimated $N_{\mathrm{H}}^{\mathrm{LOS}} = 0.80_{-0.02}^{+0.03} \times 10^{24} \ \mathrm{cm}^{-2}$ and $\Gamma = 1.68_{-0.02}^{+0.03}$ with the Borus02 model.

\item \textbf{2MASX J06561197--4919499}. We report the XMM--Newton and NuSTAR observational data for the first time. The model with one apec component can reproduce the X-ray spectra observed with the XMM--Newton and NuSTAR ($\chi_{\mathrm{red}}^{2} = 1.06$). We obtain $N_{\mathrm{H}}^{\mathrm{LOS}} = 0.97_{-0.17}^{+0.17} \times 10^{24} \ \mathrm{cm}^{-2}$ and $\Gamma = 1.58_{-0.22}^{+0.04}$ with the XClumpy model.

\item \textbf{MCG +06--16--028}. The model with one apec component gives an adequate fit to the X-ray spectra observed with the Suzaku and NuSTAR ($\chi_{\mathrm{red}}^{2} = 1.30$). We obtain $N_{\mathrm{H}}^{\mathrm{LOS}} = 0.69_{-0.04}^{+0.19} \times 10^{24} \ \mathrm{cm}^{-2}$ and $\Gamma = 1.69_{-0.11}^{+0.07}$ with the XClumpy model. $N_{\mathrm{H}}^{\mathrm{LOS}}$ and $\Gamma$ of this work agree with those of \cite{Marchesi19b}. \cite{Marchesi19b} used the Swift/XRT and NuSTAR observational data, and estimated $N_{\mathrm{H}}^{\mathrm{LOS}} = 0.82_{-0.11}^{+0.11} \times 10^{24} \ \mathrm{cm}^{-2}$ and $\Gamma = 1.83_{-0.15}^{+0.15}$ with the Borus02 model.

\item \textbf{Mrk 0078}. The model with one apec component well replicates the X-ray spectra observed with the XMM--Newton and NuSTAR ($\chi_{\mathrm{red}}^{2} = 0.97$). We obtain $N_{\mathrm{H}}^{\mathrm{LOS}} = 0.63_{-0.04}^{+0.11} \times 10^{24} \ \mathrm{cm}^{-2}$ and $\Gamma = 1.54_{-0.05}^{+0.10}$ with the XClumpy model. $\Gamma$ of this work is consistent with that of \cite{Zhao20}, whereas $N_{\mathrm{H}}^{\mathrm{LOS}}$ of this work is smaller than that of \cite{Zhao20}. \cite{Zhao20} analyzed the XMM--Newton and NuSTAR observational data, and estimated $N_{\mathrm{H}}^{\mathrm{LOS}} = 0.81_{-0.04}^{+0.19} \times 10^{24} \ \mathrm{cm}^{-2}$ and $\Gamma = 1.40^{+0.21}$ with the Borus02 model.

\item \textbf{Mrk 0622}. The model with one apec component well reproduces the X-ray spectra observed with the XMM--Newton and NuSTAR ($\chi_{\mathrm{red}}^{2} = 0.84$). We obtain $N_{\mathrm{H}}^{\mathrm{LOS}} = 0.30_{-0.02}^{+0.07} \times 10^{24} \ \mathrm{cm}^{-2}$ and $\Gamma = 1.79_{-0.11}^{+0.11}$ with the XClumpy model. $N_{\mathrm{H}}^{\mathrm{LOS}}$ and $\Gamma$ of this work agree with those of \cite{Torres21}. \cite{Torres21} used the XMM--Newton and NuSTAR observational data, and estimated $N_{\mathrm{H}}^{\mathrm{LOS}} = 0.24_{-0.04}^{+0.03} \times 10^{24} \ \mathrm{cm}^{-2}$ and $\Gamma = 1.74_{-0.13}^{+0.12}$ with the Borus02 model.

\item \textbf{NGC 2788A}. We report the NuSTAR observational data for the first time. The model can replicate the X-ray spectra observed with the Suzaku and NuSTAR ($\chi_\mathrm{red}^{2} = 1.03$). We obtain $N_{\mathrm{H}}^{\mathrm{LOS}} = 2.31_{-0.76}^{+0.52} \times 10^{24} \ \mathrm{cm}^{-2}$ and $\Gamma = 1.67_{-0.08}^{+0.11}$ with the XClumpy model.

\item \textbf{SBS 0915+556}. We report the NuSTAR observational data for the first time. The model provides an adequate fit to the X-ray spectra observed with the Swift/XRT and NuSTAR ($\chi_{\mathrm{red}}^{2} = 1.33$). We obtain $N_{\mathrm{H}}^{\mathrm{LOS}} = 0.09_{-0.02}^{+0.03} \times 10^{24} \ \mathrm{cm}^{-2}$ and $\Gamma = 2.02_{-0.02}^{+0.02}$ with the XClumpy model.

\item \textbf{2MASX J09235371--3141305}. The model reproduces the X-ray spectra observed with the Swift/XRT and NuSTAR ($\chi_{\mathrm{red}}^{2} = 1.05$). We obtain $N_{\mathrm{H}}^{\mathrm{LOS}} = 0.54_{-0.04}^{+0.22} \times 10^{24} \ \mathrm{cm}^{-2}$ and $\Gamma = 1.47_{-0.12}^{+0.20}$ with the XClumpy model. $N_{\mathrm{H}}^{\mathrm{LOS}}$ of this work is consistent with that of \cite{Marchesi19b}, while $\Gamma$ is smaller than that of \cite{Marchesi19b}. \cite{Marchesi19b} analyzed the Swift/XRT and NuSTAR observational data, and estimated $N_{\mathrm{H}}^{\mathrm{LOS}} = 0.63_{-0.06}^{+0.09} \times 10^{24} \ \mathrm{cm}^{-2}$ and $\Gamma = 2.01_{-0.13}^{+0.10}$ with the Borus02 model.

\item \textbf{ESO 565--G019}. We report the XMM--Newton and NuSTAR observational data for the first time. The model with one zgauss component can replicate the X-ray spectra observed with the XMM--Newton and NuSTAR ($\chi_{\mathrm{red}}^{2} = 1.20$). We obtain $N_{\mathrm{H}}^{\mathrm{LOS}} = 1.12_{-0.05}^{+0.20} \times 10^{24} \ \mathrm{cm}^{-2}$ and $\Gamma = 1.09_{-0.08}^{+0.03}$ with the XClumpy model.

\item \textbf{MCG +10--14--025}. The model can reproduce the X-ray spectra observed with the Suzaku and NuSTAR ($\chi_{\mathrm{red}}^{2} = 1.07$). We obtain $N_{\mathrm{H}}^{\mathrm{LOS}} = 1.53_{-0.09}^{+0.53} \times 10^{24} \ \mathrm{cm}^{-2}$ and $\Gamma = 1.58_{-0.32}^{+0.27}$ with the XClumpy model. $N_{\mathrm{H}}^{\mathrm{LOS}}$ and $\Gamma$ of this work agree those of \cite{Oda17}. \cite{Oda17} used the XMM--Newton, Suzaku, and NuSTAR observational data, and estimated $N_{\mathrm{H}}^{\mathrm{LOS}} = 1.31_{-0.36}^{+0.31} \times 10^{24} \ \mathrm{cm}^{-2}$ and $\Gamma = 1.63_{-0.13}^{+0.27}$ with the Ikeda model.

\item \textbf{NGC 3079}. The model can replicate the X-ray spectra observed with the XMM--Newton and NuSTAR ($\chi_{\mathrm{red}}^{2} = 1.16$). We fixed $i$ at 87\degr since the maser has been detected in this object. We obtain $N_{\mathrm{H}}^{\mathrm{LOS}} = 2.65_{-0.13}^{+0.40} \times 10^{24} \ \mathrm{cm}^{-2}$ and $\Gamma = 1.61_{-0.10}^{+0.12}$ with the XClumpy model. $N_{\mathrm{H}}^{\mathrm{LOS}}$ of this work is larger than that of \cite{Marchesi19b} and $\Gamma$ of this work is smaller than that of \cite{Marchesi19b}. \cite{Marchesi19b} analyzed the XMM--Newton and NuSTAR observational data, and estimated $N_{\mathrm{H}}^{\mathrm{LOS}} = 1.51_{-0.09}^{+0.19} \times 10^{24} \ \mathrm{cm}^{-2}$ and $\Gamma = 1.93_{-0.10}^{+0.06}$ with the Borus02 model. The reason is that $N_{\mathrm{H}}^{\mathrm{LOS}}$ and $\Gamma$ are degenerated.

\item \textbf{ESO 317--G041}. The model with one apec component gives an adequate fit to the X-ray spectra observed with the XMM--Newton and NuSTAR ($\chi_{\mathrm{red}}^{2} = 1.32$). We obtain $N_{\mathrm{H}}^{\mathrm{LOS}} = 0.85_{-0.12}^{+0.16} \times 10^{24} \ \mathrm{cm}^{-2}$ and $\Gamma = 1.40_{-0.12}^{+0.26}$ with the XClumpy model. $N_{\mathrm{H}}^{\mathrm{LOS}}$ and $\Gamma$ of this work is consistent with those of \cite{Marchesi19a}. \cite{Marchesi19a} used the XMM--Newton and NuSTAR observational data, and estimated $N_{\mathrm{H}}^{\mathrm{LOS}} = 0.71_{-0.15}^{+0.14} \times 10^{24} \ \mathrm{cm}^{-2}$ and $\Gamma = 1.56_{-0.16}^{+0.20}$ with the Borus02 model.

\item \textbf{SDSS J103315.71+525217.8}. The model can reproduce the X-ray spectra observed with the XMM--Newton and NuSTAR ($\chi_{\mathrm{red}}^{2} = 1.15$). We obtain $N_{\mathrm{H}}^{\mathrm{LOS}} = 0.71_{-0.12}^{+0.16} \times 10^{24} \ \mathrm{cm}^{-2}$ and $\Gamma = 1.55_{-0.18}^{+0.25}$ with the XClumpy model. $N_{\mathrm{H}}^{\mathrm{LOS}}$ and $\Gamma$ of this work agree with those of \cite{Marchesi19a}. \cite{Marchesi19a} analyzed the XMM--Newton and NuSTAR observational data, and estimated $N_{\mathrm{H}}^{\mathrm{LOS}} = 1.17_{-0.32}^{+0.38} \times 10^{24} \ \mathrm{cm}^{-2}$ and $\Gamma = 1.40$ (fixed) with the Borus02 model.

\item \textbf{NGC 3393}. The model can replicate the X-ray spectra observed with the XMM--Newton and NuSTAR ($\chi_{\mathrm{red}}^{2} = 1.14$). We fixed $i$ at 87\degr because the maser has been detected in this object. We obtain $N_{\mathrm{H}}^{\mathrm{LOS}} = 2.05_{-0.21}^{+0.30} \times 10^{24} \ \mathrm{cm}^{-2}$ and $\Gamma = 1.76_{-0.18}^{+0.18}$ with the XClumpy model. $N_{\mathrm{H}}^{\mathrm{LOS}}$ and $\Gamma$ of this work agree with those of \cite{Marchesi19b}. \cite{Marchesi19b} used the XMM--Newton and NuSTAR observational data, and estimated $N_{\mathrm{H}}^{\mathrm{LOS}} = 2.58_{-0.87}^{+1.32} \times 10^{24} \ \mathrm{cm}^{-2}$ and $\Gamma = 1.72_{-0.13}^{+0.22}$ with the Borus02 model.

\item \textbf{NGC 4102}. The model with one apec component can reproduce the X-ray spectra observed with the XMM--Newton and NuSTAR ($\chi_{\mathrm{red}}^{2} = 1.12$). We obtain $N_{\mathrm{H}}^{\mathrm{LOS}} = 0.67_{-0.14}^{+0.02} \times 10^{24} \ \mathrm{cm}^{-2}$ and $\Gamma = 1.56_{-0.07}^{+0.01}$ with the XClumpy. $N_{\mathrm{H}}^{\mathrm{LOS}}$ of this work is consistent with that of \cite{Marchesi19b}, whereas $\Gamma$ of this work is smaller than that of \cite{Marchesi19b}. \cite{Marchesi19b} analyzed the XMM--Newton and NuSTAR observational data, and estimated $N_{\mathrm{H}}^{\mathrm{LOS}} = 0.62_{-0.08}^{+0.09} \times 10^{24} \ \mathrm{cm}^{-2}$ and $\Gamma = 1.75_{-0.11}^{+0.09}$ with the Borus02 model.

\item \textbf{NGC 4180}. We report the NuSTAR observational data for the first time. The model well replicates the X-ray spectra observed with the Chandra and NuSTAR ($\chi_{\mathrm{red}}^{2} = 0.80$). We obtain $N_{\mathrm{H}}^{\mathrm{LOS}} = 1.84_{-1.16}^{+1.91} \times 10^{24} \ \mathrm{cm}^{-2}$ and $\Gamma = 1.61_{-0.17}^{+0.05}$ with the XClumpy model.

\item \textbf{ESO 323--G032}. The model with one apec component well reproduces the X-ray spectra observed with the XMM--Newton and NuSTAR ($\chi_{\mathrm{red}}^{2} = 0.77$). We obtain $N_{\mathrm{H}}^{\mathrm{LOS}} = 1.27_{-0.45}^{+0.17} \times 10^{24} \ \mathrm{cm}^{-2}$ and $\Gamma = 1.89_{-0.27}^{+0.12}$ with the XClumpy model. $N_{\mathrm{H}}^{\mathrm{LOS}}$ and $\Gamma$ of this work agree with those of \cite{Torres21}. \cite{Torres21} used the XMM--Newton and NuSTAR observational data, and estimated $N_{\mathrm{H}}^{\mathrm{LOS}} = 1.75_{-0.49}^{+1.46} \times 10^{24} \ \mathrm{cm}^{-2}$ and $\Gamma = 2.02_{-0.30}^{+0.13}$ with the Borus02 model.

\item \textbf{NGC 4945}. The model with one zgauss component provides an adequate fit to the X-ray spectra observed with the Suzaku and NuSTAR ($\chi_{\mathrm{red}}^{2} = 1.21$). We fixed $i$ at 87\degr since the maser has been detected in this object. We obtain $N_{\mathrm{H}}^{\mathrm{LOS}} = 3.40_{-0.08}^{+0.11} \times 10^{24} \ \mathrm{cm}^{-2}$ and $\Gamma = 1.61_{-0.02}^{+0.03}$ with the XClumpy model. $N_{\mathrm{H}}^{\mathrm{LOS}}$ and $\Gamma$ of this work are smaller than those of \cite{Marchesi19a}. \cite{Marchesi19b} analyzed the XMM--Newton and NuSTAR observational data, and estimated $N_{\mathrm{H}}^{\mathrm{LOS}} = 3.97_{-0.15}^{+0.15} \times 10^{24} \ \mathrm{cm}^{-2}$ and $\Gamma = 1.95_{-0.05}^{+0.05}$ with the Borus02 model.

\item \textbf{IGR J14175--4641}. The model gives an adequate fit to the X-ray spectra observed with the XMM--Newton and NuSTAR ($\chi_{\mathrm{red}}^{2} = 1.22$). We obtain $N_{\mathrm{H}}^{\mathrm{LOS}} = 0.77_{-0.13}^{+0.08} \times 10^{24} \ \mathrm{cm}^{-2}$ and $\Gamma = 1.51_{-0.14}^{+0.11}$ with the XClumpy model. $N_{\mathrm{H}}^{\mathrm{LOS}}$ and $\Gamma$ of this work agree with those of \cite{Marchesi19b}. \cite{Marchesi19b} used the Swift/XRT and NuSTAR observational data, and obtained $N_{\mathrm{H}}^{\mathrm{LOS}} = 0.86_{-0.19}^{+0.25} \times 10^{24} \ \mathrm{cm}^{-2}$ and $\Gamma = 1.70_{-0.19}^{+0.20}$ with the Borus02 model.

\item \textbf{NGC 5643}. The model with two zgauss components provides an adequate fit to the X-ray spectra observed with the XMM--Newton and NuSTAR ($\chi_{\mathrm{red}}^{2} = 1.35$). We fixed $i$ at 87\degr because the maser has been detected in this object. We obtain $N_{\mathrm{H}}^{\mathrm{LOS}} = 0.63_{-0.10}^{+0.05} \times 10^{24} \ \mathrm{cm}^{-2}$ and $\Gamma = 1.25_{-0.07}^{+0.06}$ with the XClumpy model. $N_{\mathrm{H}}^{\mathrm{LOS}}$ and $\Gamma$ of this work are smaller than those of \cite{Marchesi19b}. \cite{Marchesi19b} analyzed the XMM--Newton and NuSTAR observational data, and estimated $N_{\mathrm{H}}^{\mathrm{LOS}} = 2.69_{-0.65}^{+1.88} \times 10^{24} \ \mathrm{cm}^{-2}$ and $\Gamma = 1.55_{-0.15}^{+0.13}$ with the Borus02 model.

\item \textbf{NGC 5728}. The model with one apec component well replicates the X-ray spectra observed with the XMM--Newton and NuSTAR ($\chi_{\mathrm{red}}^{2} = 0.99$). We fixed $i$ at 87\degr since the maser has been detected in this object. We obtain $N_{\mathrm{H}}^{\mathrm{LOS}} = 1.36_{-0.11}^{+0.08} \times 10^{24} \ \mathrm{cm}^{-2}$ and $\Gamma = 1.55_{-0.05}^{+0.07}$ with the XClumpy model. $N_{\mathrm{H}}^{\mathrm{LOS}}$ of this work is larger than that of \cite{Marchesi19b} and $\Gamma$ of this work is smaller than that of \cite{Marchesi19b}. \cite{Marchesi19b} used the Chandra and NuSTAR observational data, and obtained $N_{\mathrm{H}}^{\mathrm{LOS}} = 0.97_{-0.03}^{+0.05} \times 10^{24} \ \mathrm{cm}^{-2}$ and $\Gamma = 1.81_{-0.04}^{+0.07}$ with the Borus02 model.

\item \textbf{CGCG 164--019}. The model provides an adequate fit to the X-ray spectra observed with the Swift/XRT and NuSTAR ($\chi_{\mathrm{red}}^{2} = 1.29$). We obtain $N_{\mathrm{H}}^{\mathrm{LOS}} = 0.69_{-0.06}^{+0.55} \times 10^{24} \ \mathrm{cm}^{-2}$ and $\Gamma = 1.87_{-0.23}^{+0.23}$ with the XClumpy model. $N_{\mathrm{H}}^{\mathrm{LOS}}$ and $\Gamma$ of this work are consistent with those of \cite{Marchesi19b}. \cite{Marchesi19b} analyzed the Swift/XRT and NuSTAR observational data, and estimated $N_{\mathrm{H}}^{\mathrm{LOS}} = 1.37_{-0.45}^{+0.57} \times 10^{24} \ \mathrm{cm}^{-2}$ and $\Gamma = 1.79_{-0.32}^{+0.36}$ with the Borus02 model.

\item \textbf{ESO 137--G034}. The model with one apec component can reproduce the X-ray spectra observed with the XMM--Newton and NuSTAR ($\chi_{\mathrm{red}}^{2} = 1.13$). We obtain $N_{\mathrm{H}}^{\mathrm{LOS}} = 2.87_{-0.20}^{+0.20} \times 10^{24} \ \mathrm{cm}^{-2}$ and $\Gamma = 1.85_{-0.11}^{+0.09}$ with the XClumpy model. $N_{\mathrm{H}}^{\mathrm{LOS}}$ of this work agrees with that of \cite{Georgantopoulos19}. \cite{Georgantopoulos19} used the NuSTAR data, and estimated $N_{\mathrm{H}}^{\mathrm{LOS}} = 2.90_{-0.54}^{+0.86} \times 10^{24} \ \mathrm{cm}^{-2}$ with the MYTorus model.

\item \textbf{NGC 6232}. The model can replicate the X-ray spectra observed with the Swift and NuSTAR ($\chi_{\mathrm{red}}^{2} = 1.13$). We obtain $N_{\mathrm{H}}^{\mathrm{LOS}} = 1.10_{-0.59}^{+1.43} \times 10^{24} \ \mathrm{cm}^{-2}$ and $\Gamma = 1.31_{-0.14}^{+0.57}$ with the XClumpy model. $N_{\mathrm{H}}^{\mathrm{LOS}}$ and $\Gamma$ of this work are consistent with those of \cite{Marchesi19b}. \cite{Marchesi19b} analyzed the Swift/XRT and NuSTAR observational data, and estimated $N_{\mathrm{H}}^{\mathrm{LOS}} = 0.63_{-0.24}^{+0.13} \times 10^{24} \ \mathrm{cm}^{-2}$ and $\Gamma = 1.40$ (fixed) with the Borus02 model.

\item \textbf{ESO 138--G001}. We report the NuSTAR observational data for the first time. The model with one zgauss component provides an adequate fit to the X-ray spectra observed with the XMM--Newton and NuSTAR ($\chi_{\mathrm{red}}^{2} = 1.39$). We obtain $N_{\mathrm{H}}^{\mathrm{LOS}} = 0.45_{-0.03}^{+0.05} \times 10^{24} \ \mathrm{cm}^{-2}$ and $\Gamma = 1.45_{-0.03}^{+0.06}$ with the XClumpy model.

\item \textbf{NGC 6240}. The model with one zgauss component can reproduce the X-ray spectra observed with the XMM--Newton and NuSTAR ($\chi_{\mathrm{red}}^{2} = 1.07$). We fixed $i$ at \degr because the maser has been detected in this object. We obtain $N_{\mathrm{H}}^{\mathrm{LOS}} = 1.45_{-0.05}^{+0.15} \times 10^{24} \ \mathrm{cm}^{-2}$ and $\Gamma = 1.71_{-0.09}^{+0.06}$ with the XClumpy model. $\Gamma$ of this work agrees with that of \cite{Marchesi19b}, while $N_{\mathrm{H}}^{\mathrm{LOS}}$ of this work is larger than that of \cite{Marchesi19b}. \cite{Marchesi19b} used the XMM--Newton and NuSTAR observational data, and estimated $N_{\mathrm{H}}^{\mathrm{LOS}} = 1.10_{-0.08}^{+0.10} \times 10^{24} \ \mathrm{cm}^{-2}$ and $\Gamma = 1.75_{-0.04}^{+0.03}$ with the Borus02 model.

\item \textbf{NGC 6552}. The model with one apec component gives an adequate fit to the X-ray spectra observed with the XMM--Newton and NuSTAR  ($\chi_{\mathrm{red}}^{2} = 1.29$). We obtain $N_{\mathrm{H}}^{\mathrm{LOS}} = 0.96_{-0.05}^{+0.55} \times 10^{24} \ \mathrm{cm}^{-2}$ and $\Gamma = 2.25_{-0.10}^{+0.07}$ with the XClumpy model. $N_{\mathrm{H}}^{\mathrm{LOS}}$ of this work is smaller than that of \cite{Torres21} and $\Gamma$ of this work is larger than that of \cite{Torres21}. \cite{Torres21} analyzed the XMM--Newton and NuSTAR observational data, and estimated $N_{\mathrm{H}}^{\mathrm{LOS}} = 2.18_{-0.35}^{+0.38} \times 10^{24} \ \mathrm{cm}^{-2}$ and $\Gamma = 1.76_{-0.12}^{+0.08}$ with the Borus02 model.

\item \textbf{2MASX J20145928+2523010}. We report the NuSTAR data for the first time. The model can replicate the X-ray spectra observed with the XMM--Newton and NuSTAR ($\chi_{\mathrm{red}}^{2} = 1.18$). We obtain $N_{\mathrm{H}}^{\mathrm{LOS}} = 0.02_{-0.01}^{+0.01} \times 10^{24} \ \mathrm{cm}^{-2}$ and $\Gamma = 1.42_{-0.09}^{+0.03}$ with the XClumpy model.

\item \textbf{ESO 464--G016}. The model with one apec component well reproduces the X-ray spectra observed with the XMM--Newton and NuSTAR ($\chi_{\mathrm{red}}^{2} = 0.97$). We obtain $N_{\mathrm{H}}^{\mathrm{LOS}} = 0.82_{-0.10}^{+0.19} \times 10^{24} \ \mathrm{cm}^{-2}$ and $\Gamma = 1.54_{-0.15}^{+0.13}$. $N_{\mathrm{H}}^{\mathrm{LOS}}$ and $\Gamma$ of this work are consistent with those of \cite{Marchesi19b}. \cite{Marchesi19b} used the XMM--Newton and NuSTAR observational data, and estimated $N_{\mathrm{H}}^{\mathrm{LOS}} = 0.86_{-0.17}^{+0.25} \times 10^{24} \ \mathrm{cm}^{-2}$ and $\Gamma = 1.67_{-0.27}^{+0.29}$ with the Borus02 model.

\item \textbf{NGC 7130}. The model with one apec component can replicate the X-ray spectra observed with the Suzaku and NuSTAR ($\chi_{\mathrm{red}}^{2} = 1.03$). We obtain $N_{\mathrm{H}}^{\mathrm{LOS}} = 2.32_{-0.24}^{+1.95} \times 10^{24} \ \mathrm{cm}^{-2}$ and $\Gamma = 1.59_{-0.24}^{+0.09}$ with the XClumpy model. $N_{\mathrm{H}}^{\mathrm{LOS}}$ and $\Gamma$ of this work agree with those of \cite{Marchesi19b}. \cite{Marchesi19b} analyzed the Chandra and NuSTAR observational data, and obtained $N_{\mathrm{H}}^{\mathrm{LOS}} = 3.43_{-1.26}^{+6.57} \times 10^{24} \ \mathrm{cm}^{-2}$ and $\Gamma = 1.40$ (fixed) with the Borus02 model.

\item \textbf{NGC 7212}. The model provides an adequate fit to the X-ray spectra observed with the XMM--Newton and NuSTAR ($\chi_{\mathrm{red}}^{2} = 1.38$). We obtain $N_{\mathrm{H}}^{\mathrm{LOS}} = 1.27_{-0.69}^{+0.54} \times 10^{24} \ \mathrm{cm}^{-2}$ and $\Gamma = 2.12_{-0.29}^{+0.02}$. $N_{\mathrm{H}}^{\mathrm{LOS}}$ and $\Gamma$ of this work are consistent with those of \cite{Marchesi19b}. \cite{Marchesi19b} used the XMM--Newton and NuSTAR observational data, and estimated $N_{\mathrm{H}}^{\mathrm{LOS}} = 1.94_{-0.43}^{+0.49} \times 10^{24} \ \mathrm{cm}^{-2}$ and $\Gamma = 1.99_{-0.22}^{+0.30}$ with the Borus02 model.

\item \textbf{ESO 406--G004}. We report the NuSTAR data for the first time. The model gives an adequate fit to the X-ray spectra observed with the Swift/XRT and NuSTAR ($\chi_{\mathrm{red}}^{2} = 1.33$). We obtain $N_{\mathrm{H}}^{\mathrm{LOS}} = 0.52_{-0.14}^{+5.82} \times 10^{24} \ \mathrm{cm}^{-2}$ and $\Gamma = 1.10^{+0.28}$ with the XClumpy model.

\item \textbf{NGC 7479}. The model provides an adequate fit to the X-ray spectra observed with the XMM--Newton and NuSTAR ($\chi_{\mathrm{red}}^{2} = 1.21$). We fixed $i$ at 87\degr since the maser has been detected in this object. We obtain $N_{\mathrm{H}}^{\mathrm{LOS}} = 6.07_{-0.80}^{+3.61} \times 10^{24} \ \mathrm{cm}^{-2}$ and $\Gamma = 1.85_{-0.16}^{+0.12}$ with the XClumpy model. $\Gamma$ of this work agrees with that of \cite{Marchesi19a}, whereas $N_{\mathrm{H}}^{\mathrm{LOS}}$ is larger than that of \cite{Marchesi19a}. \cite{Marchesi19a} analyzed the Swift/XRT and NuSTAR observational data, and estimated $N_{\mathrm{H}}^{\mathrm{LOS}} = 1.32_{-0.28}^{+0.24} \times 10^{24} \ \mathrm{cm}^{-2}$ and $\Gamma = 1.64_{-0.19}^{+0.14}$ with the Borus02 model.

\item \textbf{SWIFT J2307.9+2245}. We report the XMM--Newton and NuSTAR simultaneous observational data for the first time. The model well reproduces the X-ray spectra observed with the XMM--Newton and NuSTAR ($\chi_{\mathrm{red}}^{2} = 0.98$). We obtain $N_{\mathrm{H}}^{\mathrm{LOS}} = 1.62_{-0.74}^{+1.26} \times 10^{24} \ \mathrm{cm}^{-2}$ and $\Gamma = 1.86_{-0.35}^{+0.03}$ with the XClumpy model.

\item \textbf{NGC 7582}. The model with one zgauss component provides an adequate fit to the X-ray spectra observed with the XMM--Newton and NuSTAR ($\chi_{\mathrm{red}}^{2} = 1.33$). We obtain $N_{\mathrm{H}}^{\mathrm{LOS}} = 0.32_{-0.02}^{+0.01} \times 10^{24} \ \mathrm{cm}^{-2}$ and $\Gamma = 1.44_{-0.01}^{+0.04}$ with the XClumpy model. $\Gamma$ of this work is smaller than that of \cite{Marchesi19b}. \cite{Marchesi19b} used the Swift/XRT and NuSTAR observational data, and estimated $N_{\mathrm{H}}^{\mathrm{LOS}} = 10.0 \times 10^{24} \ \mathrm{cm}^{-2}$ (fixed) and $\Gamma = 1.96_{-0.05}^{+0.03}$. This is because of the time variability of the object. In fact, \cite{Piconcelli07} analyzed the multi period XMM--Newton observational data and found that $N_{\mathrm{H}}^{\mathrm{LOS}}$ and $\Gamma$ are variable.

\item \textbf{NGC 7682}. We report the NuSTAR data for the first time. The model with one apec component well reproduces the X-ray spectra observed with the XMM--Newton and NuSTAR ($\chi_{\mathrm{red}}^{2} = 0.98$). We obtain $N_{\mathrm{H}}^{\mathrm{LOS}} = 0.25_{-0.02}^{+0.20} \times 10^{24} \ \mathrm{cm}^{-2}$ and $\Gamma = 1.27_{-0.07}^{+0.25}$ with the XClumpy model.
\end{enumerate}
\clearpage

\bigskip
\bibliographystyle{aasjournal}
\bibliography{tanimoto}

\begin{thebibliography}{}
\expandafter\ifx\csname natexlab\endcsname\relax\def\natexlab#1{#1}\fi
\providecommand{\url}[1]{\href{#1}{#1}}
\providecommand{\dodoi}[1]{doi:~\href{http://doi.org/#1}{\nolinkurl{#1}}}
\providecommand{\doeprint}[1]{\href{http://ascl.net/#1}{\nolinkurl{http://ascl.net/#1}}}
\providecommand{\doarXiv}[1]{\href{https://arxiv.org/abs/#1}{\nolinkurl{https://arxiv.org/abs/#1}}}

\bibitem[{{Ananna} {et~al.}(2020){Ananna}, {Treister}, {Urry}, {Ricci},
  {Hickox}, {Padmanabhan}, {Marchesi}, \& {Kirkpatrick}}]{Ananna20}
{Ananna}, T.~T., {Treister}, E., {Urry}, C.~M., {et~al.} 2020, \apj, 889, 17,
  \dodoi{10.3847/1538-4357/ab5aef}

\bibitem[{{Ananna} {et~al.}(2019){Ananna}, {Treister}, {Urry}, {Ricci},
  {Kirkpatrick}, {LaMassa}, {Buchner}, {Civano}, {Tremmel}, \&
  {Marchesi}}]{Ananna19}
---. 2019, \apj, 871, 240, \dodoi{10.3847/1538-4357/aafb77}

\bibitem[{{Ananna} {et~al.}(2022){Ananna}, {Weigel}, {Trakhtenbrot}, {Koss},
  {Urry}, {Ricci}, {Hickox}, {Treister}, {Bauer}, {Ueda}, {Mushotzky}, {Ricci},
  {Oh}, {Mejia-Restrepo}, {Den Brok}, {Stern}, {Powell}, {Caglar}, {Ichikawa},
  {Wong}, {Harrison}, \& {Schawinski}}]{Ananna22}
{Ananna}, T.~T., {Weigel}, A.~K., {Trakhtenbrot}, B., {et~al.} 2022, arXiv
  e-prints, arXiv:2201.05603.
\newblock \doarXiv{2201.05603}

\bibitem[{{Anders} \& {Grevesse}(1989)}]{Anders89}
{Anders}, E., \& {Grevesse}, N. 1989, Geochimica et Cosmochimica Acta, 53, 197,
  \dodoi{10.1016/0016-7037(89)90286-X}

\bibitem[{{Ar{\'e}valo} {et~al.}(2014){Ar{\'e}valo}, {Bauer}, {Puccetti},
  {Walton}, {Koss}, {Boggs}, {Brandt}, {Brightman}, {Christensen}, {Comastri},
  {Craig}, {Fuerst}, {Gandhi}, {Grefenstette}, {Hailey}, {Harrison}, {Luo},
  {Madejski}, {Madsen}, {Marinucci}, {Matt}, {Saez}, {Stern}, {Stuhlinger},
  {Treister}, {Urry}, \& {Zhang}}]{Arevalo14}
{Ar{\'e}valo}, P., {Bauer}, F.~E., {Puccetti}, S., {et~al.} 2014, \apj, 791,
  81, \dodoi{10.1088/0004-637X/791/2/81}

\bibitem[{{Arnaud}(1996)}]{Arnaud96}
{Arnaud}, K.~A. 1996, in Astronomical Data Analysis Software and Systems V,
  Vol. 101, 17

\bibitem[{{Balokovi{\'c}} {et~al.}(2014){Balokovi{\'c}}, {Comastri},
  {Harrison}, {Alexand er}, {Ballantyne}, {Bauer}, {Boggs}, {Brandt},
  {Brightman}, \& {Christensen}}]{Balokovic14}
{Balokovi{\'c}}, M., {Comastri}, A., {Harrison}, F.~A., {et~al.} 2014, \apj,
  794, 111, \dodoi{10.1088/0004-637X/794/2/111}

\bibitem[{{Balokovi{\'c}} {et~al.}(2018){Balokovi{\'c}}, {Brightman},
  {Harrison}, {Comastri}, {Ricci}, {Buchner}, {Gandhi}, {Farrah}, \&
  {Stern}}]{Balokovic18}
{Balokovi{\'c}}, M., {Brightman}, M., {Harrison}, F.~A., {et~al.} 2018, \apj,
  854, 42, \dodoi{10.3847/1538-4357/aaa7eb}

\bibitem[{{Barthelmy} {et~al.}(2005){Barthelmy}, {Barbier}, {Cummings},
  {Fenimore}, {Gehrels}, {Hullinger}, {Krimm}, {Markwardt}, {Palmer},
  {Parsons}, {Sato}, {Suzuki}, {Takahashi}, {Tashiro}, \&
  {Tueller}}]{Barthelmy05}
{Barthelmy}, S.~D., {Barbier}, L.~M., {Cummings}, J.~R., {et~al.} 2005, \ssr,
  120, 143, \dodoi{10.1007/s11214-005-5096-3}

\bibitem[{{Bauer} {et~al.}(2015){Bauer}, {Ar{\'e}valo}, {Walton}, {Koss},
  {Puccetti}, {Gandhi}, {Stern}, {Alexander}, {Balokovi{\'c}}, \&
  {Boggs}}]{Bauer15}
{Bauer}, F.~E., {Ar{\'e}valo}, P., {Walton}, D.~J., {et~al.} 2015, \apj, 812,
  116, \dodoi{10.1088/0004-637X/812/2/116}

\bibitem[{{Baumgartner} {et~al.}(2013){Baumgartner}, {Tueller}, {Markwardt},
  {Skinner}, {Barthelmy}, {Mushotzky}, {Evans}, \& {Gehrels}}]{Baumgartner13}
{Baumgartner}, W.~H., {Tueller}, J., {Markwardt}, C.~B., {et~al.} 2013, The
  Astrophysical Journal Supplement Series, 207, 19,
  \dodoi{10.1088/0067-0049/207/2/19}

\bibitem[{{Bian} \& {Gu}(2007)}]{Bian07}
{Bian}, W., \& {Gu}, Q. 2007, \apj, 657, 159, \dodoi{10.1086/510708}

\bibitem[{{Boorman} {et~al.}(2018){Boorman}, {Gandhi}, {Balokovi{\'c}},
  {Brightman}, {Harrison}, {Ricci}, \& {Stern}}]{Boorman18}
{Boorman}, P.~G., {Gandhi}, P., {Balokovi{\'c}}, M., {et~al.} 2018, \mnras,
  477, 3775, \dodoi{10.1093/mnras/sty861}

\bibitem[{{Boorman} {et~al.}(2016){Boorman}, {Gandhi}, {Alexander}, {Annuar},
  {Ballantyne}, {Bauer}, {Boggs}, {Brandt}, {Brightman}, {Christensen},
  {Craig}, {Farrah}, {Hailey}, {Harrison}, {H{\"o}nig}, {Koss}, {LaMassa},
  {Masini}, {Ricci}, {Risaliti}, {Stern}, \& {Zhang}}]{Boorman16}
{Boorman}, P.~G., {Gandhi}, P., {Alexander}, D.~M., {et~al.} 2016, \apj, 833,
  245, \dodoi{10.3847/1538-4357/833/2/245}

\bibitem[{{Brightman} \& {Nandra}(2011)}]{Brightman11}
{Brightman}, M., \& {Nandra}, K. 2011, \mnras, 413, 1206,
  \dodoi{10.1111/j.1365-2966.2011.18207.x}

\bibitem[{{Buchner} {et~al.}(2019){Buchner}, {Brightman}, {Nandra}, {Nikutta},
  \& {Bauer}}]{Buchner19}
{Buchner}, J., {Brightman}, M., {Nandra}, K., {Nikutta}, R., \& {Bauer}, F.~E.
  2019, \aap, 629, A16, \dodoi{10.1051/0004-6361/201834771}

\bibitem[{{Burrows} {et~al.}(2005){Burrows}, {Hill}, {Nousek}, {Kennea},
  {Wells}, {Osborne}, {Abbey}, {Beardmore}, {Mukerjee}, {Short}, {Chincarini},
  {Campana}, {Citterio}, {Moretti}, {Pagani}, {Tagliaferri}, {Giommi},
  {Capalbi}, {Tamburelli}, {Angelini}, {Cusumano}, {Br{\"a}uninger}, {Burkert},
  \& {Hartner}}]{Burrows05}
{Burrows}, D.~N., {Hill}, J.~E., {Nousek}, J.~A., {et~al.} 2005, \ssr, 120,
  165, \dodoi{10.1007/s11214-005-5097-2}

\bibitem[{{Eguchi} {et~al.}(2009){Eguchi}, {Ueda}, {Terashima}, {Mushotzky}, \&
  {Tueller}}]{Eguchi09}
{Eguchi}, S., {Ueda}, Y., {Terashima}, Y., {Mushotzky}, R., \& {Tueller}, J.
  2009, \apj, 696, 1657, \dodoi{10.1088/0004-637X/696/2/1657}

\bibitem[{{Fabbiano} {et~al.}(2017){Fabbiano}, {Elvis}, {Paggi}, {Karovska},
  {Maksym}, {Raymond}, {Risaliti}, \& {Wang}}]{Fabbiano17}
{Fabbiano}, G., {Elvis}, M., {Paggi}, A., {et~al.} 2017, \apjl, 842, L4,
  \dodoi{10.3847/2041-8213/aa7551}

\bibitem[{{Fabian} {et~al.}(2009){Fabian}, {Vasudevan}, {Mushotzky}, {Winter},
  \& {Reynolds}}]{Fabian09}
{Fabian}, A.~C., {Vasudevan}, R.~V., {Mushotzky}, R.~F., {Winter}, L.~M., \&
  {Reynolds}, C.~S. 2009, \mnras, 394, L89,
  \dodoi{10.1111/j.1745-3933.2009.00617.x}

\bibitem[{{Furui} {et~al.}(2016){Furui}, {Fukazawa}, {Odaka}, {Kawaguchi},
  {Ohno}, \& {Hayashi}}]{Furui16}
{Furui}, S., {Fukazawa}, Y., {Odaka}, H., {et~al.} 2016, \apj, 818, 164,
  \dodoi{10.3847/0004-637X/818/2/164}

\bibitem[{{Garmire} {et~al.}(2003){Garmire}, {Bautz}, {Ford}, {Nousek}, \&
  {Ricker}}]{Garmire03}
{Garmire}, G.~P., {Bautz}, M.~W., {Ford}, P.~G., {Nousek}, J.~A., \& {Ricker},
  George~R., J. 2003, in Society of Photo-Optical Instrumentation Engineers
  (SPIE) Conference Series, Vol. 4851, X-Ray and Gamma-Ray Telescopes and
  Instruments for Astronomy., ed. J.~E. {Truemper} \& H.~D. {Tananbaum},
  28--44, \dodoi{10.1117/12.461599}

\bibitem[{{Gehrels} {et~al.}(2004){Gehrels}, {Chincarini}, {Giommi}, {Mason},
  {Nousek}, {Wells}, {White}, {Barthelmy}, {Burrows}, {Cominsky}, {Hurley},
  {Marshall}, {M{\'e}sz{\'a}ros}, {Roming}, {Angelini}, {Barbier}, {Belloni},
  {Campana}, {Caraveo}, {Chester}, {Citterio}, {Cline}, {Cropper}, {Cummings},
  {Dean}, {Feigelson}, {Fenimore}, {Frail}, {Fruchter}, {Garmire}, {Gendreau},
  {Ghisellini}, {Greiner}, {Hill}, {Hunsberger}, {Krimm}, {Kulkarni}, {Kumar},
  {Lebrun}, {Lloyd- Ronning}, {Markwardt}, {Mattson}, {Mushotzky}, {Norris},
  {Osborne}, {Paczynski}, {Palmer}, {Park}, {Parsons}, {Paul}, {Rees},
  {Reynolds}, {Rhoads}, {Sasseen}, {Schaefer}, {Short}, {Smale}, {Smith},
  {Stella}, {Tagliaferri}, {Takahashi}, {Tashiro}, {Townsley}, {Tueller},
  {Turner}, {Vietri}, {Voges}, {Ward}, {Willingale}, {Zerbi}, \&
  {Zhang}}]{Gehrels04}
{Gehrels}, N., {Chincarini}, G., {Giommi}, P., {et~al.} 2004, \apj, 611, 1005,
  \dodoi{10.1086/422091}

\bibitem[{{Georgantopoulos} \& {Akylas}(2019)}]{Georgantopoulos19}
{Georgantopoulos}, I., \& {Akylas}, A. 2019, \aap, 621, A28,
  \dodoi{10.1051/0004-6361/201833038}

\bibitem[{{Gilli} {et~al.}(2007){Gilli}, {Comastri}, \& {Hasinger}}]{Gilli07}
{Gilli}, R., {Comastri}, A., \& {Hasinger}, G. 2007, \aap, 463, 79,
  \dodoi{10.1051/0004-6361:20066334}

\bibitem[{{Guainazzi} {et~al.}(2005){Guainazzi}, {Fabian}, {Iwasawa}, {Matt},
  \& {Fiore}}]{Guainazzi05}
{Guainazzi}, M., {Fabian}, A.~C., {Iwasawa}, K., {Matt}, G., \& {Fiore}, F.
  2005, \mnras, 356, 295, \dodoi{10.1111/j.1365-2966.2004.08448.x}

\bibitem[{{Guainazzi} {et~al.}(2004){Guainazzi}, {Rodriguez-Pascual}, {Fabian},
  {Iwasawa}, \& {Matt}}]{Guainazzi04}
{Guainazzi}, M., {Rodriguez-Pascual}, P., {Fabian}, A.~C., {Iwasawa}, K., \&
  {Matt}, G. 2004, \mnras, 355, 297, \dodoi{10.1111/j.1365-2966.2004.08317.x}

\bibitem[{{Guainazzi} {et~al.}(2016){Guainazzi}, {Risaliti}, {Awaki},
  {Arevalo}, {Bauer}, {Bianchi}, {Boggs}, {Brand t}, {Brightman},
  {Christensen}, {Craig}, {Forster}, {Hailey}, {Harrison}, {Koss},
  {Longinotti}, {Markwardt}, {Marinucci}, {Matt}, {Reynolds}, {Ricci}, {Stern},
  {Svoboda}, {Walton}, \& {Zhang}}]{Guainazzi16}
{Guainazzi}, M., {Risaliti}, G., {Awaki}, H., {et~al.} 2016, \mnras, 460, 1954,
  \dodoi{10.1093/mnras/stw1033}

\bibitem[{{Harrison} {et~al.}(2013){Harrison}, {Craig}, {Christensen},
  {Hailey}, {Zhang}, {Boggs}, {Stern}, {Cook}, {Forster}, {Giommi},
  {Grefenstette}, {Kim}, {Kitaguchi}, {Koglin}, {Madsen}, {Mao}, {Miyasaka},
  {Mori}, {Perri}, {Pivovaroff}, {Puccetti}, {Rana}, {Westergaard}, {Willis},
  {Zoglauer}, {An}, {Bachetti}, {Barri{\`e}re}, {Bellm}, {Bhalerao},
  {Brejnholt}, {Fuerst}, {Liebe}, {Markwardt}, {Nynka}, {Vogel}, {Walton},
  {Wik}, {Alexander}, {Cominsky}, {Hornschemeier}, {Hornstrup}, {Kaspi},
  {Madejski}, {Matt}, {Molendi}, {Smith}, {Tomsick}, {Ajello}, {Ballantyne},
  {Balokovi{\'c}}, {Barret}, {Bauer}, {Blandford}, {Brandt}, {Brenneman},
  {Chiang}, {Chakrabarty}, {Chenevez}, {Comastri}, {Dufour}, {Elvis}, {Fabian},
  {Farrah}, {Fryer}, {Gotthelf}, {Grindlay}, {Helfand}, {Krivonos}, {Meier},
  {Miller}, {Natalucci}, {Ogle}, {Ofek}, {Ptak}, {Reynolds}, {Rigby},
  {Tagliaferri}, {Thorsett}, {Treister}, \& {Urry}}]{Harrison13}
{Harrison}, F.~A., {Craig}, W.~W., {Christensen}, F.~E., {et~al.} 2013, \apj,
  770, 103, \dodoi{10.1088/0004-637X/770/2/103}

\bibitem[{{Harrison} {et~al.}(2016){Harrison}, {Aird}, {Civano}, {Lansbury},
  {Mullaney}, {Ballantyne}, {Alexander}, {Stern}, {Ajello}, {Barret}, {Bauer},
  {Balokovi{\'c}}, {Brandt}, {Brightman}, {Boggs}, {Christensen}, {Comastri},
  {Craig}, {Del Moro}, {Forster}, {Gandhi}, {Giommi}, {Grefenstette}, {Hailey},
  {Hickox}, {Hornstrup}, {Kitaguchi}, {Koglin}, {Luo}, {Madsen}, {Mao},
  {Miyasaka}, {Mori}, {Perri}, {Pivovaroff}, {Puccetti}, {Rana}, {Treister},
  {Walton}, {Westergaard}, {Wik}, {Zappacosta}, {Zhang}, \&
  {Zoglauer}}]{Harrison16}
{Harrison}, F.~A., {Aird}, J., {Civano}, F., {et~al.} 2016, \apj, 831, 185,
  \dodoi{10.3847/0004-637X/831/2/185}

\bibitem[{{H{\"o}nig} \& {Beckert}(2007)}]{Honig07}
{H{\"o}nig}, S.~F., \& {Beckert}, T. 2007, \mnras, 380, 1172,
  \dodoi{10.1111/j.1365-2966.2007.12157.x}

\bibitem[{{H{\"o}nig} {et~al.}(2006){H{\"o}nig}, {Beckert}, {Ohnaka}, \&
  {Weigelt}}]{Honig06}
{H{\"o}nig}, S.~F., {Beckert}, T., {Ohnaka}, K., \& {Weigelt}, G. 2006, \aap,
  452, 459, \dodoi{10.1051/0004-6361:20054622}

\bibitem[{{Hopkins} {et~al.}(2006){Hopkins}, {Hernquist}, {Cox}, {Di Matteo},
  {Robertson}, \& {Springel}}]{Hopkins06}
{Hopkins}, P.~F., {Hernquist}, L., {Cox}, T.~J., {et~al.} 2006, The
  Astrophysical Journal Supplement Series, 163, 1, \dodoi{10.1086/499298}

\bibitem[{{Ikeda} {et~al.}(2009){Ikeda}, {Awaki}, \& {Terashima}}]{Ikeda09}
{Ikeda}, S., {Awaki}, H., \& {Terashima}, Y. 2009, \apj, 692, 608,
  \dodoi{10.1088/0004-637X/692/1/608}

\bibitem[{{Ishisaki} {et~al.}(2007){Ishisaki}, {Maeda}, {Fujimoto}, {Ozaki},
  {Ebisawa}, {Takahashi}, {Ueda}, {Ogasaka}, {Ptak}, {Mukai}, {Hamaguchi},
  {Hirayama}, {Kotani}, {Kubo}, {Shibata}, {Ebara}, {Furuzawa}, {Iizuka},
  {Inoue}, {Mori}, {Okada}, {Yokoyama}, {Matsumoto}, {Nakajima}, {Yamaguchi},
  {Anabuki}, {Tawa}, {Nagai}, {Katsuda}, {Hayashida}, {Bamba}, {Miller},
  {Sato}, \& {Yamasaki}}]{Ishisaki07}
{Ishisaki}, Y., {Maeda}, Y., {Fujimoto}, R., {et~al.} 2007, Publications of the
  Astronomical Society of Japan, 59, 113, \dodoi{10.1093/pasj/59.sp1.S113}

\bibitem[{{Jansen} {et~al.}(2001){Jansen}, {Lumb}, {Altieri}, {Clavel}, {Ehle},
  {Erd}, {Gabriel}, {Guainazzi}, {Gondoin}, {Much}, {Munoz}, {Santos},
  {Schartel}, {Texier}, \& {Vacanti}}]{Jansen01}
{Jansen}, F., {Lumb}, D., {Altieri}, B., {et~al.} 2001, \aap, 365, L1,
  \dodoi{10.1051/0004-6361:20000036}

\bibitem[{{Kawakatu} {et~al.}(2020){Kawakatu}, {Wada}, \&
  {Ichikawa}}]{Kawakatu20}
{Kawakatu}, N., {Wada}, K., \& {Ichikawa}, K. 2020, \apj, 889, 84,
  \dodoi{10.3847/1538-4357/ab5f60}

\bibitem[{{Kawamuro} {et~al.}(2019){Kawamuro}, {Izumi}, \&
  {Imanishi}}]{Kawamuro19}
{Kawamuro}, T., {Izumi}, T., \& {Imanishi}, M. 2019, \pasj, 71, 68,
  \dodoi{10.1093/pasj/psz045}

\bibitem[{{Kawamuro} {et~al.}(2020){Kawamuro}, {Izumi}, {Onishi}, {Imanishi},
  {Nguyen}, \& {Baba}}]{Kawamuro20}
{Kawamuro}, T., {Izumi}, T., {Onishi}, K., {et~al.} 2020, \apj, 895, 135,
  \dodoi{10.3847/1538-4357/ab8b62}

\bibitem[{{Kawamuro} {et~al.}(2016){Kawamuro}, {Ueda}, {Tazaki}, {Terashima},
  \& {Mushotzky}}]{Kawamuro16b}
{Kawamuro}, T., {Ueda}, Y., {Tazaki}, F., {Terashima}, Y., \& {Mushotzky}, R.
  2016, \apj, 831, 37, \dodoi{10.3847/0004-637X/831/1/37}

\bibitem[{{Konami} {et~al.}(2012){Konami}, {Matsushita}, {Gandhi}, \&
  {Tamagawa}}]{Konami12}
{Konami}, S., {Matsushita}, K., {Gandhi}, P., \& {Tamagawa}, T. 2012, \pasj,
  64, 117, \dodoi{10.1093/pasj/64.5.117}

\bibitem[{{Kormendy} \& {Ho}(2013)}]{Kormendy13}
{Kormendy}, J., \& {Ho}, L.~C. 2013, Annual Review of Astronomy and
  Astrophysics, 51, 511, \dodoi{10.1146/annurev-astro-082708-101811}

\bibitem[{{Koss} {et~al.}(2017){Koss}, {Trakhtenbrot}, {Ricci}, {Lamperti},
  {Oh}, {Berney}, {Schawinski}, {Balokovi{\'c}}, {Baronchelli}, {Crenshaw},
  {Fischer}, {Gehrels}, {Harrison}, {Hashimoto}, {Hogg}, {Ichikawa}, {Masetti},
  {Mushotzky}, {Sartori}, {Stern}, {Treister}, {Ueda}, {Veilleux}, \&
  {Winter}}]{Koss17}
{Koss}, M., {Trakhtenbrot}, B., {Ricci}, C., {et~al.} 2017, \apj, 850, 74,
  \dodoi{10.3847/1538-4357/aa8ec9}

\bibitem[{{Koss} {et~al.}(2016){Koss}, {Assef}, {Balokovi{\'c}}, {Stern},
  {Gandhi}, {Lamperti}, {Alexander}, {Ballantyne}, {Bauer}, {Berney}, {Brandt},
  {Comastri}, {Gehrels}, {Harrison}, {Lansbury}, {Markwardt}, {Ricci},
  {Rivers}, {Schawinski}, {Trakhtenbrot}, {Treister}, \& {Urry}}]{Koss16}
{Koss}, M.~J., {Assef}, R., {Balokovi{\'c}}, M., {et~al.} 2016, \apj, 825, 85,
  \dodoi{10.3847/0004-637X/825/2/85}

\bibitem[{{Koyama} {et~al.}(2007){Koyama}, {Tsunemi}, {Dotani}, {Bautz},
  {Hayashida}, {Tsuru}, {Matsumoto}, {Ogawara}, {Ricker}, {Doty}, {Kissel},
  {Foster}, {Nakajima}, {Yamaguchi}, {Mori}, {Sakano}, {Hamaguchi},
  {Nishiuchi}, {Miyata}, {Torii}, {Namiki}, {Katsuda}, {Matsuura}, {Miyauchi},
  {Anabuki}, {Tawa}, {Ozaki}, {Murakami}, {Maeda}, {Ichikawa}, {Prigozhin},
  {Boughan}, {Lamarr}, {Miller}, {Burke}, {Gregory}, {Pillsbury}, {Bamba},
  {Hiraga}, {Senda}, {Katayama}, {Kitamoto}, {Tsujimoto}, {Kohmura}, {Tsuboi},
  \& {Awaki}}]{Koyama07}
{Koyama}, K., {Tsunemi}, H., {Dotani}, T., {et~al.} 2007, Publications of the
  Astronomical Society of Japan, 59, 23, \dodoi{10.1093/pasj/59.sp1.S23}

\bibitem[{{Krolik} \& {Begelman}(1988)}]{Krolik88}
{Krolik}, J.~H., \& {Begelman}, M.~C. 1988, \apj, 329, 702,
  \dodoi{10.1086/166414}

\bibitem[{{LaMassa} {et~al.}(2019){LaMassa}, {Yaqoob}, {Boorman}, {Tzanavaris},
  {Levenson}, {Gandhi}, {Ptak}, \& {Heckman}}]{LaMassa19}
{LaMassa}, S.~M., {Yaqoob}, T., {Boorman}, P.~G., {et~al.} 2019, \apj, 887,
  173, \dodoi{10.3847/1538-4357/ab552c}

\bibitem[{{Laor} \& {Draine}(1993)}]{Laor93}
{Laor}, A., \& {Draine}, B.~T. 1993, \apj, 402, 441, \dodoi{10.1086/172149}

\bibitem[{{Liu} \& {Li}(2014)}]{Liu14}
{Liu}, Y., \& {Li}, X. 2014, \apj, 787, 52, \dodoi{10.1088/0004-637X/787/1/52}

\bibitem[{{Liu} \& {Li}(2015)}]{Liu15}
---. 2015, \mnras, 448, L53, \dodoi{10.1093/mnrasl/slu198}

\bibitem[{{Madsen} {et~al.}(2017){Madsen}, {Beardmore}, {Forster}, {Guainazzi},
  {Marshall}, {Miller}, {Page}, \& {Stuhlinger}}]{Madsen17}
{Madsen}, K.~K., {Beardmore}, A.~P., {Forster}, K., {et~al.} 2017, \aj, 153, 2,
  \dodoi{10.3847/1538-3881/153/1/2}

\bibitem[{{Marchesi} {et~al.}(2019{\natexlab{a}}){Marchesi}, {Ajello}, {Zhao},
  {Comastri}, {La Parola}, \& {Segreto}}]{Marchesi19a}
{Marchesi}, S., {Ajello}, M., {Zhao}, X., {et~al.} 2019{\natexlab{a}}, \apj,
  882, 162, \dodoi{10.3847/1538-4357/ab340a}

\bibitem[{{Marchesi} {et~al.}(2019{\natexlab{b}}){Marchesi}, {Ajello}, {Zhao},
  {Marcotulli}, {Balokovi{\'c}}, {Brightman}, {Comastri}, {Cusumano},
  {Lanzuisi}, {La Parola}, {Segreto}, \& {Vignali}}]{Marchesi19b}
---. 2019{\natexlab{b}}, \apj, 872, 8, \dodoi{10.3847/1538-4357/aafbeb}

\bibitem[{{Markwardt} {et~al.}(2005){Markwardt}, {Tueller}, {Skinner},
  {Gehrels}, {Barthelmy}, \& {Mushotzky}}]{Markwardt05}
{Markwardt}, C.~B., {Tueller}, J., {Skinner}, G.~K., {et~al.} 2005, \apjl, 633,
  L77, \dodoi{10.1086/498569}

\bibitem[{{Merloni} {et~al.}(2003){Merloni}, {Heinz}, \& {di
  Matteo}}]{Merloni03}
{Merloni}, A., {Heinz}, S., \& {di Matteo}, T. 2003, \mnras, 345, 1057,
  \dodoi{10.1046/j.1365-2966.2003.07017.x}

\bibitem[{{Mitsuda} {et~al.}(2007){Mitsuda}, {Bautz}, {Inoue}, {Kelley},
  {Koyama}, {Kunieda}, {Makishima}, {Ogawara}, {Petre}, {Takahashi}, {Tsunemi},
  {White}, {Anabuki}, {Angelini}, {Arnaud}, {Awaki}, {Bamba}, {Boyce}, {Brown},
  {Chan}, {Cottam}, {Dotani}, {Doty}, {Ebisawa}, {Ezoe}, {Fabian}, {Figueroa},
  {Fujimoto}, {Fukazawa}, {Furusho}, {Furuzawa}, {Gendreau}, {Griffiths},
  {Haba}, {Hamaguchi}, {Harrus}, {Hasinger}, {Hatsukade}, {Hayashida}, {Henry},
  {Hiraga}, {Holt}, {Hornschemeier}, {Hughes}, {Hwang}, {Ishida}, {Ishisaki},
  {Isobe}, {Itoh}, {Iyomoto}, {Kahn}, {Kamae}, {Katagiri}, {Kataoka},
  {Katayama}, {Kawai}, {Kilbourne}, {Kinugasa}, {Kissel}, {Kitamoto}, {Kohama},
  {Kohmura}, {Kokubun}, {Kotani}, {Kotoku}, {Kubota}, {Madejski}, {Maeda},
  {Makino}, {Markowitz}, {Matsumoto}, {Matsumoto}, {Matsuoka}, {Matsushita},
  {McCammon}, {Mihara}, {Misaki}, {Miyata}, {Mizuno}, {Mori}, {Mori}, {Morii},
  {Moseley}, {Mukai}, {Murakami}, {Murakami}, {Mushotzky}, {Nagase}, {Namiki},
  {Negoro}, {Nakazawa}, {Nousek}, {Okajima}, {Ogasaka}, {Ohashi}, {Oshima},
  {Ota}, {Ozaki}, {Ozawa}, {Parmar}, {Pence}, {Porter}, {Reeves}, {Ricker},
  {Sakurai}, {Sanders}, {Senda}, {Serlemitsos}, {Shibata}, {Soong}, {Smith},
  {Suzuki}, {Szymkowiak}, {Takahashi}, {Tamagawa}, {Tamura}, {Tamura},
  {Tanaka}, {Tashiro}, {Tawara}, {Terada}, {Terashima}, {Tomida}, {Torii},
  {Tsuboi}, {Tsujimoto}, {Tsuru}, {Turner}, {Ueda}, {Ueno}, {Ueno}, {Uno},
  {Urata}, {Watanabe}, {Yamamoto}, {Yamaoka}, {Yamasaki}, {Yamashita},
  {Yamauchi}, {Yamauchi}, {Yaqoob}, {Yonetoku}, \& {Yoshida}}]{Mitsuda07}
{Mitsuda}, K., {Bautz}, M., {Inoue}, H., {et~al.} 2007, Publications of the
  Astronomical Society of Japan, 59, S1, \dodoi{10.1093/pasj/59.sp1.S1}

\bibitem[{{Miyaji} {et~al.}(2019){Miyaji}, {Herrera-Endoqui}, {Krumpe},
  {Hanzawa}, {Shogaki}, {Matsuura}, {Tanimoto}, {Ueda}, {Ishigaki}, {Barrufet},
  {Brunner}, {Matsuhara}, {Goto}, {Takagi}, {Pearson}, {Burgarella}, {Oi},
  {Malkan}, {Toba}, {White}, \& {Hanami}}]{Miyaji19}
{Miyaji}, T., {Herrera-Endoqui}, M., {Krumpe}, M., {et~al.} 2019, \apjl, 884,
  L10, \dodoi{10.3847/2041-8213/ab46bc}

\bibitem[{{Murphy} \& {Yaqoob}(2009)}]{Murphy09}
{Murphy}, K.~D., \& {Yaqoob}, T. 2009, \mnras, 397, 1549,
  \dodoi{10.1111/j.1365-2966.2009.15025.x}

\bibitem[{{Nenkova} {et~al.}(2008{\natexlab{a}}){Nenkova}, {Sirocky},
  {Ivezi{\'c}}, \& {Elitzur}}]{Nenkova08a}
{Nenkova}, M., {Sirocky}, M.~M., {Ivezi{\'c}}, {\v{Z}}., \& {Elitzur}, M.
  2008{\natexlab{a}}, \apj, 685, 147, \dodoi{10.1086/590482}

\bibitem[{{Nenkova} {et~al.}(2008{\natexlab{b}}){Nenkova}, {Sirocky},
  {Nikutta}, {Ivezi{\'c}}, \& {Elitzur}}]{Nenkova08b}
{Nenkova}, M., {Sirocky}, M.~M., {Nikutta}, R., {Ivezi{\'c}}, {\v{Z}}., \&
  {Elitzur}, M. 2008{\natexlab{b}}, \apj, 685, 160, \dodoi{10.1086/590483}

\bibitem[{{Netzer} {et~al.}(2005){Netzer}, {Lemze}, {Kaspi}, {George},
  {Turner}, {Lutz}, {Boller}, \& {Chelouche}}]{Netzer05}
{Netzer}, H., {Lemze}, D., {Kaspi}, S., {et~al.} 2005, \apj, 629, 739,
  \dodoi{10.1086/431474}

\bibitem[{{Oda} {et~al.}(2017){Oda}, {Tanimoto}, {Ueda}, {Imanishi},
  {Terashima}, \& {Ricci}}]{Oda17}
{Oda}, S., {Tanimoto}, A., {Ueda}, Y., {et~al.} 2017, \apj, 835, 179,
  \dodoi{10.3847/1538-4357/835/2/179}

\bibitem[{{Odaka} {et~al.}(2011){Odaka}, {Aharonian}, {Watanabe}, {Tanaka},
  {Khangulyan}, \& {Takahashi}}]{Odaka11}
{Odaka}, H., {Aharonian}, F., {Watanabe}, S., {et~al.} 2011, \apj, 740, 103,
  \dodoi{10.1088/0004-637X/740/2/103}

\bibitem[{{Odaka} {et~al.}(2016){Odaka}, {Yoneda}, {Takahashi}, \&
  {Fabian}}]{Odaka16}
{Odaka}, H., {Yoneda}, H., {Takahashi}, T., \& {Fabian}, A. 2016, \mnras, 462,
  2366, \dodoi{10.1093/mnras/stw1764}

\bibitem[{{Ogawa} {et~al.}(2021){Ogawa}, {Ueda}, {Tanimoto}, \&
  {Yamada}}]{Ogawa21}
{Ogawa}, S., {Ueda}, Y., {Tanimoto}, A., \& {Yamada}, S. 2021, \apj, 906, 84,
  \dodoi{10.3847/1538-4357/abccce}

\bibitem[{{Ogawa} {et~al.}(2019){Ogawa}, {Ueda}, {Yamada}, {Tanimoto}, \&
  {Kawaguchi}}]{Ogawa19}
{Ogawa}, S., {Ueda}, Y., {Yamada}, S., {Tanimoto}, A., \& {Kawaguchi}, T. 2019,
  \apj, 875, 115, \dodoi{10.3847/1538-4357/ab0e08}

\bibitem[{{Oh} {et~al.}(2018){Oh}, {Koss}, {Markwardt}, {Schawinski},
  {Baumgartner}, {Barthelmy}, {Cenko}, {Gehrels}, {Mushotzky}, {Petulante},
  {Ricci}, {Lien}, \& {Trakhtenbrot}}]{Oh18}
{Oh}, K., {Koss}, M., {Markwardt}, C.~B., {et~al.} 2018, \apjs, 235, 4,
  \dodoi{10.3847/1538-4365/aaa7fd}

\bibitem[{{Panessa} {et~al.}(2020){Panessa}, {Castangia}, {Malizia}, {Bassani},
  {Tarchi}, {Bazzano}, \& {Ubertini}}]{Panessa20}
{Panessa}, F., {Castangia}, P., {Malizia}, A., {et~al.} 2020, \aap, 641, A162,
  \dodoi{10.1051/0004-6361/201937407}

\bibitem[{{Piconcelli} {et~al.}(2007){Piconcelli}, {Bianchi}, {Guainazzi},
  {Fiore}, \& {Chiaberge}}]{Piconcelli07}
{Piconcelli}, E., {Bianchi}, S., {Guainazzi}, M., {Fiore}, F., \& {Chiaberge},
  M. 2007, \aap, 466, 855, \dodoi{10.1051/0004-6361:20066439}

\bibitem[{{Piconcelli} {et~al.}(2011){Piconcelli}, {Bianchi}, {Vignali},
  {Jim{\'e}nez-Bail{\'o}n}, \& {Fiore}}]{Piconcelli11}
{Piconcelli}, E., {Bianchi}, S., {Vignali}, C., {Jim{\'e}nez-Bail{\'o}n}, E.,
  \& {Fiore}, F. 2011, \aap, 534, A126, \dodoi{10.1051/0004-6361/201117462}

\bibitem[{{Puccetti} {et~al.}(2014){Puccetti}, {Comastri}, {Fiore},
  {Ar{\'e}valo}, {Risaliti}, {Bauer}, {Brandt}, {Stern}, {Harrison},
  {Alexander}, {Boggs}, {Christensen}, {Craig}, {Gandhi}, {Hailey}, {Koss},
  {Lansbury}, {Luo}, {Madejski}, {Matt}, {Walton}, \& {Zhang}}]{Puccetti14}
{Puccetti}, S., {Comastri}, A., {Fiore}, F., {et~al.} 2014, \apj, 793, 26,
  \dodoi{10.1088/0004-637X/793/1/26}

\bibitem[{{Puccetti} {et~al.}(2016){Puccetti}, {Comastri}, {Bauer}, {Brand t},
  {Fiore}, {Harrison}, {Luo}, {Stern}, {Urry}, {Alexander}, {Annuar},
  {Ar{\'e}valo}, {Balokovi{\'c}}, {Boggs}, {Brightman}, {Christensen}, {Craig},
  {Gand hi}, {Hailey}, {Koss}, {La Massa}, {Marinucci}, {Ricci}, {Walton},
  {Zappacosta}, \& {Zhang}}]{Puccetti16}
{Puccetti}, S., {Comastri}, A., {Bauer}, F.~E., {et~al.} 2016, \aap, 585, A157,
  \dodoi{10.1051/0004-6361/201527189}

\bibitem[{{Ricci} {et~al.}(2015){Ricci}, {Ueda}, {Koss}, {Trakhtenbrot},
  {Bauer}, \& {Gandhi}}]{Ricci15}
{Ricci}, C., {Ueda}, Y., {Koss}, M.~J., {et~al.} 2015, \apj, 815, L13,
  \dodoi{10.1088/2041-8205/815/1/L13}

\bibitem[{{Ricci} {et~al.}(2017{\natexlab{a}}){Ricci}, {Trakhtenbrot}, {Koss},
  {Ueda}, {Schawinski}, {Oh}, {Lamperti}, {Mushotzky}, {Treister}, {Ho},
  {Weigel}, {Bauer}, {Paltani}, {Fabian}, {Xie}, \& {Gehrels}}]{Ricci17c}
{Ricci}, C., {Trakhtenbrot}, B., {Koss}, M.~J., {et~al.} 2017{\natexlab{a}},
  \nat, 549, 488, \dodoi{10.1038/nature23906}

\bibitem[{{Ricci} {et~al.}(2017{\natexlab{b}}){Ricci}, {Bauer}, {Treister},
  {Schawinski}, {Privon}, {Blecha}, {Arevalo}, {Armus}, {Harrison}, {Ho},
  {Iwasawa}, {Sanders}, \& {Stern}}]{Ricci17b}
{Ricci}, C., {Bauer}, F.~E., {Treister}, E., {et~al.} 2017{\natexlab{b}},
  \mnras, 468, 1273, \dodoi{10.1093/mnras/stx173}

\bibitem[{{Ricci} {et~al.}(2017{\natexlab{c}}){Ricci}, {Trakhtenbrot}, {Koss},
  {Ueda}, {Del Vecchio}, {Treister}, {Schawinski}, {Paltani}, {Oh}, {Lamperti},
  {Berney}, {Gand hi}, {Ichikawa}, {Bauer}, {Ho}, {Asmus}, {Beckmann}, {Soldi},
  {Balokovi{\'c}}, {Gehrels}, \& {Markwardt}}]{Ricci17a}
{Ricci}, C., {Trakhtenbrot}, B., {Koss}, M.~J., {et~al.} 2017{\natexlab{c}},
  \apjs, 233, 17, \dodoi{10.3847/1538-4365/aa96ad}

\bibitem[{{Ricci} {et~al.}(2018){Ricci}, {Ho}, {Fabian}, {Trakhtenbrot},
  {Koss}, {Ueda}, {Lohfink}, {Shimizu}, {Bauer}, {Mushotzky}, {Schawinski},
  {Paltani}, {Lamperti}, {Treister}, \& {Oh}}]{Ricci18}
{Ricci}, C., {Ho}, L.~C., {Fabian}, A.~C., {et~al.} 2018, \mnras, 480, 1819,
  \dodoi{10.1093/mnras/sty1879}

\bibitem[{{Ricci} {et~al.}(2021){Ricci}, {Privon}, {Pfeifle}, {Armus},
  {Iwasawa}, {Torres-Alb{\`a}}, {Satyapal}, {Bauer}, {Treister}, {Ho}, {Aalto},
  {Ar{\'e}valo}, {Barcos-Mu{\~n}oz}, {Charmandaris}, {Diaz-Santos}, {Evans},
  {Gao}, {Inami}, {Koss}, {Lansbury}, {Linden}, {Medling}, {Sanders}, {Song},
  {Stern}, {U}, {Ueda}, \& {Yamada}}]{Ricci21}
{Ricci}, C., {Privon}, G.~C., {Pfeifle}, R.~W., {et~al.} 2021, \mnras,
  \dodoi{10.1093/mnras/stab2052}

\bibitem[{{Rivers} {et~al.}(2015){Rivers}, {Balokovi{\'c}}, {Ar{\'e}valo},
  {Bauer}, {Boggs}, {Brandt}, {Brightman}, {Christensen}, {Craig}, {Gandhi},
  {Hailey}, {Harrison}, {Koss}, {Ricci}, {Stern}, {Walton}, \&
  {Zhang}}]{Rivers15}
{Rivers}, E., {Balokovi{\'c}}, M., {Ar{\'e}valo}, P., {et~al.} 2015, \apj, 815,
  55, \dodoi{10.1088/0004-637X/815/1/55}

\bibitem[{{Str{\"u}der} {et~al.}(2001){Str{\"u}der}, {Briel}, {Dennerl},
  {Hartmann}, {Kendziorra}, {Meidinger}, {Pfeffermann}, {Reppin}, {Aschenbach},
  {Bornemann}, {Br{\"a}uninger}, {Burkert}, {Elender}, {Freyberg}, {Haberl},
  {Hartner}, {Heuschmann}, {Hippmann}, {Kastelic}, {Kemmer}, {Kettenring},
  {Kink}, {Krause}, {M{\"u}ller}, {Oppitz}, {Pietsch}, {Popp}, {Predehl},
  {Read}, {Stephan}, {St{\"o}tter}, {Tr{\"u}mper}, {Holl}, {Kemmer}, {Soltau},
  {St{\"o}tter}, {Weber}, {Weichert}, {von Zanthier}, {Carathanassis}, {Lutz},
  {Richter}, {Solc}, {B{\"o}ttcher}, {Kuster}, {Staubert}, {Abbey}, {Holland},
  {Turner}, {Balasini}, {Bignami}, {La Palombara}, {Villa}, {Buttler},
  {Gianini}, {Lain{\'e}}, {Lumb}, \& {Dhez}}]{Struder01}
{Str{\"u}der}, L., {Briel}, U., {Dennerl}, K., {et~al.} 2001, \aap, 365, L18,
  \dodoi{10.1051/0004-6361:20000066}

\bibitem[{{Tanimoto} {et~al.}(2016){Tanimoto}, {Ueda}, {Kawamuro}, \&
  {Ricci}}]{Tanimoto16}
{Tanimoto}, A., {Ueda}, Y., {Kawamuro}, T., \& {Ricci}, C. 2016, Publications
  of the Astronomical Society of Japan, 68, S26, \dodoi{10.1093/pasj/psw008}

\bibitem[{{Tanimoto} {et~al.}(2018){Tanimoto}, {Ueda}, {Kawamuro}, {Ricci},
  {Awaki}, \& {Terashima}}]{Tanimoto18}
{Tanimoto}, A., {Ueda}, Y., {Kawamuro}, T., {et~al.} 2018, \apj, 853, 146,
  \dodoi{10.3847/1538-4357/aaa47c}

\bibitem[{{Tanimoto} {et~al.}(2019){Tanimoto}, {Ueda}, {Odaka}, {Kawaguchi},
  {Fukazawa}, \& {Kawamuro}}]{Tanimoto19}
{Tanimoto}, A., {Ueda}, Y., {Odaka}, H., {et~al.} 2019, \apj, 877, 95,
  \dodoi{10.3847/1538-4357/ab1b20}

\bibitem[{{Tanimoto} {et~al.}(2020){Tanimoto}, {Ueda}, {Odaka}, {Ogawa},
  {Yamada}, {Kawaguchi}, \& {Ichikawa}}]{Tanimoto20}
---. 2020, \apj, 897, 2, \dodoi{10.3847/1538-4357/ab96bc}

\bibitem[{{Toba} {et~al.}(2020){Toba}, {Yamada}, {Ueda}, {Ricci}, {Terashima},
  {Nagao}, {Wang}, {Tanimoto}, \& {Kawamuro}}]{Toba20}
{Toba}, Y., {Yamada}, S., {Ueda}, Y., {et~al.} 2020, \apj, 888, 8,
  \dodoi{10.3847/1538-4357/ab5718}

\bibitem[{{Torres-Alb{\`a}} {et~al.}(2021){Torres-Alb{\`a}}, {Marchesi},
  {Zhao}, {Ajello}, {Silver}, {Ananna}, {Balokovi{\'c}}, {Boorman}, {Comastri},
  {Gilli}, {Lanzuisi}, {Murphy}, {Urry}, \& {Vignali}}]{Torres21}
{Torres-Alb{\`a}}, N., {Marchesi}, S., {Zhao}, X., {et~al.} 2021, \apj, 922,
  252, \dodoi{10.3847/1538-4357/ac1c73}

\bibitem[{{Tueller} {et~al.}(2008){Tueller}, {Mushotzky}, {Barthelmy},
  {Cannizzo}, {Gehrels}, {Markwardt}, {Skinner}, \& {Winter}}]{Tueller08}
{Tueller}, J., {Mushotzky}, R.~F., {Barthelmy}, S., {et~al.} 2008, \apj, 681,
  113, \dodoi{10.1086/588458}

\bibitem[{{Tueller} {et~al.}(2010){Tueller}, {Baumgartner}, {Markwardt},
  {Skinner}, {Mushotzky}, {Ajello}, {Barthelmy}, {Beardmore}, {Brandt},
  {Burrows}, {Chincarini}, {Campana}, {Cummings}, {Cusumano}, {Evans},
  {Fenimore}, {Gehrels}, {Godet}, {Grupe}, {Holland}, {Kennea}, {Krimm},
  {Koss}, {Moretti}, {Mukai}, {Osborne}, {Okajima}, {Pagani}, {Page}, {Palmer},
  {Parsons}, {Schneider}, {Sakamoto}, {Sambruna}, {Sato}, {Stamatikos},
  {Stroh}, {Ukwata}, \& {Winter}}]{Tueller10}
{Tueller}, J., {Baumgartner}, W.~H., {Markwardt}, C.~B., {et~al.} 2010, The
  Astrophysical Journal Supplement Series, 186, 378,
  \dodoi{10.1088/0067-0049/186/2/378}

\bibitem[{{Tully} {et~al.}(2009){Tully}, {Rizzi}, {Shaya}, {Courtois},
  {Makarov}, \& {Jacobs}}]{Tully09}
{Tully}, R.~B., {Rizzi}, L., {Shaya}, E.~J., {et~al.} 2009, \aj, 138, 323,
  \dodoi{10.1088/0004-6256/138/2/323}

\bibitem[{{Turner} {et~al.}(2001){Turner}, {Abbey}, {Arnaud}, {Balasini},
  {Barbera}, {Belsole}, {Bennie}, {Bernard}, {Bignami}, {Boer}, {Briel},
  {Butler}, {Cara}, {Chabaud}, {Cole}, {Collura}, {Conte}, {Cros}, {Denby},
  {Dhez}, {Di Coco}, {Dowson}, {Ferrando}, {Ghizzardi}, {Gianotti}, {Goodall},
  {Gretton}, {Griffiths}, {Hainaut}, {Hochedez}, {Holland}, {Jourdain},
  {Kendziorra}, {Lagostina}, {Laine}, {La Palombara}, {Lortholary}, {Lumb},
  {Marty}, {Molendi}, {Pigot}, {Poindron}, {Pounds}, {Reeves}, {Reppin},
  {Rothenflug}, {Salvetat}, {Sauvageot}, {Schmitt}, {Sembay}, {Short},
  {Spragg}, {Stephen}, {Str{\"u}der}, {Tiengo}, {Trifoglio}, {Tr{\"u}mper},
  {Vercellone}, {Vigroux}, {Villa}, {Ward}, {Whitehead}, \& {Zonca}}]{Turner01}
{Turner}, M.~J.~L., {Abbey}, A., {Arnaud}, M., {et~al.} 2001, \aap, 365, L27,
  \dodoi{10.1051/0004-6361:20000087}

\bibitem[{{Ueda} {et~al.}(2014){Ueda}, {Akiyama}, {Hasinger}, {Miyaji}, \&
  {Watson}}]{Ueda14}
{Ueda}, Y., {Akiyama}, M., {Hasinger}, G., {Miyaji}, T., \& {Watson}, M.~G.
  2014, \apj, 786, 104, \dodoi{10.1088/0004-637X/786/2/104}

\bibitem[{{Ueda} {et~al.}(2003){Ueda}, {Akiyama}, {Ohta}, \& {Miyaji}}]{Ueda03}
{Ueda}, Y., {Akiyama}, M., {Ohta}, K., \& {Miyaji}, T. 2003, \apj, 598, 886,
  \dodoi{10.1086/378940}

\bibitem[{{Ueda} {et~al.}(1998){Ueda}, {Takahashi}, {Inoue}, {Tsuru}, {Sakano},
  {Ishisaki}, {Ogasaka}, {Makishima}, {Yamada}, {Ohta}, \& {Akiyama}}]{Ueda98}
{Ueda}, Y., {Takahashi}, T., {Inoue}, H., {et~al.} 1998, \nat, 391, 866,
  \dodoi{10.1038/36047}

\bibitem[{{Ueda} {et~al.}(1999){Ueda}, {Takahashi}, {Inoue}, {Tsuru}, {Sakano},
  {Ishisaki}, {Ogasaka}, {Makishima}, {Yamada}, {Akiyama}, \& {Ohta}}]{Ueda99}
---. 1999, \apj, 518, 656, \dodoi{10.1086/307291}

\bibitem[{{Ueda} {et~al.}(2007){Ueda}, {Eguchi}, {Terashima}, {Mushotzky},
  {Tueller}, {Markwardt}, {Gehrels}, {Hashimoto}, \& {Potter}}]{Ueda07}
{Ueda}, Y., {Eguchi}, S., {Terashima}, Y., {et~al.} 2007, \apj, 664, L79,
  \dodoi{10.1086/520576}

\bibitem[{{Uematsu} {et~al.}(2021){Uematsu}, {Ueda}, {Tanimoto}, {Kawamuro},
  {Setoguchi}, {Ogawa}, {Yamada}, \& {Odaka}}]{Uematsu21}
{Uematsu}, R., {Ueda}, Y., {Tanimoto}, A., {et~al.} 2021, arXiv e-prints,
  arXiv:2103.11224.
\newblock \doarXiv{2103.11224}

\bibitem[{{van den Bosch}(2016)}]{Bosch16}
{van den Bosch}, R. C.~E. 2016, \apj, 831, 134,
  \dodoi{10.3847/0004-637X/831/2/134}

\bibitem[{{Weisskopf} {et~al.}(2002){Weisskopf}, {Brinkman}, {Canizares},
  {Garmire}, {Murray}, \& {Van Speybroeck}}]{Weisskopf02}
{Weisskopf}, M.~C., {Brinkman}, B., {Canizares}, C., {et~al.} 2002, \pasp, 114,
  1, \dodoi{10.1086/338108}

\bibitem[{{Willingale} {et~al.}(2013){Willingale}, {Starling}, {Beardmore},
  {Tanvir}, \& {O'Brien}}]{Willingale13}
{Willingale}, R., {Starling}, R.~L.~C., {Beardmore}, A.~P., {Tanvir}, N.~R., \&
  {O'Brien}, P.~T. 2013, \mnras, 431, 394, \dodoi{10.1093/mnras/stt175}

\bibitem[{{Yamada} {et~al.}(2021){Yamada}, {Ueda}, {Tanimoto}, {Imanishi},
  {Toba}, {Ricci}, \& {Privon}}]{Yamada21}
{Yamada}, S., {Ueda}, Y., {Tanimoto}, A., {et~al.} 2021, \apjs, 257, 61,
  \dodoi{10.3847/1538-4365/ac17f5}

\bibitem[{{Yamada} {et~al.}(2020){Yamada}, {Ueda}, {Tanimoto}, {Oda},
  {Imanishi}, {Toba}, \& {Ricci}}]{Yamada20}
---. 2020, \apj, 897, 107, \dodoi{10.3847/1538-4357/ab94b1}

\bibitem[{{Zhao} {et~al.}(2019{\natexlab{a}}){Zhao}, {Marchesi}, \&
  {Ajello}}]{Zhao19b}
{Zhao}, X., {Marchesi}, S., \& {Ajello}, M. 2019{\natexlab{a}}, \apj, 871, 182,
  \dodoi{10.3847/1538-4357/aaf80b}

\bibitem[{{Zhao} {et~al.}(2020){Zhao}, {Marchesi}, {Ajello}, {Balokovi{\'c}},
  \& {Fischer}}]{Zhao20}
{Zhao}, X., {Marchesi}, S., {Ajello}, M., {Balokovi{\'c}}, M., \& {Fischer}, T.
  2020, \apj, 894, 71, \dodoi{10.3847/1538-4357/ab879d}

\bibitem[{{Zhao} {et~al.}(2019{\natexlab{b}}){Zhao}, {Marchesi}, {Ajello},
  {Marcotulli}, {Cusumano}, {La Parola}, \& {Vignali}}]{Zhao19a}
{Zhao}, X., {Marchesi}, S., {Ajello}, M., {et~al.} 2019{\natexlab{b}}, \apj,
  870, 60, \dodoi{10.3847/1538-4357/aaf1a0}

\end{thebibliography}
\end{document}